\documentclass[useAMS,usenatbib]{mn2e}

\usepackage{graphicx,epsfig,times,color,amsmath}

\usepackage{subfig}

\usepackage{amssymb} 

\usepackage{enumerate}

\usepackage{url}

\def\Msun{\mbox{~M$_\odot$}}
\def\Msunns{\mbox{M$_\odot$}}
\def\kms{\mbox{~km~s$^{-1}$}}
\def\kmsns{\mbox{km~s$^{-1}$}}
\def\pc{\mbox{~pc}}
\def\kpc{\mbox{~kpc}}

\def\Mpch{\mbox{$~h^{-1}$ Mpc}}

\def\M2vir{M_{\rm 2vir}}

\def\LCDM{$\Lambda$CDM }

\def\mathnew{\mathsurround=0pt}
\def\simov#1#2{\lower .5pt\vbox{\baselineskip0pt
    \lineskip-.5pt\ialign{$\mathnew#1\hfil##\hfil$\crcr#2\crcr\sim\crcr}}}

\def\lesssim{\mathrel{\mathpalette\simov <}}

\title[Regulation of stellar mass assembly by radiation feedback]{Low-mass galaxy assembly in simulations: regulation of early star formation by radiation from massive stars}
\author[S. Trujillo-Gomez et al.]
{Sebastian Trujillo-Gomez$^{1}$\thanks{st@astronomy.nmsu.edu},
Anatoly Klypin$^1$,
Pedro Col\'{i}n$^2$,
Daniel Ceverino$^3$, \and  
Kenza S. Arraki$^1$ and 
Joel Primack$^4$\\
$^1$Astronomy Department, New Mexico State University, Las Cruces, NM 88003, USA\\
$^2$Centro de Radioastronomia y Astrof\'{i}sica. Universidad Nacional Aut\'{o}noma de M\'{e}xico, A.P. 72-3 (Xangari), Morelia, Michoac\'{a}n 58089, M\'{e}xico\\
$^3$Departamento de F\'{i}sica Te\'{o}rica, Universidad Aut\'{o}noma de Madrid, 28049 Madrid, Spain \\
$^4$Department of Physics, University of California at Santa Cruz, Santa Cruz, CA 95064, USA  }

\begin{document}

\date{Submitted to MNRAS, November 2013}

\pagerange{\pageref{firstpage}--\pageref{lastpage}} \pubyear{2013}

\maketitle

\label{firstpage}

\begin{abstract}

Despite recent success in forming realistic present-day galaxies, simulations still form the bulk of their stars earlier than observations indicate. We investigate the process of stellar mass assembly in low-mass field galaxies, 
a dwarf and a typical spiral, focusing on the effects of radiation from young stellar clusters on the star formation histories. We implement a novel model of star formation (SF) with a deterministic low efficiency per free-fall time, as observed in molecular clouds. Stellar feedback is based on observations of   star-forming regions, and includes radiation pressure from massive stars, photoheating in H\,{\sc ii} regions, supernovae, and stellar winds.  We find that stellar radiation has a strong effect on the formation of low-mass galaxies, especially at $z>1$, where it efficiently suppresses SF by dispersing cold and dense gas,  preventing runaway growth of the stellar component.  This behaviour is evident in a variety of observations but had so far eluded analytical and numerical models without radiation feedback. 
Compared to supernovae alone, radiation feedback reduces the SF rate by a factor of $\sim 100$ at $z \la 2$, 
yielding rising SF histories which reproduce recent observations of Local Group dwarfs. Stellar radiation also produces bulgeless spiral galaxies and may be responsible for excess thickening of the stellar disc. 
The galaxies also feature rotation curves and baryon fractions in  excellent agreement with current data. Lastly, the dwarf galaxy shows a very slow  reduction of the central dark matter density caused by radiation feedback over the last $\sim 7~\rm{Gyr}$ of cosmic evolution. 

\end{abstract}


\section{Introduction}
\label{sec:intro}

In the last few years, the \LCDM  cosmological paradigm has been
extensively and successfully tested at various scales and now
provides a well defined framework in which to study the process of
galaxy formation. 
As observations of galaxies become more detailed
and reach farther into the past, the theory is confronted with a large
number of constraints both at present and at high
redshift. Cosmological simulations of individual galaxies have only
very recently reached the level of detail and resolution necessary to properly include
the physical processes that shape galaxies. Although it now seems
possible to fine-tune physical models to produce individual simulated
galaxies that resemble the Milky Way (MW) and present day dwarfs
\citep[e.g.,][]{governato10,guedes11,Brook11,Shen13}, much progress is yet to come in our
understanding of galaxy formation. Two essential properties of galaxies are absent from 
numerical models \citep[see, however,][]{Brook12}. First, the observed fraction of the universal baryons that condense to form stars in  
galaxies is very small and has a steep mass dependence \citep[e.g,][]{Trujillo-Gomez11}. Second, there is a sharp contrast 
between the growth of dark matter (DM) haloes and the growth of galaxies within them \citep{Weinmann12}. Observational estimates 
of the stellar mass growth of galaxies show that low-mass galaxies formed most of their stars 
recently, while massive ones assembled a large fraction of their stellar mass at high redshift \citep[e.g.,][]{Salim07,Firmani10b,Behroozi12,Leitner12,Moster12}. 

Several lines of evidence
including lensing, satellite kinematics and semi-empirical models indicate that galaxy formation is 
a very inefficient process, where
most of the cosmological baryons (i.e., the primordial gas) do not condense into galaxies
\citep[e.g.,][]{fukugita98,Hoekstra05,mandelbaum06,Jiang07,guo10,behroozi10,
Trujillo-Gomez11,Rodriguez-Puebla12,Leauthaud12}. Previous
works have outlined the need for efficient stellar feedback to expel
gas from galaxies. However, this process should have a strong mass
dependence in order to reproduce the observed cold baryon fractions of
all galaxies. A dependence of the baryon fraction
on morphology has also been suggested
\citep[e.g.,][]{dutton10a,Trujillo-Gomez11}. After decades of producing galaxies with too many stars, several simulations have recently succeeded at matching observed stellar masses \citep[e.g.,][]{guedes11,Munshi13,DiCintio14}. These typically rely on increasing the effect of supernovae feedback by boosting the energy yield or by extending the disipantionless phase of the expansion to several million years at the scales of molecular clouds. However, it is uncertain whether these large superheated bubbles actually occur in nature. 


As observations of the high-redshift universe improved, it was realized that simulated galaxies assemble their 
stars much earlier than evidence suggests. Studies of the assembly
of the stellar component of galaxies have found evidence of the phenomenon termed
``downsizing'' . Direct measurements of the specific star
formation rates (sSFR) in dwarf galaxies indicate that they increase with
decreasing stellar mass, both at present and out to $z \sim 2$ \citep[e.g.,][]{Baldry04,Noeske07a,Salim07,Rodighiero10,Karim11,Whitaker12}. 
Stellar mass
growth tracks obtained from semi-empirical models also produce sSFRs
that are higher for lower mass galaxies \citep{Firmani10b,Behroozi12}. This implies
that dwarfs assembled their stellar component later than massive galaxies, and the smaller their 
mass, the
greater the delay 
\citep[e.g.,][]{Firmani10b,Avila-Reese11}. \citet{Avila-Reese11}, \citet{deRossi13}, and \citep{GonzalezSamaniego13} show that
cosmological galaxy formation simulations that include only supernovae feedback are incapable of producing rising
star formation histories in dwarfs, even when strong outflows are imposed. In addition, \citet{Firmani10a} show that
semi-numerical models of disc formation with outflows and re-accretion that are tuned to reproduce the present-day stellar
mass-halo mass relation give rise to an increasing specific star formation rate as a
function of mass, opposite to the observed trend. This problem is also present in
semi-analytic models \citep[e.g.,][]{Somerville08}, which obtain a
population of dwarf galaxies that is old and
quenched. This problem is not limited to low-mass haloes. Fine-tuned simulations of MW-like 
galaxies also form most of their stars at high redshift. For instance, the {\it Eris2} simulation is 
intended to reproduce the Milky Way at present but has a stellar mass $M_* = 1.5\times10^{10}\Msun$ 
already in place at $z=3$ \citep{Shen12}. This is more than ten times larger than in the observational inferences \citep[e.g.,][]{Behroozi12,Moster12}. Clearly, the models currently lack a mechanism 
to decouple the
galaxy growth from the dark matter halo growth \citep{Weinmann12}. 

Recently, forms of feedback that are fundamentally different from
supernovae have been shown to be essential to galaxy
formation. \citet{Murray10} analyzed the dynamical effects of several forms of
stellar feedback on parent molecular clouds. In their models they
include momentum input from ionized gas in H\,{\sc ii} regions, shocked stellar winds, hot gas pressure, 
protostellar
jets and cosmic rays. \citet{Murray10} conclude that radiation pressure (RP) on dust
grains is likely
to be the
dominant form of feedback in star-forming galaxies. A variety of other
studies have reached the same conclusions, placing the combination of
radiation pressure and photoionization of gas by massive stars as the
dominant mechanism for disruption of molecular clouds and internal
regulation of the star formation process
\citep[e.g.,][]{KrumholzMatzner09,Indebetouw09,Murray10,Andrews11,Lopez11,Pellegrini11,Hopkins11}. Radiation
pressure alone might also be the only mechanism that explains galactic
fountains and the warm gas outflows observed in absorption in high
redshift galaxies \citep{Murray11}. In addition, recent numerical work by
\citet{Krumholz12} shows that radiation feedback fully accounts
for the large gas velocity dispersions measured in young star clusters in the
MW. There are at least three reasons why
radiative feedback is an essential ingredient of the galaxy formation
process. First, observations show that molecular clouds begin to disperse shortly after the O stars form and before the first supernovae explode and deposit their energy
into the gas \citep{Kawamura09}. Second, the total energy output of a stellar cluster is
dominated by radiation. The rate of radiative energy output by O and B stars is $\sim 200$ times 
larger
than the average power injected by supernovae and stellar winds during the lifetime of the most
massive stars. Third, it is difficult to explain the large gas
turbulence values observed in star-forming regions without including
the momentum input by radiation \citep{Murray10}. 

There have been few attempts to incorporate radiation pressure from
young stellar clusters in numerical models of galaxy formation. In most cases, the effect
of radiation is crudely modeled as a simple increase in the
thermal energy output of massive stars \citep[similar to supernovae (SN) thermal feedback;][]{Brook11,Maccio12} 
or by
imparting kinetic energy to the gas while temporarily decoupling it
from hydrodynamic forces \citep[e.g.,][]{Oppenheimer08,Oppenheimer10,Genel12}. In the most 
detailed
approach, \citet{Hopkins11} performed high resolution Smoothed Particle Hydrodynamics (SPH) simulations
of isolated galaxy models to follow the effects of radiation pressure
using self-consistent star cluster identification and optical depth
calculation. They show that radiative feedback may be the mechanism
responsible for controlling the amount of gas available for star
formation in a manner that becomes independent of the sub-grid star
formation parameters. In their isolated models this resulted in a
drastic reduction of the SF efficiency.

\citet{Hopkins11} show that simulated isolated galaxies
can self-regulate their star formation with radiative feedback alone,
without the need for supernova explosions. In a recent paper, \citet{Ceverino13} include the effect of radiation pressure and photoionization from UV photons in cosmological simulations of a high-redshift Milky Way progenitor. Their simulations with radiation feedback show a reduction of the star formation rate by a factor of $\sim 2-3$ at $z=3$ compared to a simulation with supernovae energy alone, with radiation momentum playing the dominant role and gas photoionization having only a secondary effect. These works point toward a new paradigm of galaxy formation, 
where
radiation from massive stars is
responsible for regulating star formation and powering galactic winds,
while the properties of the inter-stellar medium (ISM) are controlled by the energy from
supernovae. 


In this paper we use cosmological simulations with a new, realistic star formation model and radiative feedback from massive stars to investigate the assembly of baryons in low-mass galaxies. Unless specified otherwise, we employ the term ``radiative feedback" to refer to both the  effect of scattering of stellar radiation by gas and dust, and to the thermal pressure of gas that is photoionized by stellar UV photons. Hereon we use the term ``radiation pressure" to denote only the pressure due to scattering of photons on gas and dust. The paper is organised as follows. Section~\ref
{sec:model} describes the code and the new model of star formation and feedback and Section~\ref{sec:analytics} shows estimates of the effects of feedback on star-forming gas. Section~\ref{sec:simulations} summarizes the properties of the simulations and Section~\ref{sec:results} presents our results. In Section~\ref{sec:conclusions} we summarize our results.

\section{Physical model}
\label{sec:model}


To perform our simulations we used the adaptive mesh refinement $N$-body+hydrodynamics 
code {\it hydroART} \citep{Kravtsov97,Kravtsov99}. The code is adaptive in both space and time, 
achieving higher resolution in regions of higher mass density. The physical model in the code 
includes many relevant physical processes such as cooling due to metals and molecules down to 
$300~\rm{K}$, a homogeneous ultraviolet (UV) background, gas self-shielding,  as well as advection of metals. We now describe the motivation and implementation of the processes of star formation and feedback.

\subsection{Stellar feedback}
\label{sec:feedback}

Our stellar feedback model includes contributions from four dominant terms:  radiation pressure, photoionization heating, type II supernovae, and shocked stellar winds.

\begin{figure}
 \includegraphics[width=0.49\textwidth]{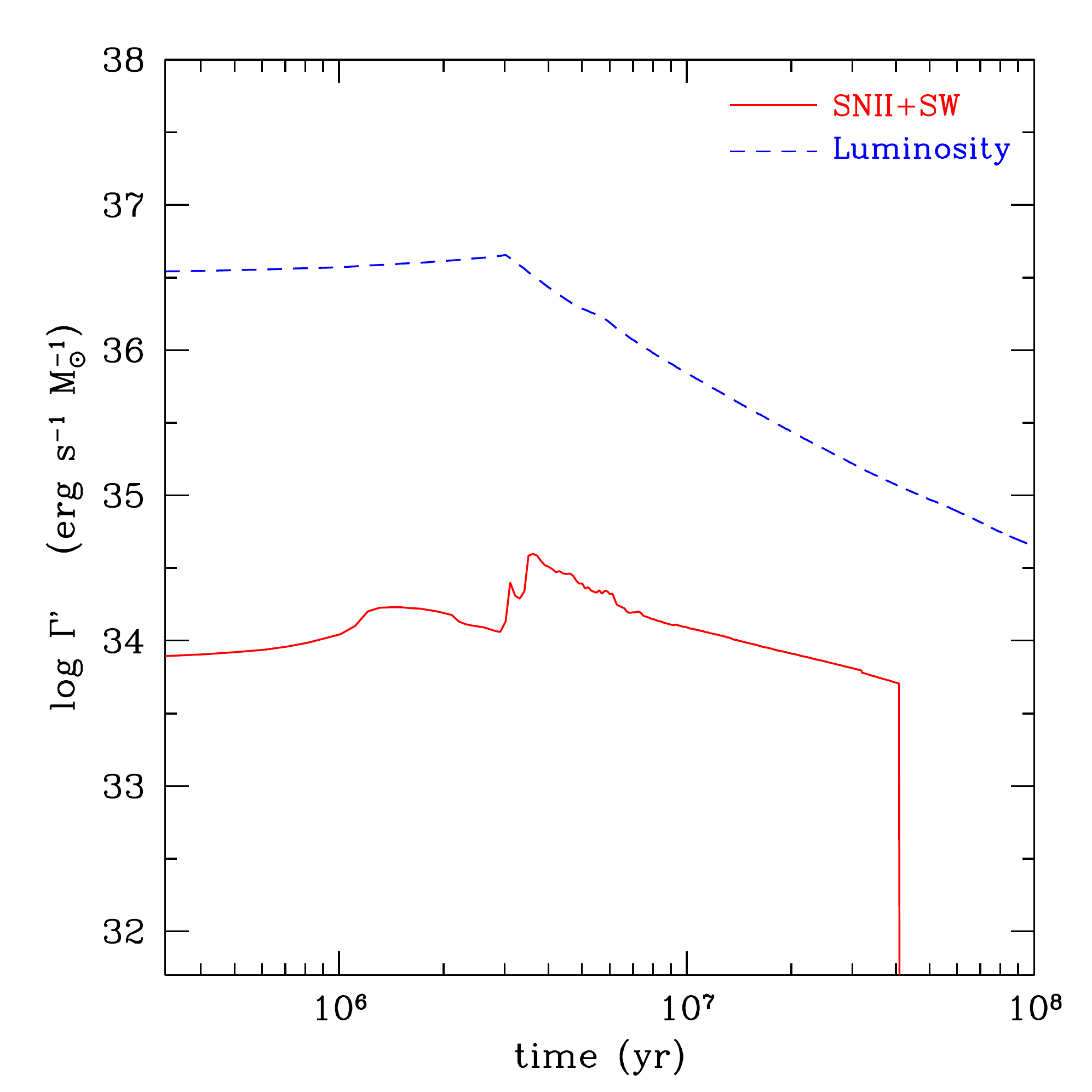}
 \caption{Specific power output by a single burst of star formation as a function of time as calculated by 
{\sc starburst99}. The dashed line shows the specific stellar luminosity per unit stellar mass, while the solid line is the specific mechanical power from type II supernovae and stellar 
winds. The radiative 
energy is at least two hundred times larger than the combined energy from supernova explosions 
and stellar winds during the initial $\sim 3$~Myr and declines quickly thereafter. Most galaxy formation simulations do not include 
this contribution to stellar feedback.} 
 \label{sb}
\end{figure}

{\it Radiation Pressure.} The treatment of stellar radiation pressure follows \citet{Murray10}, 
\citet{Hopkins11}, \citet{Agertz12}, and \citet{Ceverino13}, where the total momentum of the radiation field (in the form of UV and optical photons),
\begin{equation} 
\dot{p}_{\rm rad}(t) = \frac{L(t)}{c} , 
\end{equation}
is coupled to the gas and dust as a result of scattering and/or absorption. 
Figure~\ref{sb} shows the total luminosity output from a burst of star formation obtained using {\sc starburst99} \citep{starburst99}. During the initial $40$ million years, the total energy radiated by a star cluster is $1.8\times10^{50}~\rm{erg}\Msun^{-1}$, or about ten times larger than the mechanical energy released by supernovae and stellar winds. The total momentum of the radiation couples to the gas through two separate processes. Initially, a photon of any wavelength may be scattered or absorbed by an atom. In addition, UV and optical photons may also be absorbed by dust. If a photon is absorbed by a dust grain, it will be re-emitted as a lower energy infrared photon. Depending on the density structure of the star-forming cloud, and on the optical depth of dust in the infrared, the photon may be absorbed and re-emitted multiple times. Through collisions between atoms and grains, this process will enhance the amount of momentum transferred to the  star-forming cloud. Thus, the total momentum imparted to the gas will be
\begin{equation} 
\dot{p}(t) \approx (1 - e^{-\tau_{\rm UV}} + \tau_{\rm IR}) \frac{L(t)}{c} .
\label{eq:fullradmomentum}
\end{equation}
Because the gas immediately surrounding
a newly formed star cluster reaches high densities, and dust is an
efficient absorber of UV photons, we assume $1 - \exp(-\tau_{\rm
UV}) \approx 1$. The effect of radiation momentum on the gas will vanish 
once the expanding gas shell grows larger than the size of a giant molecular cloud (GMC) and its optical depth decreases \citep
{Murray10}.

There is some debate in the literature over the value of the infrared optical depth of the gas due to dust. While $\tau_{\rm IR}$ can reach large values ($\sim 50-100$) in galaxy models without radiative transfer 
\citep{Murray10,Hopkins11},
self-regulation seems to make the properties of the ISM nearly independent of
the strength of the radiative forcing. However, in simplified radiation
hydrodynamics simulations, \citet{Krumholz12} find that although there is
enough momentum in the radiation field, once the
radiation pushes on the gas strongly enough for radiative
Raleigh-Taylor instabilities to set in, radiation becomes dynamically
ineffective and the Eddington ratio of the gas saturates at unity. They obtain modest values $1 < \tau_{\rm 
IR} < 10$. Most recently, \citet{Davis14} show that using a more accurate radiative transfer approximation allows the radiation to be trapped, resulting in large optical depths that scale linearly with the column density estimated without radiative transfer. The local metallicity of the gas is also expected to affect the dust content and hence its absorption properties. Due to the current uncertainty in the optical depth $\tau_{\rm IR}$, we 
explore a broad range of values in our numerical experiments.

{\it Photoionization Heating.} In addition to imparting momentum through scattering, UV radiation ionizes the gas surrounding a young cluster, creating an H\,{\sc ii} region. The observed extent of \emph{evolved} H\,{\sc ii} regions ranges from $\sim 10 - 100~\rm{pc}$, with temperatures of $\sim 10^4~\rm{K}$, and densities $50 - 1000~\rm{cm}^{-3}$ \citep{Lopez13}. Thus, the thermal pressure of gas due to photoionization within the H{\sc ii} region is  $P/\rm{k}_{B} \sim 10^6 - 10^7~\rm{K~cm}^{-3}$ a few million years after the first massive stars form \citep{Lopez13}. However, the density of ionized gas is expected to be higher at the time of formation of the first O and B stars and thus the thermal pressure should be considerably larger during the initial stages\footnote{Ultra-compact H\,{\sc ii} regions are thought to be the progenitors of evolved H\,{\sc ii} regions. They are much more compact and dense than their evolved counterparts, with $n \ga 10^5~\rm{cm}^{-3}$ \citep{Hoare07}. }. Since the region is overpressured with respect to its surrounding gas, it will expand and become rarefied, causing a decrease in the thermal gas pressure with time.  

Very few galaxy formation simulations, including those that attempt a full accounting of stellar feedback \citep[e.g.,][]{Agertz12,Hopkins13}, include the extra pressure from photoheating explicitly. However, in high resolution simulations of star formation in isolated molecular clouds, \citet{Dale12} and \citet{Colin13} find that ionization heating alone is sufficient to destroy massive clouds and to regulate the star formation efficiency.

{\it SNe and Stellar Winds.} As is commonly done in simulations of galaxy formation, we include the mechanical energy from unresolved type II supernovae shockwaves and shocked stellar winds. Following the model in \citep{Ceverino10}, this energy is injected in the gas as thermal energy. As Figure~\ref{sb} shows, there is a $3~{\rm Myr}$ delay after the formation of a cluster and before the first supernovae explode. During this period, stellar winds provide an amount of power equivalent to the SNII contribution. In addition, the contribution of type Ia supernovae is modeled as an injection of thermal energy with a broad distribution in time that peaks at a stellar age of $1~\rm{Gyr}$. Since the energy deposition rate is about 3 orders of magnitude smaller than the SNII contribution, it should be dynamically unimportant for galaxies with ongoing star formation. 
\\

We implement each feedback component in the code as follows:

\begin{enumerate}[1.]

\item The rate of momentum injection from the radiation field into the gas is
\begin{equation} 
\dot{p}_{\rm rad}(t) \equiv  \tau_{\rm tot} \frac{L(t)}{c},
\label{eq:radmomentum}
\end{equation} where $\tau_{\rm tot} \equiv 1 + \tau_{\rm IR}$, and $L(t)$
is the total luminosity of the star cluster as a function of time estimated
using {\sc starburst99} for a Chabrier IMF. Assuming conservation of momentum, we obtain the pressure on the surrounding gas by computing the momentum flux through the cell interface:
\begin{equation} 
  P_{\rm rad}(t) = \eta \ \frac{\dot{p}_{\rm rad}(t)}{6 \Delta x^2} ,
\label{eq:radpressure}
\end{equation} 
where the term $6 \Delta x^2$ is the surface area of the cell where the stellar particle is located. We set the parameter $\eta = 3$ to approximately match the correction used by \citet{Agertz12}, which accounts for numerical momentum loss when a star particle moves across grid cells. In general, the model is independent of resolution since by construction the total momentum of the radiation is conserved as it flows out of the cell. A detailed discussion of convergence of the model can be found in \citet{Ceverino13}. We adopt the initial specific radiant energy injection rate $L_0/M_* = 3.66\times10^{36}~\rm{erg}~\rm{s}
^{-1}\Msun^{-1}$, and a time dependence modeled as shown in Figure~\ref{sb}. We assume a fixed value $\tau_{\rm
UV} = 1$ wherever gas has the conditions for stars to form. The radiation pressure is added as a non-thermal pressure in every
star-forming cell until the gas density in the cell drops below a fixed threshold value, $n_{\rm th}$. Since our typical resolution limits the 
highest gas density in SF regions, we cannot estimate the optical depth directly using the column density of the gas. Instead, we select a fixed $\tau_{\rm IR}$ for each simulation run. The values are selected to span the range of estimates in the literature.

\item The dynamical effect of heating due to photoionization (i.e., photoheating) is included by adding an extra non-thermal pressure to the cell containing the young stellar particle during the initial $\sim 1~\rm{Myr}$ of the lifetime of the particle. The fiducial value of this pressure is chosen to be $P_{{\rm PH},0}/\rm{k}_B = 10^6~\rm{K~cm}^{-3}$ in the initial $100~{\rm kyr}$ and rapidly declines to $\sim 10^4~\rm{K~cm}^{-3}$ after $800~\rm{kyr}$ to mimic the drop in density as the H\,{\sc ii} region grows. The time dependence is modelled as
\begin{equation}
  P_{\rm PH}(t) = 
  \begin{cases} 
    P_{{\rm PH},0} & \text{for } t \leq 100~{\rm kyr},\\ \\
    \frac{P_{{\rm PH},0}}{(t/100~{\rm kyr})^{2.5}} & \text{for } 100~{\rm kyr} < t < 1~{\rm Myr},\\ 
  \end{cases}
\end{equation}  
In our simulations we find that the dynamical effect of this pressure does not depend on the numerical implementation of the steep decline. Note that the chosen value of the initial pressure is close to that measured in \emph{evolved} H\,{\sc ii} regions. In this sense, it is the most conservative estimate of the dynamical effect of photoheating due to massive stars.

\item Thermal energy from SNII is deposited in the cell containing a star particle of age less than $40~\rm{Myr}$ at a constant rate $
\Gamma'  = 1.59\times10^{34}~\rm{erg}~\rm{s}^{-1}\Msun^{-1}$ obtained by modelling a simple stellar population using the stellar evolution code {\sc starburst99} for a Chabrier initial mass function (IMF). The energy is injected for a period of $40~\rm{Myr}$ without 
artificially preventing gas cooling. The SNIa thermal energy injection rate is about 3 orders of magnitude lower than the rate for SNII. It peaks $1~\rm{Gyr}$ after the formation of a star particle and declines over several billion years. Further details of the SN feedback employed here can be found in \citet{Ceverino09}.

\end{enumerate}

Although the model described above is an extrapolation from sub-parsec scales to the 
tens-of-parsec scales resolved in our simulations, it incorporates several features that are 
fundamentally different from sub-grid feedback implementations. As mentioned above, 
the power contained in the radiation field is hundreds of times larger than in SN explosions 
and stellar winds. In addition, radiative forcing occurs as 
soon as the star cluster forms and drops rapidly once supernovae begin to explode. Moreover, the model contains essentially no free parameters and is based on physical principles. The parameters that cannot be modelled are taken directly from observations. Thus, our implementation of stellar feedback is a step 
towards a physical rather than a phenomenological approach to forming galaxies in numerical 
simulations.

\subsection{Star formation}

To find a suitable model of star 
formation we again turn toward observations of star formation in molecular clouds. \citet
{Krumholz&Tan07} showed that at scales of $1-100\pc$ molecular clouds form stars at a rate that 
is proportional to the gas density divided by the free-fall time. For reasons that are yet not fully 
understood, this process is very slow, with only about $1-3$ per cent of the gas consumed in one free-
fall time in galactic molecular clouds. For the most active star-forming complexes in the Milky Way 
-- which are responsible for one third of the total star formation -- \citet{Murray11b} find that the 
mass-weighted efficiency per free-fall time may be as high as $8$ per cent. Since our simulations resolve regions a few times larger than the 
size of typical molecular clouds, we allow cold ($T<1000~\rm{K}$) gas to form stars once its density 
exceeds the value $n_{\rm SF} = 7~\rm{cm}^{-3}$. At such densities, most gas in the simulation has very low temperatures around a few hundred Kelvin. Stars
are formed at a rate
\begin{equation}
 \frac{\rm{d}\rho_*}{{\rm d} t} = \epsilon_{\rm ff} \frac{\rho_{\rm gas}}{t_{\rm ff}} ,
 \label{eq:SFeq}
\end{equation}
where $t_{\rm{ff}} \approx 5~\rm{Myr}$ is the typical free-fall time of observed molecular clouds, and $\epsilon_{\rm ff}$ is the observationally constrained efficiency per free-fall time. Following \citet{Ceverino09}, we include the effect of runaway stars. This is implemented by assigning to $33$ per cent of the newly created stellar particles velocities sampled from a random exponential distribution with a characteristic velocity of $17 \kms$ and random orientations.

\subsubsection{The impact of SF efficiency on feedback}

High resolution simulations require a way to limit the number of star particles 
to relieve the computational burden of $N$-body calculations. This 
is typically done either by creating star particles stochastically with low probabilities or by limiting 
the minimum mass that a star particle may have. In most high resolution simulations star particles are created by stochastically sampling equation~(\ref{eq:SFeq}) using $\epsilon_{\rm ff} \sim 1-10$ per cent \citep[e.g.,][]{Stinson06,Governato07,Brooks07,Ceverino09,governato10,Stinson10,guedes11,Christensen12b,Governato12,Brook12,Calura12,Munshi13,Brook13,Ceverino13,Shen13}. By construction, this approach will produce individual star formation events with large ratios of stellar mass to gas mass to compensate for the events where no stars are allowed to form. Only in the limit of high probability for each star formation event, does this ratio approach the observed value, $M_{\rm star}/M_{\rm gas} = \epsilon_{\rm ff}\Delta t/t_{\rm ff}$, with $\epsilon_{\rm ff} \approx 1-8$ per cent. For instance, \citet{Christensen12b} adopt a model of star formation in molecular hydrogen in their simulations but convert 33\% of the local gas mass into stars in each SF event. More recently, \citet{Brook12} use a similar stochastic recipe, converting about $20$ per cent of gas mass into star in each event. These numbers are a factor of $\sim 2-3$ larger than the average observed efficiency of star clusters in the Milky Way, $\epsilon_{\rm ff} \approx 8$ per cent \citep{Murray11b}. 

Although the typical $M_{\rm star}/M_{\rm gas}$ values of individual star formation events in simulations are unrealistically high, they increase the strength of the stellar feedback and enhance gas blowouts from star-forming regions. Moreover, this artificial efficiency sometimes requires adjustement by a factor of two for haloes of different mass in order to obtain stellar masses that fall on the \citet{Moster12} stellar-mass-halo-mass relation \citep{Brook12}. Besides artificially inflating the local star formation efficiency, the stochastic approach also inflates the stars-to-gas mass ratio, which in turn boosts the energy released by feedback. As a consequence, the energy release will be highly inhomogeneous and concentrated in a few high efficiency regions. To avoid these issues in our model, we remove the lower mass limit for star particles and allow the mass to be 
controlled solely by the observed efficiency for \emph{each} star formation event, $\epsilon_{\rm ff}$. 

Unfortunately, using deterministic star formation in cosmological simulations is prohibitely expensive because the long gas consumption times yield particles with very small 
masses. The typical distribution of initial star particle masses in our runs has a narrow peak at $\sim 200 \Msun$, with $99$ per cent of the particles in the range $50 < M/{\rm M}_{\odot} < 3000$. To avoid the computational cost of tracking large numbers of particles, we implemented a novel method for reducing the number of old stellar particles in the simulation. In this procedure, the mass contained in stars is resampled by periodically 
removing a small fraction (typically $\lesssim 10$ per cent) of old (age $> 40~\rm{Myr}$) star particles while redistributing their total mass and metals uniformly among the remaining 
particles of the same age. This method preserves the total stellar mass of the galaxy while enforcing the observed \emph{local} efficiency for each SF event. Appendix~\ref{sec:appendix} compares the relation between surface density of star formation and cold gas surface density obtained using our SF model with observations at kiloparsec scales.

In the following section we estimate the effect of a local model of star formation on the disruption of star-forming clouds. We show that the efficiency of star formation is crucial in controlling the total energy and momentum output from young stars, which in turn determines whether the cloud expands and is dispersed by stellar feedback or collapses further to continue fueling star formation.

\section{Analytical estimates of the effects of stellar feedback}
\label{sec:analytics}

In this section, we provide analytical estimates of the effects of thermal and radiative feedback on 
the gas surrounding a young star cluster. We examine three conditions necessary for stellar feedback to have an effect on galaxy formation: (1) the overpressuring of gas due to SN heating, (2) the outward pressure from radiation, photoheating, SN and shocked stellar winds vs. the confining pressure, and (3) the fate of the gas once the cloud is dispersed. 

\subsection{Heating}

\begin{figure}
 \includegraphics[width=0.49\textwidth]{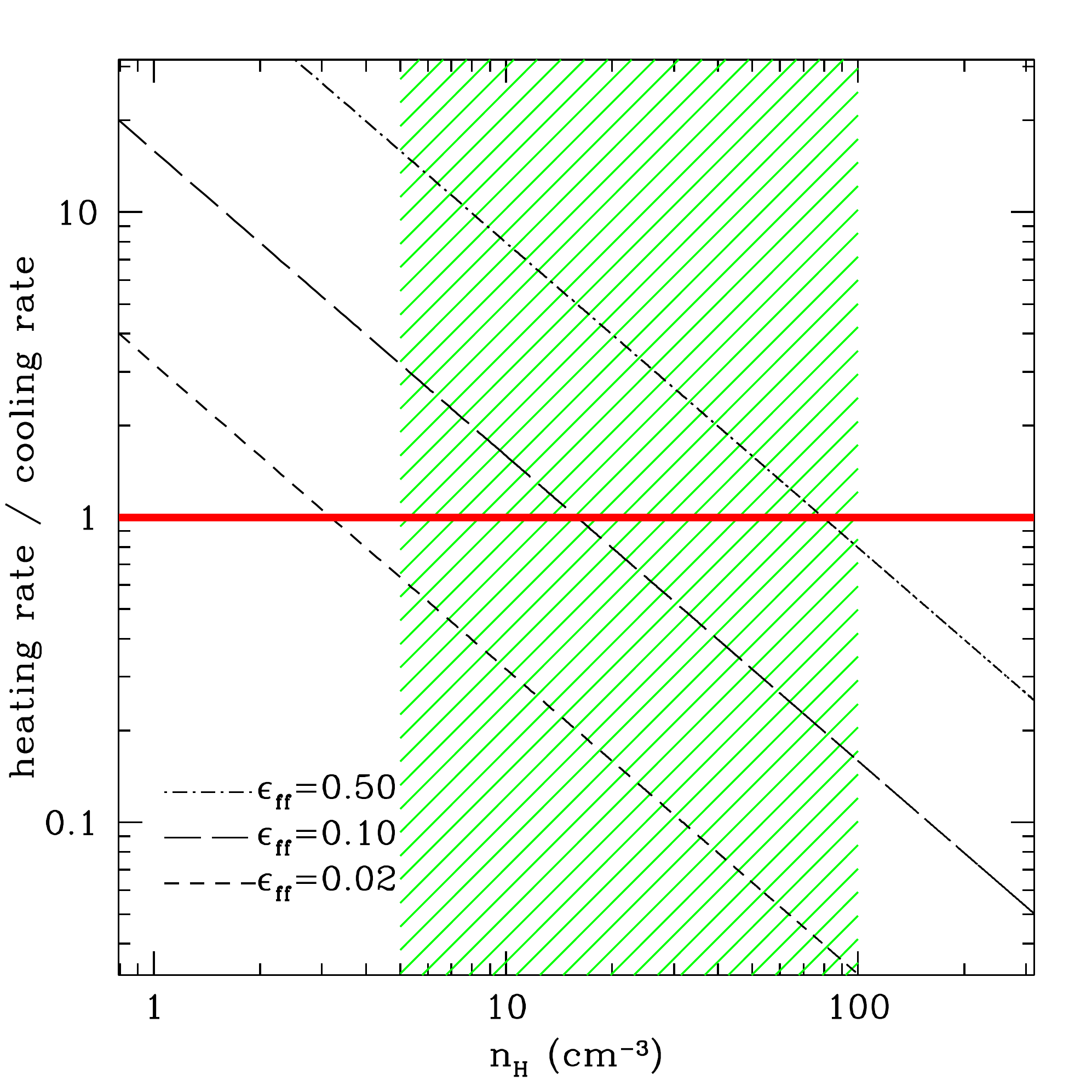}
 \caption{Ratio of heating rate due to SNII and shocked stellar winds to the cooling rate of cold gas as a function of gas density at $T \approx 100$ K for different values of the star 
formation efficiency, $\epsilon_{\rm ff}$. The hatched area shows the typical range in  star formation density thresholds in high resolution galaxy simulations. Only for values above the horizontal solid line is thermal 
feedback able to overpressure the gas before the energy is radiated away.} 
 \label{heating}
\end{figure}

For stellar feedback to be effective in dispersing the birth gas cloud, the outward  gas pressure gradient must overcome self-gravity.
In the case of supernovae feedback, this is accomplished by injecting thermal energy to increase the gas temperature and pressure.
For the cloud temperature to increase, the heating rate must be larger than the cooling rate of the gas due to radiative losses. The radiative cooling rate is $\Lambda = n_{\rm H}^2 \Lambda'$, and the heating rate is $\Gamma = \rho_{*}\Gamma'$, where $n_{\rm H}$ and $\rho_*$ are the number density of Hydrogen atoms and the mass density of young stars respectively. Using equation~(\ref{eq:SFeq}) we obtain the ratio of supernovae heating to radiative cooling during one free-fall time:
\begin{equation}
 \frac{\Gamma}{\Lambda} = \frac{\rho_{*}\Gamma'}{n_{\rm H}^2\Lambda'} = \frac{\epsilon_{\rm ff}}{n_{\rm H}}\left( \frac{\mu_{\rm H}m_{\rm H}\Gamma'}{\Lambda'} \right), 
\end{equation}
which depends only on the gas density and the star formation efficiency. Figure~\ref{heating} shows the ratio of the heating rate to the cooling rate as a function of gas density for cold, solar metallicity  
gas. We take $\Gamma' = 1.59\times10^{34}~\rm{erg}~\rm{s}^{-1}\Msun^{-1}$ (which includes the contribution from shocked stellar winds), and assume that the gas is initially cold ($T \sim 100$ K), which gives $\Lambda' \approx 10^{-25}~\rm{erg}~\rm{s}^{-1}~\rm{cm}^{3}$. Evidently, realistic values of $\epsilon_
{\rm ff}$ overheat the gas \emph{only} if star formation occurs at densities below $\sim 3~ 
\rm{cm}^{-3}$. In typical high resolution galaxy formation simulations, $n_{\rm SF} \gtrsim 1~\rm{cm}^{-3}$, which implies that, for a free-fall time, the cooling rate will be larger than the heating rate and the gas will radiate its energy before it can expand. 

\subsection{Pressure gradient}

\begin{figure}
 \includegraphics[width=0.49\textwidth]{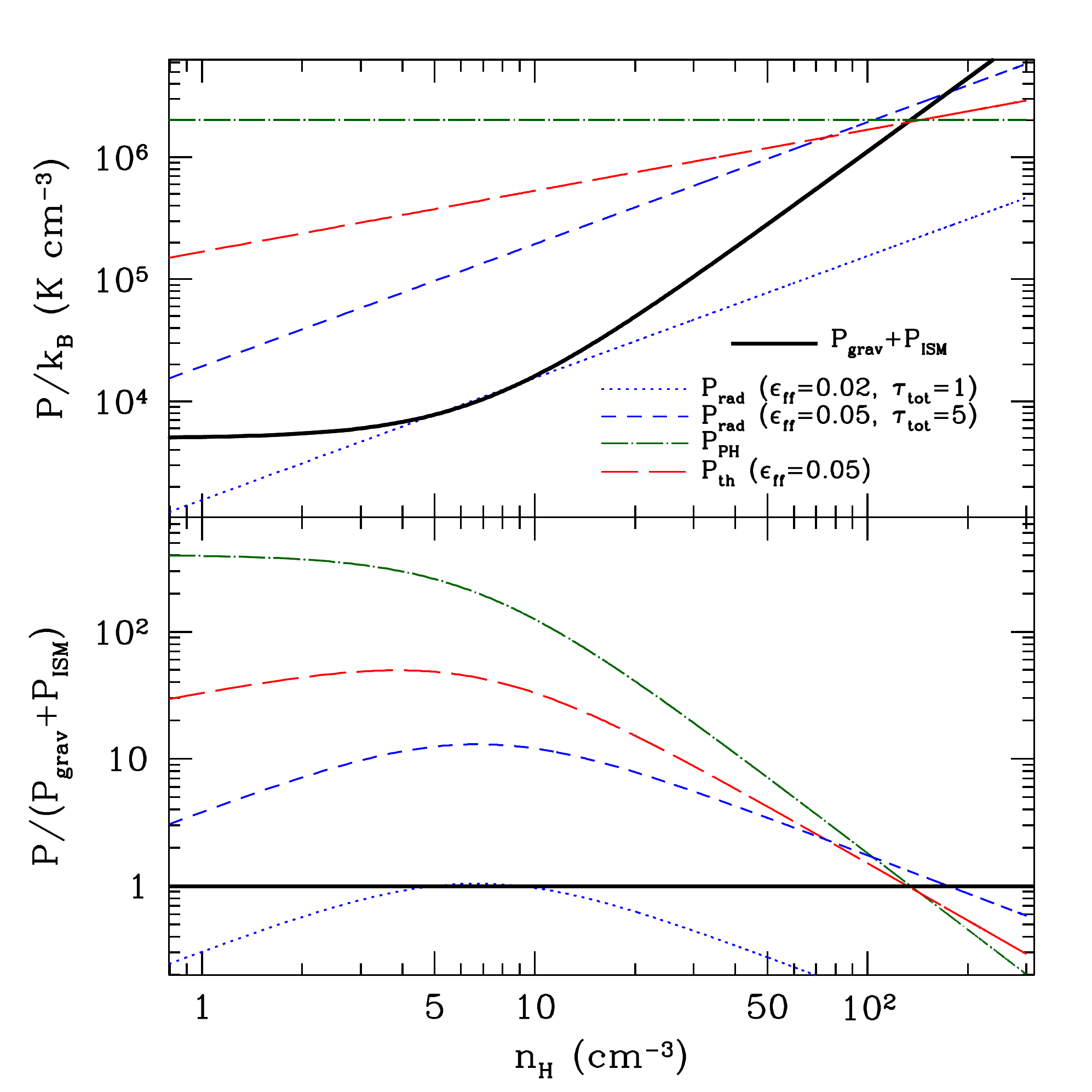}
 \caption{Different contributions to the gas pressure inside a star-forming region using a realistic local star formation model. {\it Top:} Comparison of ``blastwave" pressure, stellar radiation pressure, and the inward pressure due to self-gravity and the ambient medium at the typical resolution of our simulations ($\Delta x=50\pc$). {\it Bottom:} Ratio of the outward to the inward pressure. ``Blastwave" pressure is calculated assuming that the gas cooling is delayed as is commonly done is simulations, and that the energy injection stops after one dynamical time. For radiation pressure, the curves show the effects of different assumptions about the local star formation efficiency and the gas optical depth. At the gas densities resolved in our simulations ($n_{\rm H} \la 10~{\rm cm}^{-3}$), photoheating dominates over the other mechanisms, while radiation momentum overpressures gas by a factor of $1-10$ for $\tau_{\rm tot} \leq 5$. The ``blastwave" pressure is $\sim 50$ times greater than the inward pressure. 
}  
 \label{pgradient}
\end{figure}

Even if the temperature of the gas increases, stellar feedback will be ineffective unless it can pressurize the gas enough to overcome the force of gravity at the scales resolved in the simulation. Figure~\ref{pgradient} shows an estimate of the pressure in 
star-forming gas as a function of the cloud gas density compared to self-gravity and the external pressure of the ISM. The pressure due to self-gravity is $P_{\rm grav} = 4 \pi {\rm G} \rho_{\rm gas} R^2 /3$, where $R$ is the radius of the cloud. The pressure of the ISM is assumed to be $P_{\rm ISM}/k_{\rm B} = 5\times 10^3~\rm{K}~{\rm cm}^{-3}$, radiation pressure is calculated using equation~(\ref{eq:radpressure}), and the thermal pressure due to photoheating is set to $P_{\rm PH} = 2\times10^6~\rm{K}~{\rm cm}^{-3}$. In the comparison we show the effect of increasing the local efficiency of star formation and the optical depth of the gas using two cases. Assuming $\epsilon_{\rm ff} = 5$ per cent and $\tau_{\rm tot} = 5$, at densities $3 \la n_{\rm H} \la 100~{\rm cm}^{-3}$, where supernovae heating cannot exceed radiative cooling, stellar radiation overpressures the gas by a factor of up to $\sim 10$ (for $\tau_{\rm tot} \leq 5$) with respect the inward pressure from self-gravity and the ambient medium. This outward pressure gradient will cause the expansion of a low density cavity around the star cluster and halting of subsequent star 
formation. Figure~\ref{pgradient} also shows an estimate of the pressure resulting from injection of SNII thermal energy and artificially delayed gas cooling (as in the widely used ``blastwave" approximation; \citet{Stinson06}). Since in this method cooling is typically delayed for a several million years, we calculate the thermal gas pressure that results from the increase in internal energy during a free-fall time, $P_{\rm th} = M_* \Gamma' t_{\rm ff}/(\Delta x)^3$, where $M_*$ is the total mass of massive stars and $\Delta x$ is the size of the grid cell. Initially, the large continuous 
energy injection rapidly increases the temperature and pressure of the gas, reaching $T \gtrsim 4\times10^8~\rm{K}$ in the first million years after the formation of the massive stars. At these temperatures the sound speed is several hundred kilometers per second and the sound crossing time is about $\sim 1000$ times smaller than the dynamical time. Before the system can adjust, the pressure inside the region reaches values $\sim 5$ times larger than the radiation pressure and about 50 times the inward pressure. This large gradient causes the region to expand explosively, creating the gas blowouts that are ubiquitous in simulations that use ``blastwave" supernovae feedback. Once the gas density drops, the pressure gradient vanishes. In contrast, radiation pressure remains at a constant value of several times the self-gravity, allowing for a lower but continuous evacuation of the gas from star-forming regions. Morever, in contrast to supernovae feedback, radiation pressure does not heat the gas to millions of degrees in order to disperse the star-forming cloud.

\subsection{Gas escape}

\begin{figure}
 \includegraphics[width=0.49\textwidth]{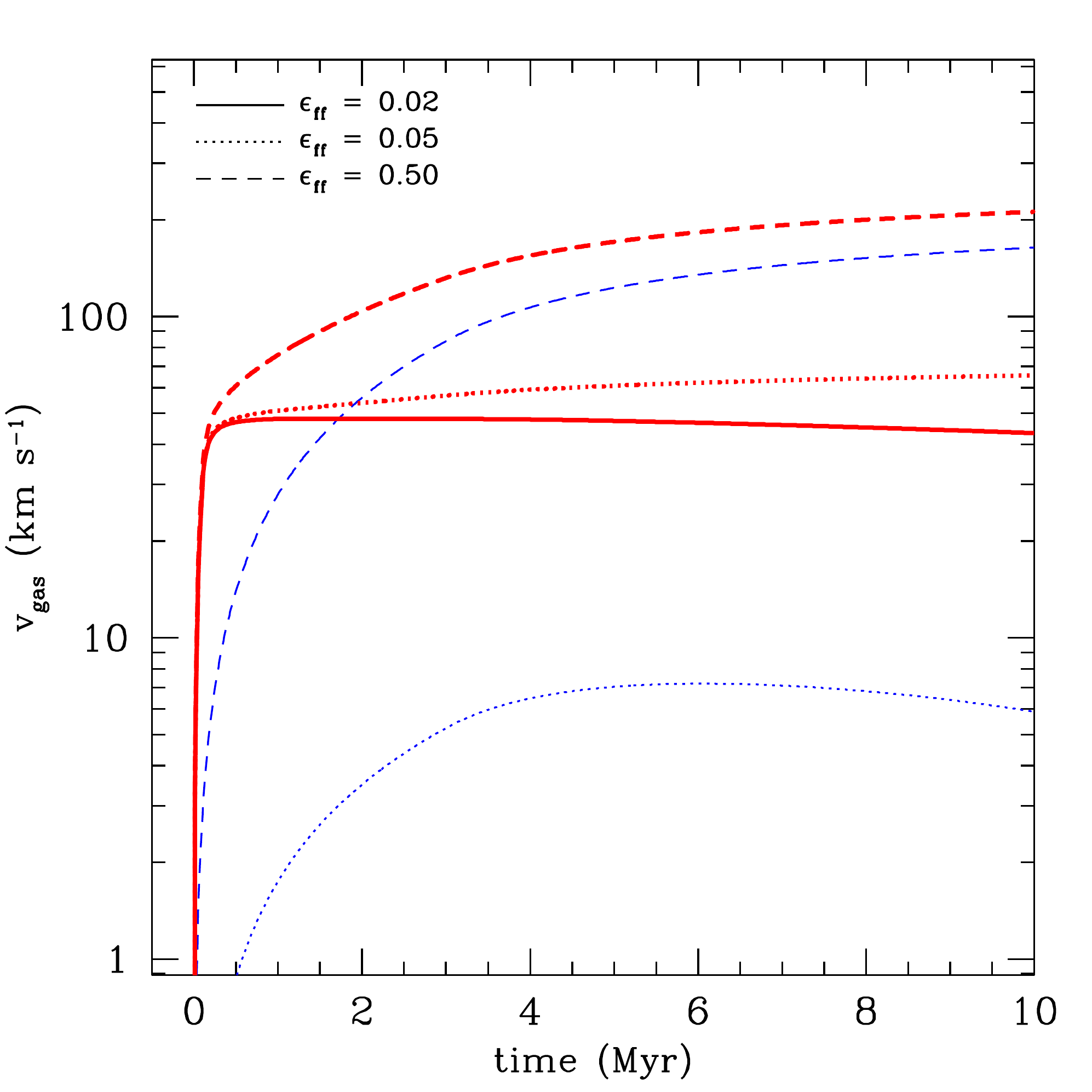}
 \caption{Gas velocity as a result of radiation feedback from an embedded star cluster. We assume that the only forces acting against the expansion are the self-gravity of the gas cloud and the confining pressure of the surrounding ISM. Solid, dotted, and dashed lines correspond to different values of the star formation efficiency, $\epsilon_{\rm ff}$, assuming an optical depth of the gas $\tau_{\rm tot} = 1$. Thin curves show radiation pressure only, while thick curves include radiation pressure and photoheating. For realistic local star formation efficiencies (i.e., $\epsilon_{\rm ff} = 0.01-0.08$), radiation momentum does not accelerate gas beyond the typical escape velocity of a low-mass galaxy. Gas escapes the galactic potential only when photoheating is included or the infrared optical depth is large.} 
 \label{pescape}
\end{figure}

Once the overpressured gas is able to overcome self-gravity and expand, its fate depends on the amount of momentum it acquires from feedback. The baryon fraction, the star formation rate, and the metallicity of the galaxy will 
depend critically on whether gas ejected from star-forming regions falls back onto the galaxy or is able to escape the galactic potential without ever returning. Figure~\ref{pescape} shows the 
ballistic velocity of the gas surrounding a stellar cluster as a function of time as a result of radiation feedback {\it only} (with no supernovae or stellar winds), as implemented in Section~\ref{sec:model}, for different 
values of the star formation efficiency, $\epsilon_{\rm ff}$. To obtain the velocity we integrate the equation of motion using the total force exerted on the gas inside a cell by radiation, self-gravity, and the ISM, $F_{\rm tot} = (P_{\rm rad} + P_{\rm PH} - P_{\rm grav} - P_{\rm ISM})(\Delta x)^2$. Figure~\ref{pescape} shows that single scattering of radiation alone accelerates gas beyond the typical escape velocities of low-mass galaxies ($\gtrsim 100\kms$) only for high star formation efficiencies, $\epsilon_{\rm ff} \gtrsim 30$ per cent. For realistic values of the efficiency, $\epsilon_{\rm ff} \approx 1-8$ per cent, either large infrared dust optical depths ($\tau_{\rm IR} \gtrsim 30$) or photoheating are necessary to eject gas from the galaxy at velocities exceeding the escape velocity. The fate of the gas subject to stellar feedback will have important implications to the fraction of baryons that stay within the galaxy and its circumgalactic gas. 


\section{Numerical Simulations}
\label{sec:simulations}

In this paper we analyse the formation of galaxies within two dark matter haloes which bracket a broad range within the mass spectrum of low-mass galaxies. One is representative of a dwarf galaxy with a mass $M_{\rm vir} = 3\times10^{10}\Msun$, and the second one is about ten times more massive and is comparable to a typical field spiral galaxy with $M_{\rm vir} = 2\times10^{11}\Msun$. The galaxies formed within these two haloes are designated ``dwarf" and ``spiral" respectively throughout the paper. To investigate the stellar mass assembly in these dark matter haloes, we performed numerical simulations with a variety of different 
stellar feedback prescriptions. All the simulations used the cosmological parameters $\Omega_{\rm m} = 
1-\Omega_{\rm \Lambda} = 0.30$, $\Omega_{\rm b} = 0.045$, $\sigma_8 = 0.8$ and 
$h=H_0/(100~\kms~\rm{Mpc}^{-1}) = 0.7$.  Each simulation ``zooms-in" on a single isolated dark matter halo inside a $10 \Mpch$ 
comoving box, achieving a maximum resolution of $40-80~h^{-1}\rm{pc}$ at $z=0$. Throughout the paper we define virial mass as the total mass (baryonic and DM) enclosed within a spherical region with an average density equal to the virial overdensity in the top-hat collapse model, $\Delta_{\rm vir}$, times the average density of the universe at that redshift. Table \ref{table1} summarizes the general properties of the DM haloes used in this work.

All the simulations in this paper use a deterministic model of star formation where stars from in cold and dense gas with $n > 7~\rm{cm}^{-3}$ and $T <1000~{\rm K}$. The gas density threshold is chosen to be smaller than the density at which the gas is artificially pressurized to prevent spurious fragmentation \citep{Truelove97}. This paper focuses on the differences between feedback models that include only standard supernovae thermal feedback versus models with radiation feedback. We dedicate most of the following discussion to a model with only thermal feedback (\verb#SN# runs) and a fiducial radiative feedback model which includes the effect of single scattering of radiation and photoheating from ionization  (\verb#ALL_1# and \verb#ALL_8#).  Table \ref{table2} shows the parameters of the feedback prescriptions used in each simulation. In addition, we test the effect of variations in the parameters of radiation feedback as explained below. 

\subsection{Duration of radiative forcing}

In addition to the models discussed above, we tested the effect of varying the duration of radiative forcing on the gas around each star particle in the simulations. In the real birth clouds of stellar clusters the radiation is initially trapped by the high column density of the gas surrounding the cluster. This forcing vanishes once the gas column density decreases enough to become optically thin to UV/optical photons. 

Our simulations do not resolve the physical column densities involved in star-forming regions, so we must estimate the time when radiation pressure becomes negligible. To do this, we set a density threshold below which the extra pressure is switched off in the cells adjacent to the star particle. We tested several models of radiative feedback using two different thresholds: $n_{\rm th} = 0.1~\rm{cm}^{-3}$ and $n_{\rm th} = 0.01~\rm{cm}^{-3}$. Lower threshold values result in a longer duration of radiative forcing up to a maximum of 40 Myr (in the \verb#_long# models).

\subsection{Infrared trapping of radiation}

In addition to the duration of radiative forcing, there is uncertainty is the optical depth of the gas to the reprocessed IR photons, $\tau_{\rm IR}$. As discussed above, observations and simulations using radiative transfer favor values between as high as $\sim 50$ \citep{Agertz12,Davis14} for young clusters. In two simulations of the \verb#dwarf# halo, we investigate varying the infrared optical depth within the range preferred by models and observations, setting $\tau_{\rm IR} =10$, and $\tau_{\rm IR} = 50$. 

\subsection{Thermal pressure due to photoheating}

As discussed in Section~\ref{sec:feedback}, there is substantial variation in the thermal pressure of gas in observed H\,{\sc ii} regions, with about an order of magnitude scatter between individual regions. Furthermore, the photoheating pressure must have been larger than that of evolved regions at the early stages of evolution, when densities were higher. To account for this uncertainties, we test a range of values of the pressure, $P_{\rm PH,0} = 10^6 - 4\times 10^7~\rm{K~cm}^{-3}$ in our numerical experiments. We refer to models in which the photoheating pressure is varied but the radiation pressure is fixed using the \verb|ALL_#| label, where the number denotes the multiplier of the fiducial value, $10^6~\rm{K~cm}^{-3}$.

\begin{table}
\centering
  \begin{tabular}{@{}lccp{1.3cm}p{1.3cm}p{1.3cm}c@{}}
   \hline \hline
   Halo & $M_{\rm vir}~(\Msunns)$ & $R_{\rm vir}~(\rm{kpc})$ & minimum\par cell size (pc~$h^{-1}$) & DM mass\par resolution $(\Msunns)$   \\
 \hline \hline
 \verb#dwarf#  & $3\times10^{10}$ & $80$ & $40$ & $9.4\times10^4$    \\
 \hline
 \verb#spiral#    & $2\times10^{11}$ & $150$ & $80$ & $7.5\times10^5$      \\
\hline \hline
\end{tabular}
\caption{Properties of the DM-only simulations at $z=0$.}
\label{table1}
\end{table}

\begin{table*}
\centering
\begin{tabular}{@{}lcccccc@{}}
   \hline \hline
   Model & $\epsilon_{\rm ff}$ & feedback  & $\tau_{\rm tot}$  & $P_{\rm PH}/{\rm k_B}~(10^6~\rm{K~cm}^{-3})$  & $n_{\rm th}~(\rm{cm}^{-3})$ &  \\
 \hline \hline
 \verb+dwSN+          & 0.02 &  SNII+SW       & -  & 0  & -     \\
 \verb+dwRP_1_long+   & 0.05 &  SNII+SW+RP    & 1  & 0  & $0.01$ \\
 \verb+dwRP_10_long+  & 0.05 &  SNII+SW+RP    & 10 & 0  & $0.01$ \\
 \verb+dwRP_50_long+  & 0.05 &  SNII+SW+RP    & 50 & 0  & $0.01$ \\
 \verb+dwALL_1+       & 0.05 &  SNII+SW+RP+PH & 1  & 1  & $0.1$ \\
 \verb+dwALL_8+       & 0.02 &  SNII+SW+RP+PH & 1  & 8  & $0.1$ \\
 \verb+dwALL_40+      & 0.02 &  SNII+SW+RP+PH & 1  & 40 & $0.1$ \\
 \verb+dwALL_8_long+  & 0.02 &  SNII+SW+RP+PH & 1  & 8  & $0.01$ \\
 \hline
 \verb+spSN+          & 0.02 &  SNII+SW       & - & -  & -   \\
 \verb+spALL_8+       & 0.02 &  SNII+SW+RP+PH & 1 & 8  & $0.1$ \\
 \verb+spALL_40+      & 0.02 &  SNII+SW+RP+PH & 1 & 40 & $0.1$ \\
 \verb+spALL_8_long+  & 0.02 &  SNII+SW+RP+PH & 1 & 8  & $0.01$ \\
\hline \hline
\end{tabular}
\caption{Parameters for feedback and star formation in the simulations. SNII+SW indicates supernovae and stellar winds, RP stands for radiation pressure, and PH denotes photoheating.} 
\label{table2}
\end{table*}

\section{Results}
\label{sec:results}

\subsection{Global properties and galaxy formation efficiency}
\label{sec:globalproperties}

Table~\ref{table3} shows the integrated properties of the simulated galaxies. Stellar and gas masses are computed within $10\kpc$ from the galaxy centre. We define the baryon fraction, $f_{\rm bar}$, as the ratio of the total baryonic mass within $10\kpc$ to the virial mass. Table~\ref{table3} shows that for the \verb#dwarf# as well as for the \verb#spiral# galaxies, the stellar mass and the ratio of gas to stellar mass, $M_{\rm gas}/M_*$, change drastically when radiation pressure and photoheating feedback are included. Models with only supernovae feedback suffer from catastrophic overcooling, which results in excessive star formation and a conversion of a large fraction of halo baryons into stars. The stellar mass of the \verb#dwSN# model is more than 6 times greater than the fiducial radiation feedback model \verb#dwALL_1#, even though \verb#dwALL_1# has a $2.5$ times greater efficiency of star formation, $\epsilon_{\rm ff}$. While not shown, simulations with radiation pressure but no photoheating also have large stellar masses at $z>1$ (see Figure~\ref{SMvstime}) showing that radiation momentum alone (with $\tau_{\rm IR} \leq 50$) is not sufficient to reduce the total stellar mass in low-mass haloes. In the low-mass spiral, the failure of standard SN feedback is more serious due to the deeper potential well. At $z \approx 2.5$, the \verb#spSN# simulation has more than $50$ times the stellar mass of the \verb#spALL_8# model. 

\begin{figure}
 \includegraphics[width=0.50\textwidth]{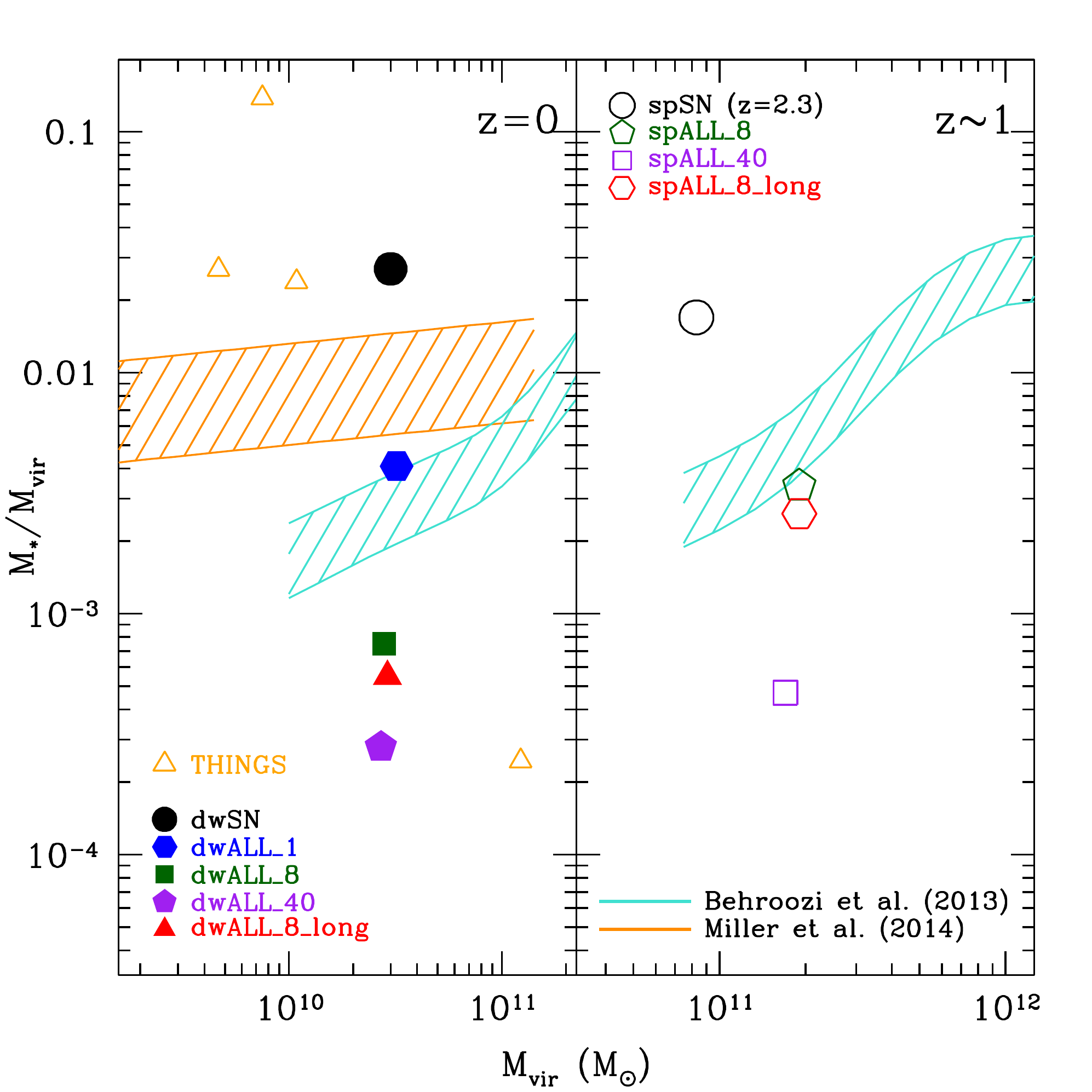} 
 \caption{Stellar-to-virial mass ratio of the simulated galaxies. {\it Left:} \texttt{dwarf} models at $z=0$. {\it Right:} \texttt{spiral} models at $z \approx 1$. For comparison, the hatched areas show the $1\sigma$ systematic envelope of the semi-empirical model by \citet{Behroozi12}, as well as the \citet{Miller14} inference from observations, and individual mass models from \citet{Oh11b}. While supernovae energy alone (\texttt{dwSN}) leads to overproduction of stars compared to Behroozi et al., including stellar radiative feedback reduces the stellar mass drastically. Note that the effect of radiation pressure is robust to large variations in the parameters $\tau_{\rm tot}$ and $n_{\rm th}$ (modulo changes in the star formation efficiency parameter, $\epsilon_{\rm ff}$). } 
 \label{behrooziplot}
\end{figure}

All the models with radiation feedback produce galaxies with stellar-to-halo mass ratios that are in good agreement with the semi-empirical models of \citet{Behroozi12} and others \citep{Moster12,Firmani10b}. For instance, using DM-only cosmological simulations and constraints from observations at many redshifts, \citet{Behroozi12} obtain a ratio $M_*/M_{\rm vir} \sim 0.0025$ for a halo of mass $M_{\rm vir} = 3\times 10^{10}\Msun$ at $z=0$. Among the dwarf simulations, the fiducial model with radiation pressure and photoheating (\verb#dwALL_1#) shows the best agreement with this value, falling within its $1\sigma$ systematic uncertainty.  This comparison is shown in Figure~\ref{behrooziplot}, where we also include estimates from other probes such as weak lensing \citep{Miller14}, and mass-modelling \citep{Oh11b}. Models with stronger radiative coupling ($10 < \tau_{\rm tot} < 50$) generally have similar stellar masses. Once photoheating is included, additional tests show that it becomes the main driver of the decrease in the stellar mass of the \verb#ALL# models. Dwarf simulations with increased photoheating pressure fall below the $1\sigma$ distribution in \citet{Behroozi12} but fit comfortably inside other estimates such as those by \citet{Moster12} and \citet{Oh11b}. However, since different observational inferences predict discrepant stellar-to-virial mass ratios, it is difficult to rule out models based on this diagnostic alone. 

The cumulative effect of radiative feedback at $z=0$ is not highly sensitive to variations in the parameters within the uncertainties in the physical model. For $M_{\rm vir} = 3\times10^{10}\Msun$, a factor of $5$ increase in forcing and a factor of $\sim 10$ change in duration of the effect produce a variation no larger than a factor of $\sim 3$ in the stellar mass for simulations with the same $\epsilon_{\rm ff}$ (compare the \verb#ALL_8#, \verb#ALL_8_long# and \verb#ALL_40# models).

Table~\ref{table3} lists the ratio of cold gas mass to stellar mass within $10 \kpc$ of the center of each galaxy. We define cold gas as gas with $T < 10^4~\rm{K}$. The cold gas fractions of galaxies with SN feedback are smaller than unity, showing that a large amount of the cold and dense gas that is available to form stars is quickly consumed. On the other hand, models with radiation pressure have gas fractions as large as $\sim 30$, which indicates that the galaxies are inefficient at converting their gas into stars. Observations of cold atomic and molecular gas in galaxies at $z=0$ reveal an anti-correlation between stellar mass and gas fraction, with galaxies of mass $M_* = 1\times10^8\Msun$ containing $M_{\rm gas, cold}/M_* \sim 4.7^{+5.3}_{-2.5}$, decreasing to $\sim 1.6$ for $M_* = 1\times10^9\Msun$ \citep{Dutton10b}. Thus, our dwarf galaxies with radiative feedback display excellent agreement with observations, showing that radiation from massive stars plays a key role in maintaining large cold gas reservoirs in low-mass galaxies. 

One essential characteristic of low-mass galaxies that simulations still struggle to reproduce is their extremely low  galaxy formation efficiency, or in other words, the amount of primordial baryons that condense to form the galaxy. For a galaxy with the mass of the Milky Way, semi-empirical models based on matching the abundance of galaxies to that of DM haloes \citep[e.g.,][]{behroozi10,Trujillo-Gomez11,Behroozi12,Moster12} predict an efficiency $M_*/M_{\rm vir} \sim 25$ per cent, whereas for a dwarf galaxy with $M_* = 10^8\Msun$ the ratio is an order of magnitude smaller, $M_*/M_{\rm vir} \sim 2.5$ per cent. The last column of Table~\ref{table3} shows the baryon fraction, $f_{\rm bar} \equiv [ M_*(<10\kpc) + M_{\rm gas}(<10\kpc) ] / M_{\rm vir}$ for all the model galaxies. 

\begin{table*}
\centering
  \begin{tabular}{@{}lcccccccc@{}}
   \hline \hline
   Model          & redshift & $M_{\rm vir}~(\Msunns)$ & $M_*~(\Msunns)$ &  $M_{\rm gas}~(\Msunns)$ &  $M_{\rm gas, cold}~(\Msunns)$ &
   $M_*/M_{\rm vir}$ & $M_{\rm gas, cold}/M_*$ & $ {\rm f}_{\rm bar}$ \\
 \hline \hline
 \verb+dwSN+     & 0 &  $3.0\times10^{10}$ & $8.1\times10^8$ & $8.1\times10^8$ & $5.5\times10^8$ & $2.7\times10^{-2}$ & $0.7$ & $0.054$\\
 \verb+dwALL_1+       & 0 &  $3.2\times10^{10}$ & $1.3\times10^8$ & $2.1\times10^9$ & $1.4\times10^9$ & $4.1\times10^{-3}$ & $10.8$ & $0.070$\\
 \verb+dwALL_8+     & 0 &   $2.8\times10^{10}$ & $2.1\times10^7$ & $5.5\times10^8$ & $1.2\times10^8$ & $7.5\times10^{-4}$ & $5.7$ & $0.020$ \\
 \verb+dwALL_40+     & 0 &  $2.7\times10^{10}$ & $7.5\times10^6$ & $1.9\times10^8$ & $1.3\times10^7$ & $2.8\times10^{-4}$ & $1.7$ & $0.007$ \\
 \verb+dwALL_8_long+  & 0 &  $2.9\times10^{10}$ & $1.6\times10^7$ & $7.4\times10^8$ & $6.2\times10^7$ & $5.5\times10^{-4}$ & $3.9$ & $0.026$ \\
 \hline
 \verb+spSN+       & $2.3$ &  $8.3\times10^{10}$ & $1.4\times10^9$ & $3.7\times10^9$ & $2.1\times10^9$ & $1.7\times10^{-2}$ & $1.5$ & $0.061$ \\
 \verb+spALL_8+     & $1.3$ &  $1.9\times10^{11}$ & $6.4\times10^8$ & $1.0\times10^{10}$ & $7.6\times10^9$    & $3.4\times10^{-3}$ & $11.9$ & $0.056$ \\
 \verb+spALL_40+    & $1.3 $ & $1.7\times10^{11}$ & $7.9\times10^7$ & $4.0\times10^9$ & $2.2\times10^9$    & $4.7\times10^{-4}$ & $27.8$ & $0.024$ \\
 \verb+spALL_8_long+ & $1.3 $ & $1.9\times10^{11}$ & $4.9\times10^8$ & $8.9\times10^9$      & $6.5\times10^9$ & $2.6\times10^{-3}$ & $13.3$ & $0.049$  \\
\hline \hline
\end{tabular}
 \caption{Global galaxy properties.} 
\label{table3}
\end{table*}

In general, stellar radiation does not seem to have a large effect on the baryon fraction of low-mass DM haloes. For both the \verb#dwarf# and the \verb#spiral# simulations, the total baryon fraction within the virial radius and within the galaxy are virtually unaffected by radiation feedback. Focusing on the dwarf simulations, we find that the total baryon fraction within the galaxy $(r < 5 \kpc)$ is $\sim 25$ per cent, and is $\sim 5$ per cent within the virial radius when only SN feedback is included. In models with radiation pressure and photoheating, the baryon fraction is relatively unchanged at $\sim 25$ per cent within the \verb#dwALL_1# galaxy and $\sim 7$ per cent within its virial radius. Surprisingly, this implies that the amount of baryons retained within the halo is not affected by feedback due to radiation from massive stars. 

\begin{figure*}
 \includegraphics[width=0.49\textwidth]{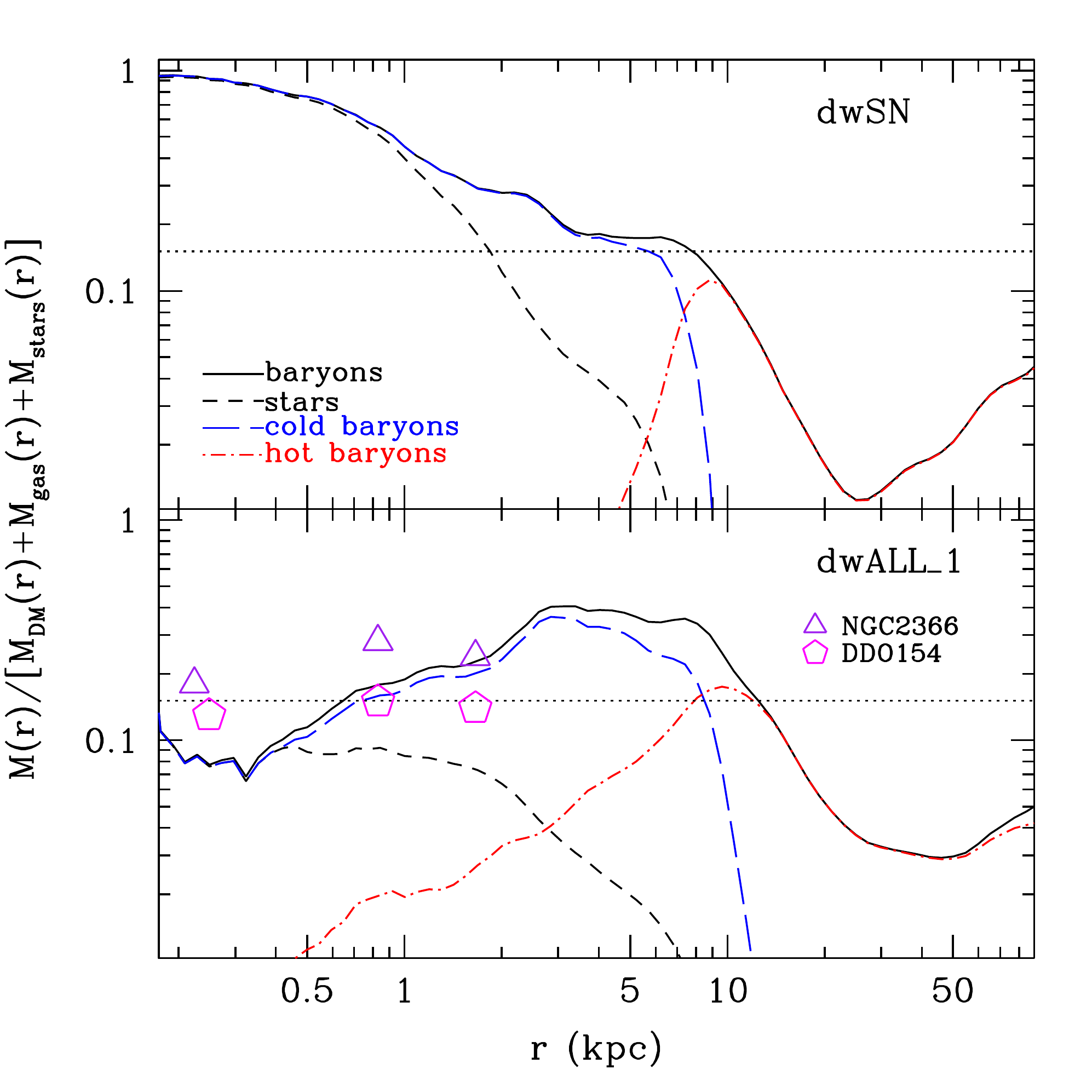} 
 \includegraphics[width=0.49\textwidth]{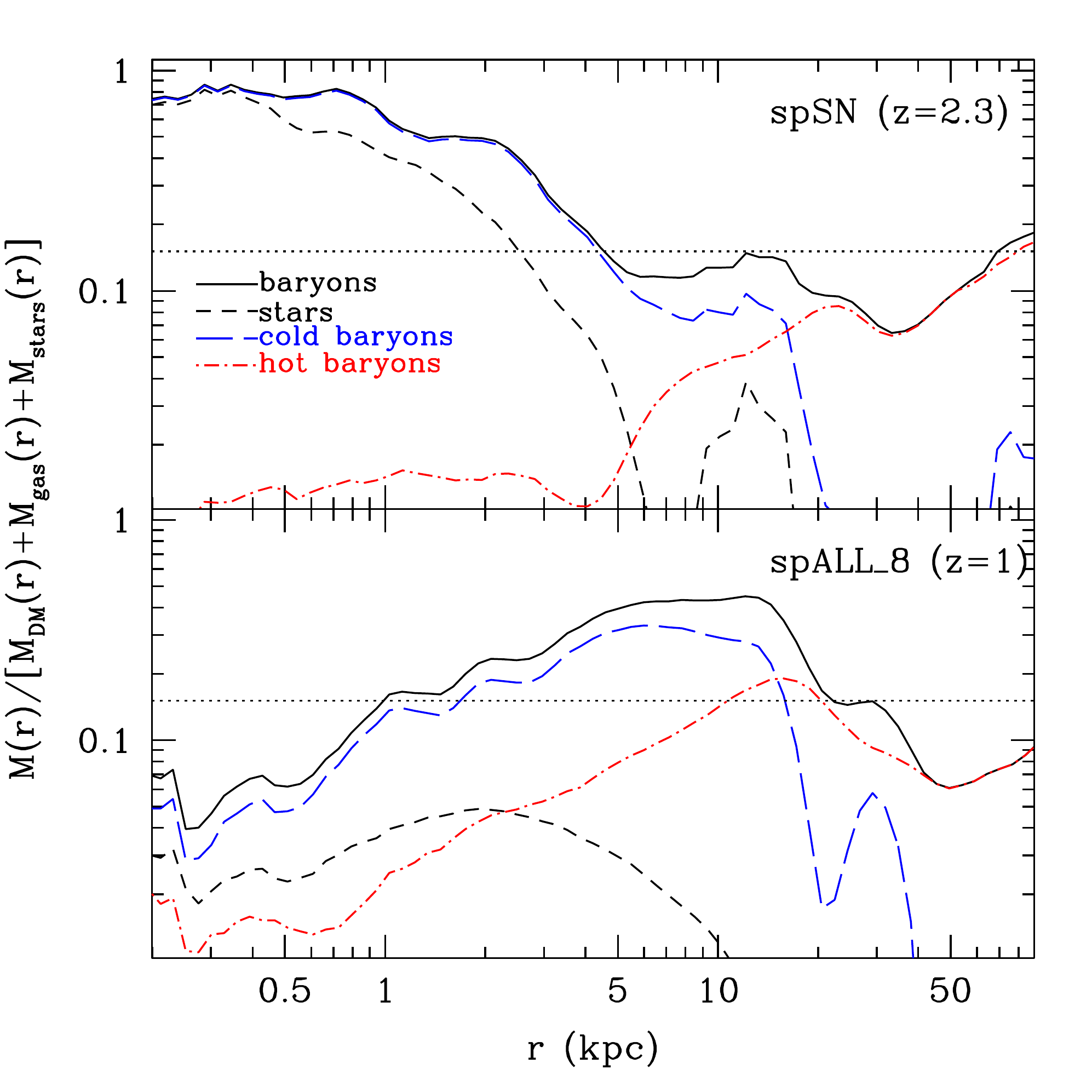}
 \caption{Baryon mass fraction radial profiles. The curves show the ratio of the mass in each baryonic component to the total mass as a function of radius, $M_{\rm bar}(r)/M_{\rm tot}(r)$. ``Cold baryons" include stars and gas with $T<10^4~\rm{K}$, whereas ``hot baryons" represents gas with $T>10^4~\rm{K}$. The dotted horizontal line represents the universal baryon fraction. {\it Left:} \texttt{dwarf} models. Open symbols show the estimated $M_{\rm bar}/M_{\rm tot}$ in the central regions of two isolated dwarf irregular galaxies from the THINGS survey \citep{Oh11a}. {\it Right:} \texttt{spiral} models. Supernovae energy cannot prevent the accumulation of cold baryons in the inner $\sim 2\kpc$. Stellar radiation pressure and photoheating reduce the mass of baryons in the galaxy by heating and expelling cold gas out into the circum-galactic medium.} 
 \label{barfracprofiles}
\end{figure*}

To investigate the physical origin of this feature, we plot in Figure~\ref{barfracprofiles} the ratio of the mass in each baryonic component to the total mass as function of radius from the centre of the galaxy, $M_{\rm bar}(r)/M_{\rm tot}(r)$. We include the baryons locked in stars, as well as the cold and hot baryons. ``Cold baryons" include stars and gas with $T < 10^4~\rm{K}$, and ``hot baryons" correspond to gas above $10^4~\rm{K}$. It is evident in the figure that although the baryon fractions do not change significantly when radiation feedback is included (see Table~\ref{table3}), the inner part of the galaxies is modified drastically. In the \verb#SN# simulations, cold baryons contribute $50-90$ per cent of the total mass in the central regions of the galaxy ($r < 2\kpc$ in the \verb#dwSN# and $r < 5\kpc$ in the \verb#spSN# model), which is up to 6 times larger than the universal fraction of baryons. Furthermore, the central baryons in these galaxies are dominated by stars. In contrast, the simulations with radiative feedback are dominated by dark matter at all radii. The central contribution of stars to the mass in these models is never greater than $8$ per cent in \verb#dwALL_1# and no larger than $6$ per cent in \verb#spALL_8#. However, cold gas contributes a larger fraction of the total baryon mass between $\sim 1$ and $\sim 10\kpc$. This reduces the average ratio $M_{\rm bar}(r)/M_{\rm tot}(r)$ to a value near $15$ per cent, or about the cosmic mean.

In conclusion, when accounting for the baryon budget within the galaxy and within the circum-galactic medium (CGM), we highlight three main consequences of stellar radiation. First, as discussed above, the total fraction of mass in baryons within the central $\sim 1-3$ kpc, where most of the star formation takes place, is reduced to values near the cosmic mean, and even below the cosmic mean for smaller radii. Second, the amount of cold gas at intermediate distances, $1 \la r \la 10~\kpc$, is much larger when radiation feedback is included, comprising about $30$ per cent of the total mass at $10\kpc$. Third, the fraction of mass in gas with $T > 10^4~\rm{K}$ increases dramatically within the \verb#dwarf# and \verb#spiral# galaxies due to radiation pressure and photoheating enhancing the effect of supernovae feedback. Together, these features indicate that the main role of radiation from massive stars is to control the amount of gas that is converted to stars within the galaxy by dispersing and heating it, and ejecting some into the halo, where it can remain in a warm, dilute phase. We note that there is also an increased amount of cold gas in the CGM of the fiducial RP simulations. In a forthcoming paper (Trujillo-Gomez et al., in prep) we analyse the properties of gas in the context of accretion and outflows.

\subsubsection{Discussion}

Our results show that including radiation feedback from young massive stars reduces the stellar mass growth in dwarf and low-mass spiral galaxies. Taken at face value, detailed comparisons with abundance matching models seem to indicate that the combination of  momentum from radiation and ionization photoheating are sufficient to prevent excessive star formation in haloes with $M_{\rm vir} \sim 3\times 10^{10}- 2\times 10^{11}\Msun$ enough to bring them into agreement with observations.

From detailed observations of the structure of nearby dwarf irregulars we can perform case-by-case comparisons. \citet{Oh11a} obtained mass models of dwarf irregulars using multi-wavelength data. Their models constrained the ``cold baryon" mass fraction $M_{\rm bar}/M_{\rm tot}$ within the optical extent of galaxies with $V_{\rm max} \approx 50 - 60 \kms$ to values $\sim 15 - 35$ per cent. Our results show that SN feedback is unable to prevent the buildup of stars in the central few kiloparsecs, which leads to large baryon fractions within the galaxy. In contrast, the fiducial radiation feedback model, with single photon scattering, reduces the mass of baryons that is locked in stars within the galaxy to values near the cosmic mean, in very good agreement with \citet{Oh11a}. In addition, a distinctive feature of the simulation with radiation feedback is that cold gas dominates the baryon budget within $10\kpc$ (see Figure~\ref{barfracprofiles}), with a ratio of cold gas mass to stellar mass that is in excellent agreement with observations (see Section~\ref{sec:globalproperties}).   

In general, the trend we find among the variations of the radiation feedback model is that a larger photoheating pressure or longer duration of radiative forcing reduces the baryon fraction further than in the fiducial model. However, both the total baryon fraction and the stellar baryon fraction within the virial radius are quite robust to the large variations in the choice of feedback and star formation parameters. Among the models with radiation pressure and photoheating, with the optical depth varying by a factor of 40, and $\epsilon_{\rm ff}$ varying by a factor of 2.5,  the baryonic and stellar mass fractions vary by less than a factor of $\sim 3$. 

It is important to note that the term ``baryon fraction" has a variety of definitions in the literature. Usually, in observations, it corresponds to the ratio of the mass of a galaxy in neutral atomic gas, molecular gas, and stars within its optical extent to the estimated total gravitational mass out to the virial radius. In our runs with full feedback (the \verb#ALL# models), cold gas dominates the baryons within the galaxy, contributing $\sim 20$ per cent of the total mass within $10\kpc$, whereas within the virial radius most baryons are in the form of warm/hot gas with $T > 10^4~{\rm K}$. Following this definition, in the fiducial full radiation feedback models the ``observed" baryon fraction is $\sim 4-5$ per cent, about 40 per cent smaller than the total baryon fraction. As shown in Figure~\ref{barfracprofiles}, this is \emph{not} due to escape of baryons from the halo, but instead it relates to an increase in the mass of warm/hot gas at $10 < r < 70\kpc$. Pressure from stellar radiation and photoheating is thus sufficient to match the observed ``cold baryon" content of low-mass galaxies estimated from observations without the need for ad hoc winds that cause the escape of large amounts of gas from the halo. These results agree with the conclusions of a companion paper, \citet{Ceverino13}, where it is shown that massive galaxies ($M_{\rm vir}(z=0) \approx 10^{12}~\Msun$) at high redshift require radiation pressure to reduce the fraction of baryons in stars to the low values obtained in constrained abundance matching models \citep[e.g.,][]{Behroozi12}. 


Other simulation works have recently reported success in reproducing the observed galaxy formation efficiency as a function of mass \citep[e.g.,][]{Brook12,Munshi13,Haas12,Vogelsberger13}. \citet{Vogelsberger13} compute the stellar-to-halo mass fraction in large volume hydrodynamic simulations that use phenomenological star formation and galactic wind prescriptions. They obtain values that are consistent with ours and with abundance matching constraints. We emphasize that for galaxies with $3\times10^{10} < M_{\rm vir}/\Msun < 2\times10^{11}$, radiative feedback fully accounts for the observed stellar-to-halo mass ratios without expelling a majority of the baryons from the halo, but instead by regulating star formation and keeping most of the gas at low densities in the circum-galactic medium (see Figure~\ref{barfracprofiles}). 


\subsection{Stellar mass assembly and star formation history}
\label{sec:assembly}

\begin{figure*}
 \includegraphics[width=0.49\textwidth]{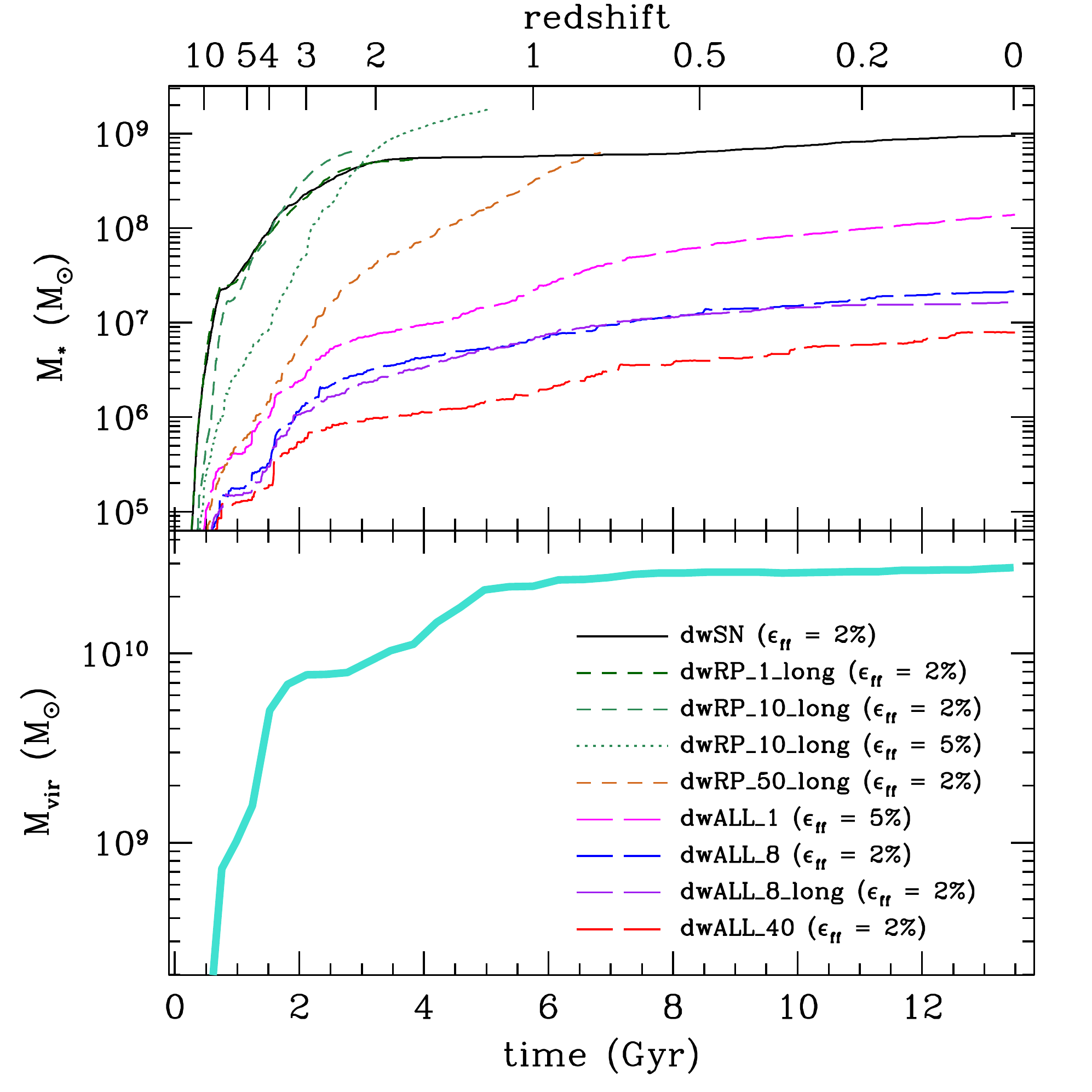} 
 \includegraphics[width=0.49\textwidth]{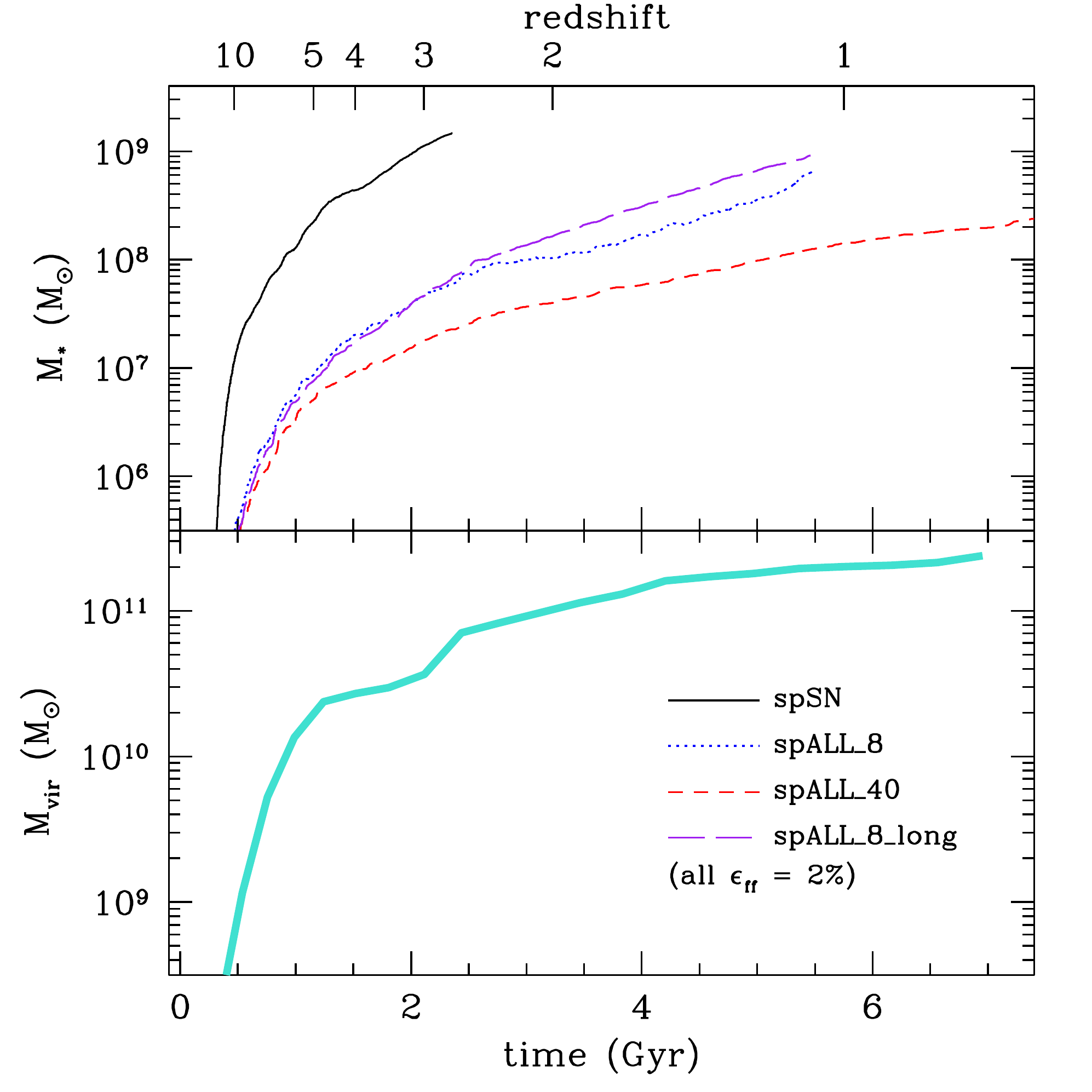} 
 \caption{Stellar and virial mass as a function of time. {\it Left:} \texttt{dwarf} models. {\it Right:} \texttt{spiral} models. Radiation momentum alone only delays the growth of stellar mass at early times. However, low-mass galaxy models that include stellar radiation pressure and photoheating show a dramatic reduction in the stellar mass growth at all epochs compared to a model with only supernovae feedback. Moreover, only galaxies that include the effect of photoheating tend to build up more than half of their stellar component after $z=1$, displaying nearly constant star formation rates in agreement with observations of Local Group dwarf irregulars. The stellar mass growth of galaxies with radiative feedback is rather insensitive to large variations in the model parameters. }
 \label{SMvstime}
\end{figure*}

In this section we present the stellar mass assembly histories of the simulated galaxies. Figure~\ref{SMvstime} compares the stellar mass as a function of time for all the models. The stellar mass assembly rate of all the dwarf galaxies with full radiative feedback is nearly constant in time with a tendency towards faster growth at late epochs ($z<1$) for models with smaller photoheating values. In contrast, the models with only SN feedback show a period of extremely fast growth in the initial $\sim 3$ Gyr, where most of the stellar mass is formed, followed by very slow growth thereafter. Including only SN energy and radiation momentum with low IR optical depth results in a small reduction of the stellar mass at high redshift followed by rapid growth at $z \sim 1$. The figure indicates that the large reduction in the present-day stellar content of models with full radiative feedback vs. models with only supernovae energy results from a reduced star formation rate since the time of the onset of star formation until $z=0$. Furthermore, photoheating pressure from young clusters is the main driver of the large suppression of growth at early times $(z>2)$, when the galaxies with SN feedback and radiation momentum grow rapidly due to the large gas accretion rates at high redshift. A combination of these two effects, the drop in the star formation rate (SFR) at all epochs, and the reduction of the early runaway star formation, result in dwarf galaxies that assemble a large fraction $(\sim 50\%)$ of their stars late $(z<1)$, in agreement with the observed phenomenon of galaxy ``downsizing" \citep[e.g.,][]{Baldry04,Noeske07a,Whitaker12}. The more massive \verb#spiral# model with stellar radiation pressure and photoheating behaves similarly to the dwarf galaxies, with a highly suppressed stellar mass growth at early times, continuing as far as the simulations have been run. 

The bottom panels of Figure~\ref{SMvstime} show the virial mass of the galaxy as a function of time. The stellar mass of the SN simulations closely follows the rate of assembly of the dark matter, both in the overall shape and in the individual episodes of fast growth due to mergers. In contrast, during the epoch of fast growth of the virial mass, stellar radiation is able to delay the conversion of newly accreted gas into stars, decoupling the growth of the galaxy from the growth of the dark matter halo. Moreover, radiative feedback also delays the episodes of stellar mass growth due to mergers by $\sim 1-2~\rm{Gyr}$ relative to the growth of the total mass. 

As discussed in Section~\ref{sec:intro}, most high resolution galaxy formation simulations to date which are able to reproduce the properties of present day galaxies suffer from the overproduction of stars at high redshift. This problem is especially pronounced for MW-mass haloes and less severe in dwarf galaxy simulations. Our \verb#spiral# halo is a prime candidate to test the growth rate of the stellar component at high redshift against observations. The top right panel of Figure~\ref{SFH} shows the semi-empirical star formation histories (SFHs) from \citet{Behroozi12} for a typical galaxy with a present-day virial mass $M_{\rm vir} = 10^{11}\Msun$, along with the $1\sigma$ scatter. The agreement with our simulations including  radiative feedback is remarkable, given that there is no fine-tuning of the parameters. The slightly larger SFR in the simulations is not significant since our galaxy has a virial mass two times larger than the Behroozi et al. value, so its SFR is expected to be a few times larger. Hence, radiative feedback from young stars seems to be an essential ingredient in regulating the growth of low-mass galaxies at early times. It is important to note that there is remarkable agreement among the models with radiative feedback. Figure~\ref{SMvstime} indicates that the assembly histories of models with $\epsilon_{\rm ff} = 2$ per cent differ at most by a factor of $\sim 3$. This is much smaller than the variation in the strength of the radiative forcing among the models. The main driver of the higher SFR in \verb#dwRP_1# is the larger local star formation efficiency. In the more massive galaxy the variations are larger possibly due to the deeper potential well. In general, the variations due to the assumed optical depth of the gas and dust are much smaller than the differences that result from the new physics of stellar radiative feedback.

The top left panel of Figure~\ref{SFH} shows the star formation histories of the models calculated in $1~\rm{Gyr}$ time bins to approximate the typical time resolution in observations. The dwarf model with standard supernovae feedback has very large star formation rates at early times when gas accretion rates are high, but the SFR gradually decreases by a factor of $\sim 30-100$ at later epochs. In contrast, star formation rates in dwarfs that include stellar radiation pressure and photoheating are suppressed by a factor of $\sim 100-500$ at $5<z<2$, and stay nearly constant or sometimes increase towards the present. The situation is similar in the \verb#spiral# models, where radiative feedback reduces the early SFR by a factor of $\sim 100-200$ at $z \approx 4$. These reduced early star formation rates in isolated dwarfs are in excellent agreement with resolved stellar population studies of nearby dwarf irregulars \citep{Weisz11a,Weisz11b} as well as with extrapolations of abundance matching models. The observed delay in the star formation at $z>1$ in low-mass galaxies may be caused by radiation pressure ejecting and dispersing cold star-forming gas out of the galaxy and into the halo, where dilute gas is stored in a reservoir that condenses back onto the galaxy after a few billion years ($z<1$) to fuel the late period of star formation. In Section~\ref{sec:gasproperties} we show that the radial distribution of gas supports this conclusion.

On the bottom left panel of Figure~\ref{SFH} we plot the ``burstiness" of the star formation histories. This is quantified by taking the ratio between the SFR calculated in narrow, $20~\rm{Myr}$ bins and the SFR calculated in broad, $1~\rm{Gyr}$ bins. All the dwarfs with full radiative feedback show large bursts that are comparable in  frequency, with typical episodes where the SFR doubles ocurring every few hundred million years, and the largest bursts increasing the SFR by a factor of $\sim 5-10$. All the simulations with supernovae feedback have much smoother star formation histories with less pronounced and less frequent bursts. The broad SFR peaks in this model likely correspond to major mergers (since we do not distinguish here between in-situ and ex-situ star formation). For the higher mass \verb#spiral# models, the bottom panel of Figure~\ref{SFH} shows that in all the runs with radiative feedback the amplitude and frequency of star bursts are similar and less intense than in the dwarf galaxy analog runs, reaching about $2-3$ times the SFR averaged over $1~\rm{Gyr}$ bins. 

\subsubsection{Discussion}
\label{sec:SFRdiscussion}

Star formation histories that peak at high redshift are ubiquitous not only in simulations of massive galaxies \citep[e.g.,][]{Governato07,Agertz11,Scannapieco12,Calura12}, but they also seem to be generic in dwarf galaxy simulations. In a simulation of a galaxy with $M_{\rm vir} \sim 2\times10^{11}\Msun$, and a feedback model tuned to match present day galaxies, \citet{Governato07} obtain a star formation history that peaks at $z \sim 1$ and declines by a factor of $2$ by $z=0$. Their physical model does not include radiation pressure. Instead, they increase the effect of thermal SN feedback by delaying gas cooling in star-forming regions using the ``blastwave" model \citep{Stinson06}. In more recent work, \citet{Christensen12b} present SPH simulations of the formation of dwarf galaxies of mass $\sim 10^{10}\Msun$, which is close to the virial mass of our dwarf halo. Even after including star formation in molecular hydrogen (which reduces the star formation efficiency in low metallicity gas), the SFR reaches a maximum in the first $\sim 6-8$ Gyr of cosmic time, with a slow decline of a factor of $\sim 2$ thereafter. Using a similar feedback prescription, \citet{Brook11} simulate the formation of a dwarf galaxy and obtain a star formation rate that peaks at $z \sim 1$ and quickly declines by a factor of ten at low redshift. In recent work, \citet{Brook12} performed SPH simulations of low mass galaxies that try to mimic the effect of radiation pressure by boosting the supernovae energy yield to unrealistic values. Their dwarf galaxies show SFHs that are nearly constant. However, their halo with $M_* \sim 10^{11}\Msun$ overproduces stars at $z>1$, with ${\rm SFR} \sim 0.5 \Msun~\rm{yr}^{-1}$ at $z=2$, a factor of $\sim 10$ larger than the prediction by \citet{Behroozi12}, and ${\rm SFR} \sim 0.1 \Msun~{\rm yr}^{-1}$ at $z=3$, a factor of $\sim 5$ larger than in \citet{Behroozi12}. Thus, increasing the efficiency of SN feedback to unrealistically high values does not seem to reduce the early star formation rate sufficiently to reproduce the effect of radiative feedback and the observed growth of low-mass galaxies. 

Resolved stellar population studies of Local Group as well as isolated dwarf irregulars \citep{Weisz11a,Weisz11b} show that their average specific star formation rates are constant in time within the uncertainties. Table~\ref{table4} shows both the SFR and the specific SFR of selected dwarf simulations averaged over the last Gyr to mimic the time resolution of the youngest populations in \citet{Weisz11a}. \citet{Weisz11a} determine an average present-day ${\rm sSFR} \approx 1 \times 10^{-10}~\rm{yr}^{-1}$ for a sample of 43 dwarf irregulars, which is in excellent agreement with the sSFRs of our fiducial dwarf model with full radiative feedback (\verb#dwALL_1#; see Table~\ref{table4}). 

In particular, for NGC2366, which resembles \verb#dwALL_1# in stellar mass and circular velocity (see Section~\ref{sec:circularvelocity}), \citet{Weisz11a} find a present-day ${\rm SFR} \approx 4\times10^{-2} \Msun~{\rm yr}^{-1}$, also in excellent agreement with the value of $2.6\times10^{-2} \Msun~{\rm yr}^{-1}$ that we obtain for this run. 

\begin{table}
\centering
  \begin{tabular}{@{}lccc@{}}
   \hline \hline
   Model          & $z$ & SFR($z$) $(\Msunns~\rm{yr}^{-1})$ & sSFR($z$) $(\rm{yr}^{-1})$    \\
 \hline \hline
 \verb+dwSN+         & 0 & $5.3\times10^{-2}$ &  $6.5\times10^{-11}$  \\
 \verb+dwALL_1+      & 0 & $2.6\times10^{-2}$ &  $2.0\times10^{-10}$  \\
 \verb+dwALL_8+      & 0 & $1.7\times10^{-3}$ &  $7.9\times10^{-11}$  \\
 \verb+dwALL_40+     & 0 & $9.9\times10^{-4}$ &  $1.3\times10^{-10}$  \\
 \verb+dwALL_8_long+ & 0 & $1.3\times10^{-3}$ &  $7.9\times10^{-11}$  \\
\hline
 \verb+spSN+         & $3.0$ & $5.6$              &  $1.9\times10^{-9}$ \\
 \verb+spALL_8+      & $1.3$ & $8.9\times10^{-1}$ &  $1.4\times10^{-9}$  \\
 \verb+spALL_40+     & $1.3$ & $9.6\times10^{-2}$ &  $1.2\times10^{-9}$ \\
 \verb+spALL_8_long+ & $1.3$ & $9.3\times10^{-1}$ &  $1.9\times10^{-9}$ \\
\hline \hline 
\end{tabular}
\caption{ Star formation rates for selected runs. Values are averaged over $1~\rm{Gyr}$.
} 
\label{table4}
\end{table}

The specific star formation rate at $z=0$ in our fiducial dwarf galaxy with stellar radiation feedback is $\sim 0.2~\rm{Gyr}^{-1}$, an order of magnitude larger than the value reported by \citet{governato10} for \verb#DG1#, a simulation tuned to produce a realistic isolated dwarf galaxy with a mass $M_{\rm vir} = 3.5\times10^{10}\Msun$. \citet{Salim07} show that the average specific star formation rates of low-mass star-forming galaxies measured using UV light and nebular emission lines are well fit by the relation ${\rm sSFR} = -0.35 (\log M_* - 10.0) - 9.83$ at $z\approx0$. For a galaxy with the stellar mass of \verb#dwALL_1#, this relation gives an average ${\rm sSFR} = 6.8^{+14.7}_{-4.6}\times10^{-10}~\rm{yr}^{-1}$, where the error bars indicate the intrinsic scatter in the galaxy population. Averaged over the last billion years, the specific star formation rate of \verb#dwALL_1# is $2\times10^{-10}~\rm{yr}^{-1}$, in excellent agreement with the observations by \citet{Salim07}. For comparison, the specific star formation rate in the \verb#dwSN# simulation is about a factor of $3$ smaller. In the simulations of the higher mass halo with radiation pressure and photoheating, the sSFR at $z \sim 1$ is $1.2-1.9\times10^{-9}~\rm{yr}^{-1}$, which is in excellent agreement with the $z = 1$ measurements by \citet{Dunne09} and \citet{Gilbank11}, adjusted to a Chabrier IMF as shown in \citet{Avila-Reese11}. Stellar radiation, with photoheating playing the dominant role, is then essential in regulating early star formation and maintaining a large supply of gas to sustain a high star formation rate at $z<1$. This process may also cause the decoupling of the growth of the DM halo from the growth of the stellar component of low-mass galaxies at high redshift. 

\begin{figure}
 \includegraphics[width=0.49\textwidth]{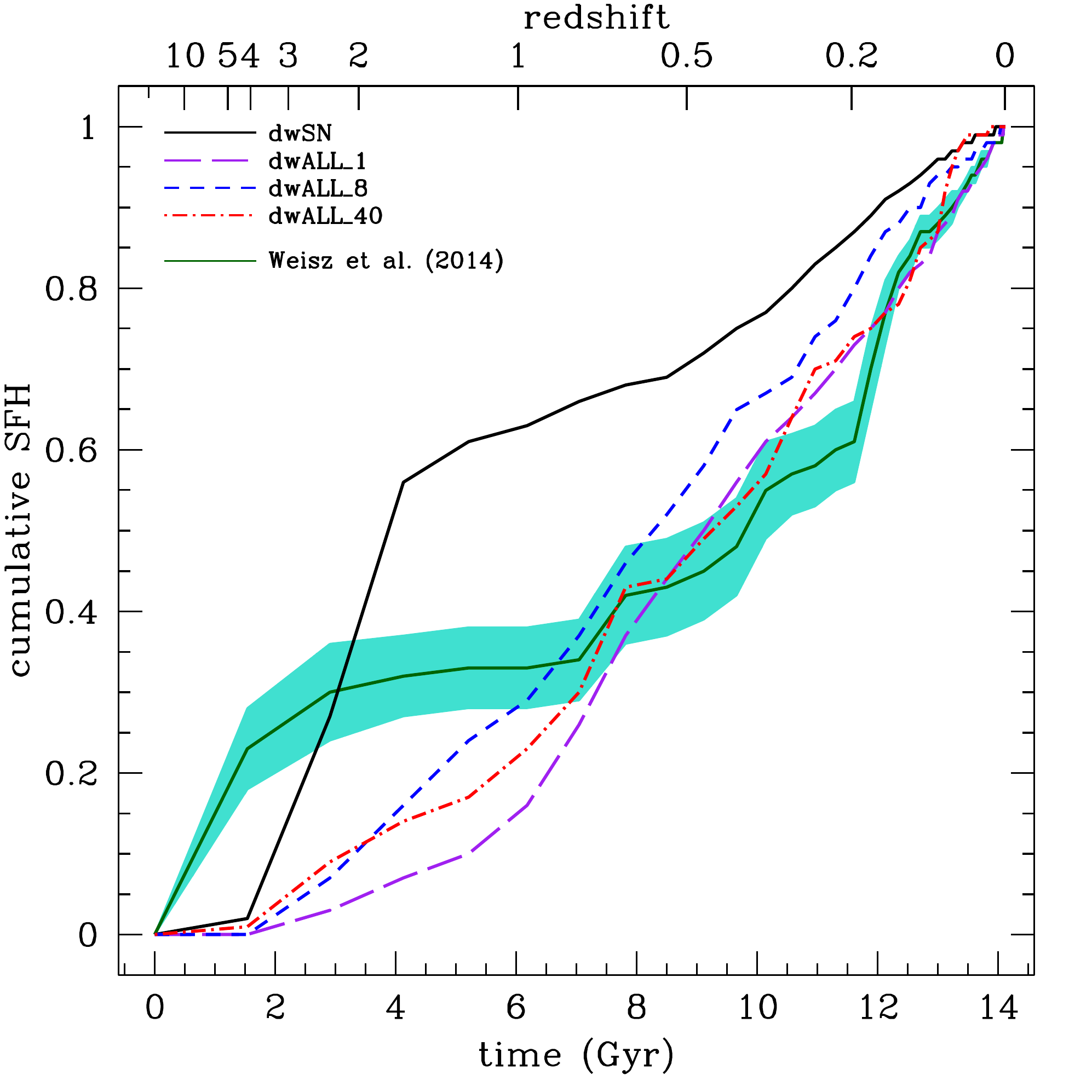} 
 \caption{Cumulative SFH of \texttt{dwarf} runs compared to Local Group dwarf irregulars. The data corresponds to the stellar mass formed prior to a given time averaged across all the dIrrs in the \citet{Weisz14} sample. The shaded area shows the standard error in the mean. Only the simulations with radiation pressure and photoheating reproduce the stellar mass growth st $z<1$. The statistical uncertainties in the observed data (not shown) are larger at $z>1$.} 
 \label{weisz14}
\end{figure}

\citet{Weisz14} assembled the most complete set of Hubble Space Telescope observations of Local Group dwarfs and performed a uniform analysis of the star formation histories. They find that dIrrs only formed about $30$ per cent of their stars by $z=1$. Figure~\ref{weisz14} shows our results compared to the average cumulative SFHs of the irregulars from \citet{Weisz14}. Clearly, SN energy alone results in larger than observed stellar masses throughout the history of the galaxy. The simulations with full radiation feedback reproduce the observed SFH at $z<1$, where the systematic uncertainties in the data are the smallest. As deeper and more complete data becomes available, the abundance of the oldest stars in dwarf galaxies should place better constraints on the role of feedback in small halos at early epochs \citep[e.g.,][]{Madau14}.

\begin{figure*}
 \includegraphics[width=0.49\textwidth]{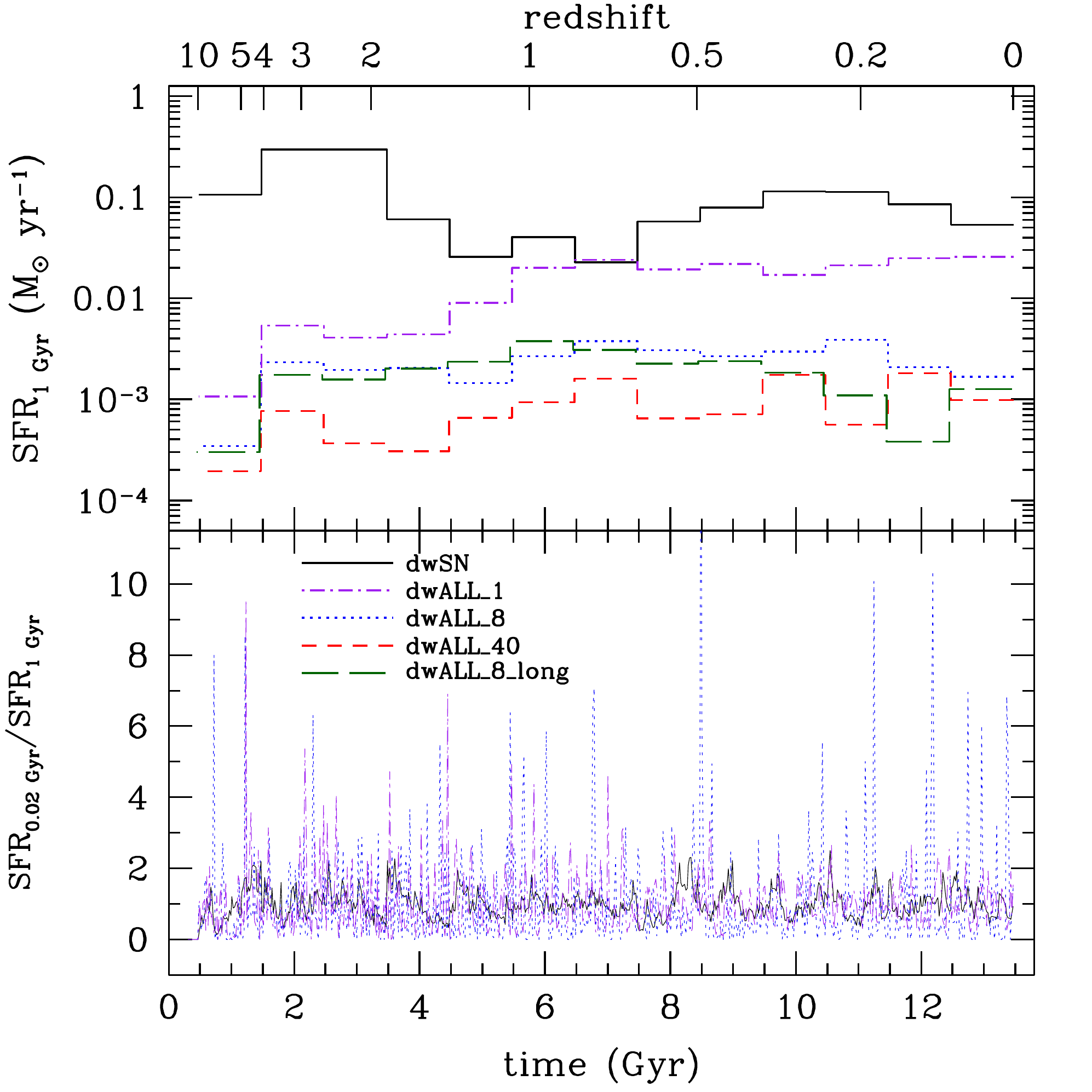} 
 \includegraphics[width=0.49\textwidth]{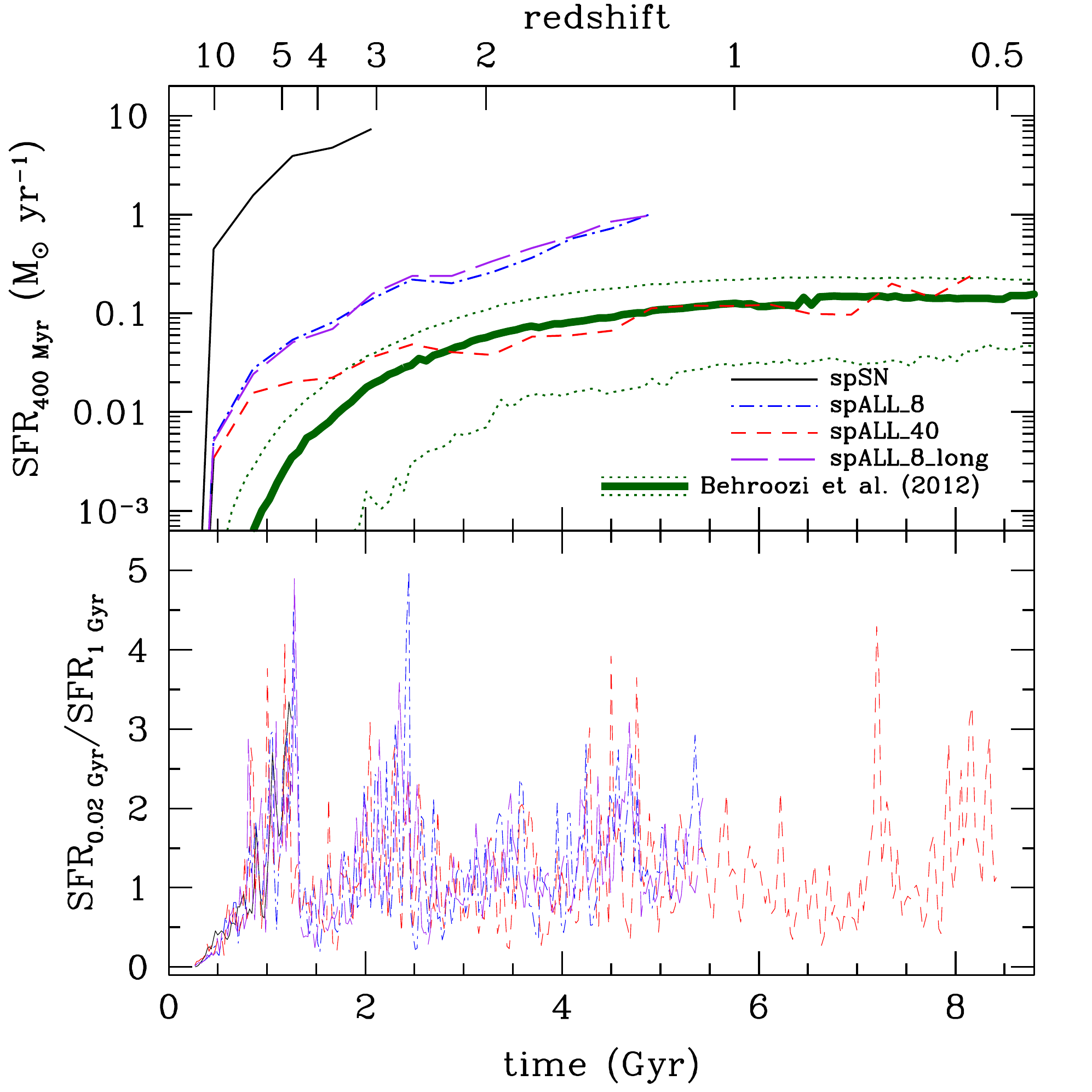} 
 \caption{Star formation histories. {\it Left:} \texttt{dwarf} models. {\it Right:} \texttt{spiral} models. The top panels show the SFHs in broad bins, while the bottom panels show the amplitude of the bursts in $20~\rm{Myr}$ bins relative to the smooth histories. The dwarf galaxy with only SN energy forms stars aggressively in the initial $\sim 3$ Gyr and its star formation rate slowly decreases thereafter. This also occurs when only radiation pressure is added if the IR optical depth is small (not shown). In contrast, including both radiation pressure and photoheating reduces the early SFR by a factor of $\ga 100$, while the galaxies are able to maintain a constant or even increasing rate of star formation until $z\sim 0$, in agreement with observations of dwarf irregulars. Simulations with radiation feedback have burstier star formation histories than models with only SNe, especially for $z<1$. The top right panel compares the SFH of \texttt{spiral} models with a semi-empirical result for a galaxy with $M_{\rm vir} = 10^{11}\Msun$ (solid curve with the $1\sigma$ scatter represented by dotted lines), showing very good agreement only when full stellar radiation feedback is included.} 
 \label{SFH}
\end{figure*}

Current observations do not reach the time resolution necessary to measure the burstiness of individual galaxies. However, 
\citet{Weisz12} fit simple models to observations and determine that galaxies with $M_* < 10^7\Msun$ have bursty star formation histories with amplitude ratios $\sim 30$, while more massive dwarfs are consistent with smooth SFHs. The bottom panel of Figure~\ref{SFH} shows the same qualitative behaviour in our models. There is a trend of increase in burstiness in the star formation histories for galaxies with stronger radiative forcing or lower stellar mass. Furthermore, the model with only SN feedback has the smoothest SFH. In sharp contrast, dwarfs with both radiation pressure and photoheating feature frequent bursts with amplitude ratios as large as $\sim 5-10$ when measured in $20~\rm{Myr}$ bins.

\subsection{Mass distribution}
\label{sec:circularvelocity}

In this section we analyze the detailed radial distribution of each mass component of the simulated galaxies. We use the circular velocity, which is a proxy for mass, and is defined as $v_{\rm circ} = \sqrt{{\rm G}M(<r)/r}$. 

\begin{figure*}
 \includegraphics[width=0.49\textwidth]{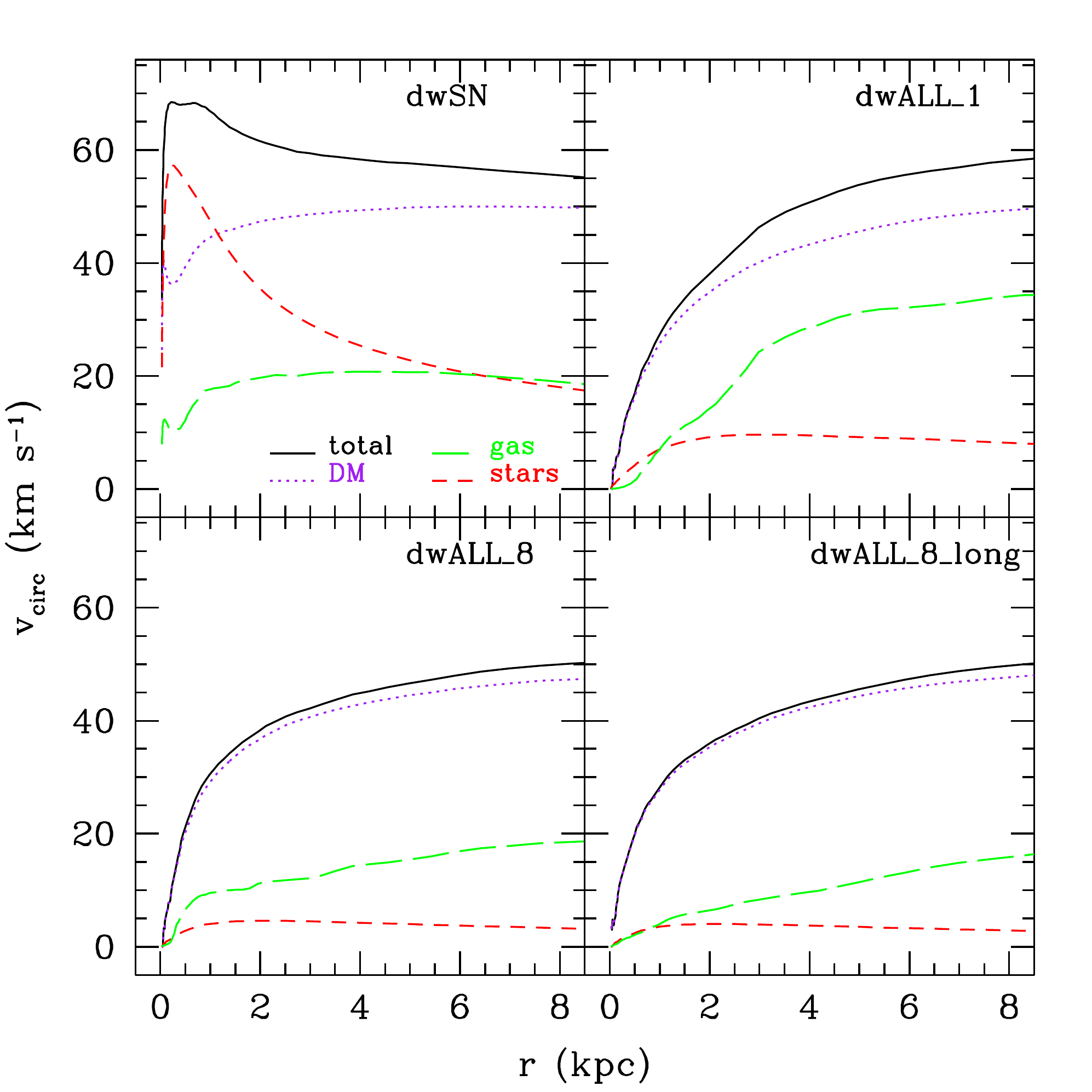} 
 \includegraphics[width=0.49\textwidth]{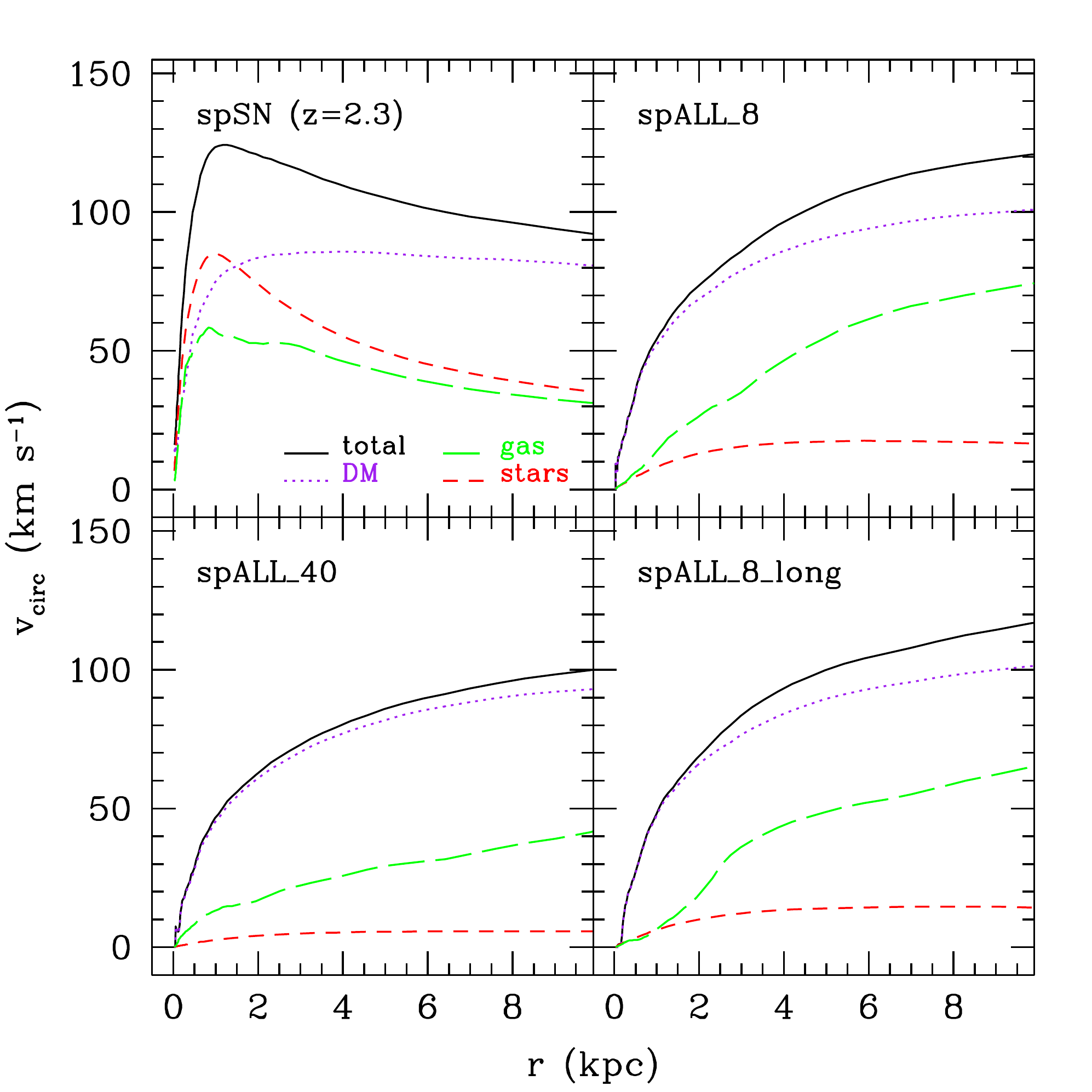} 
 \caption{Circular velocity profiles. The solid curve shows the total, the dotted line shows the DM, and the long and short dashed lines show the gas and stars respectively. {\it Left:} \texttt{dwarf} models at $z=0$. {\it Right:} \texttt{spiral} models at $z \approx 1-2$. The mass distribution in models with only SN energy is dominated by a massive stellar component in the central $1-2\kpc$. Radiative feedback ejects and dilutes star-forming gas and prevents the formation of a concentrated stellar component. There is remarkable agreement in the mass distribution of models with full radiative feedback. All produce slowly-rising, DM-dominated profiles with most of the baryons in the gas phase.} 
 \label{vcirc}
\end{figure*}

Figure~\ref{vcirc} presents the circular velocity curves of simulations with standard SN feedback versus models including full radiation feedback. The figure also shows the contributions from each component, including dark matter, stars, and gas. Both models with only supernovae feedback, \verb#dwSN# and \verb#spSN#, display a strong signature of overcooling in circular velocity curves that are sharply peaked in the central $\sim 1\kpc$ and quickly decrease at larger radii. In contrast to observed rotation curves of low-mass galaxies, these models are dynamically dominated by stars within the inner $\sim 1-2\kpc$. This occurs due to the excessive growth of a concentrated stellar component in the central region of the galaxy. In all the models with full radiative feedback the galaxies are dominated by dark matter within $\sim 10\kpc$ and the contribution of baryons to the mass is small at small radii. Both haloes with radiative feedback also have a reduced stellar component at all radii compared to models with only SN energy. The difference is more pronounced within the central $2 \kpc$, where gas is able to collapse and form stars actively unless feedback can disperse it. Figure~\ref{vcirc} demonstrates that in galaxies with $3\times10^{10} < M_{\rm vir}/\Msun < 2\times10^{11}$  radiation pressure and photoheating not only reduce the total stellar mass of the galaxy, but they preferentially prevent excessive star formation in the central kiloparsec by dispersing and blowing out the cold and dense gas that continuously flows in. Among the simulations with radiative feedback, the mass distribution is quite robust to large changes in the parameters of the radiation feedback implementation, $\tau_{\rm tot}$ and $n_{\rm th}$. In all cases, the galaxies have slowly-rising, DM-dominated circular velocity profiles where gas contributes most of the baryons at all radii. 

Circular velocity profiles allow for direct comparisons with observed rotation curves of low-mass galaxies at $z=0$. The {\sc THINGS} survey \citep{Oh11a} presented detailed observations of dwarf irregular galaxy rotation curves from H\,{\sc i} observations. Using ancillary infrared {\it Spitzer} data, they were able to decompose the mass distribution of gas-rich dwarfs. In general, these detailed models show that dwarf galaxy rotation curves are slowly rising and dominated by dark matter at any distance. In Table~\ref{rotcurvetable} we directly compare our fiducial dwarf galaxy with full radiative feedback with two galaxies from \citet{Oh11a}, NGC2366 and DDO154, which have similar maximum circular velocities to our models.

\begin{table}
\centering
  \begin{tabular}{@{}cccc@{}}
   \hline \hline
 radius ($\rm{kpc}$)  &               & $v_{\rm circ}(r)~(\kmsns)$ &   \\
   \hline
               & \verb#dwALL_1# & NGC2366 & DDO154    \\
 \hline \hline
   1.0         & 25 & 22 & 27 \\
   2.0         & 40 & 33 & 38 \\
   4.0         & 55 & 43 & 52 \\
   8.0         & 65 & 50 & 58 \\
 \hline\hline 
\end{tabular}
\caption{Circular velocity profile of the fiducial dwarf simulation with radiative feedback  compared to two observed dwarf irregulars from \citet{Oh11a}.}
\label{rotcurvetable}
\end{table}

The circular velocity profile of the fiducial model with full stellar radiation feedback is a very good match to rotation curves of observed galaxies of similar mass, NGC2366 and DDO154, except perhaps in the central $\sim 1\kpc$, where the rise in the circular velocity of the fiducial run is slightly steeper. In addition, \citet{Oh11a} determine maximum circular velocities for the stars and cold gas in NGC2366 of $\sim 15\kms$ and $28\kms$, respectively. For DDO154 they find maximum values $v_{\rm circ} \sim 7\kms$ for the stars, and $v_{\rm circ} \sim 20\kms$ for neutral atomic Hydrogen. Our fiducial model with radiative feedback has a maximum stellar $v_{\rm circ} \sim 10\kms$, and $v_{\rm circ} \sim 35\kms$ for gas at any temperature.

\subsection{Do baryons influence the DM distribution?}
\label{sec:dprofile}

There has been a debate in the literature over the past 20 years about the steepness of the inner DM density profile of galaxies. The ``cusp-core problem" originates from the standard prediction of CDM models that the density of DM should increase steeply (as $r^{-\alpha}$, with $\alpha \sim 1$) towards the center of the halo at the smallest scales that can be probed by simulations \citep{flores94,Moore94}. Ever since this prediction was made, observations have attempted to robustly obtain the density profile of the dark matter in the inner kiloparsec of disc galaxies using detailed mass modeling \citep[e.g.,][]{deblok97,deBlok01,Blais-Ouellette04,Rhee04,Simon05,Oh08,KuziodeNaray09,Adams12}. Some works obtain very shallow, constant-density ``cores", while others find steep ``cusps" consistent with CDM. Even though great effort has gone into reducing the systematics, observations using different mass tracers still yield conflicting results. Most recently, \citet{Oh11a} studied a sample of dwarf irregulars using H\,{\sc i} as a tracer and consistently obtained shallow DM slopes, $\alpha \approx 0.3$. On the other hand, using both stellar and nebular gas kinematics, \citet{Adams14} obtain a variety of steeper slope values for different galaxies, covering the range $\sim 0.5-0.8$. 

Several mechanisms have been proposed to explain the observed shallow slopes. While most require revisions of the CDM model, one relies purely on baryonic effects. This scenario involves the ``heating" of the DM due to feedback processes. Analytical models and simulations with strong supernovae ``blastwave" feedback have recently shown that gas blowouts in low-mass galaxies have the potential to alter the distribution of dark matter in the central regions. \citet{Mashchenko08}, \citet{Pontzen12a}, \citet{Governato12} and \citet{Teyssier13} argue that large and frequent bursts of star formation which cause gas blowouts and a rapid oscillation in the potential should transform an initially cuspy dark matter distribution into a shallow core. However, those simulations are different from the ones presented here in two respects. First, none of these works include radiative feedback from massive stars, and second, the star formation histories of simulated dwarf galaxies that include SN feedback with delayed cooling usually peak at $z>1$, whereas the star formation rate of the simulations presented here peaks at $z<1$ (see Section~\ref{sec:assembly}). In fully cosmological simulations the core is already in place at $z \approx 3$ \citep{Governato12,Pontzen12a} or even earlier \citep{Mashchenko08}, so this raises the question of how the mass distribution of the galaxy will respond to radiative feedback and its effect of reducing the SFR a high redshift. 

\begin{figure*}
 \includegraphics[width=0.49\textwidth]{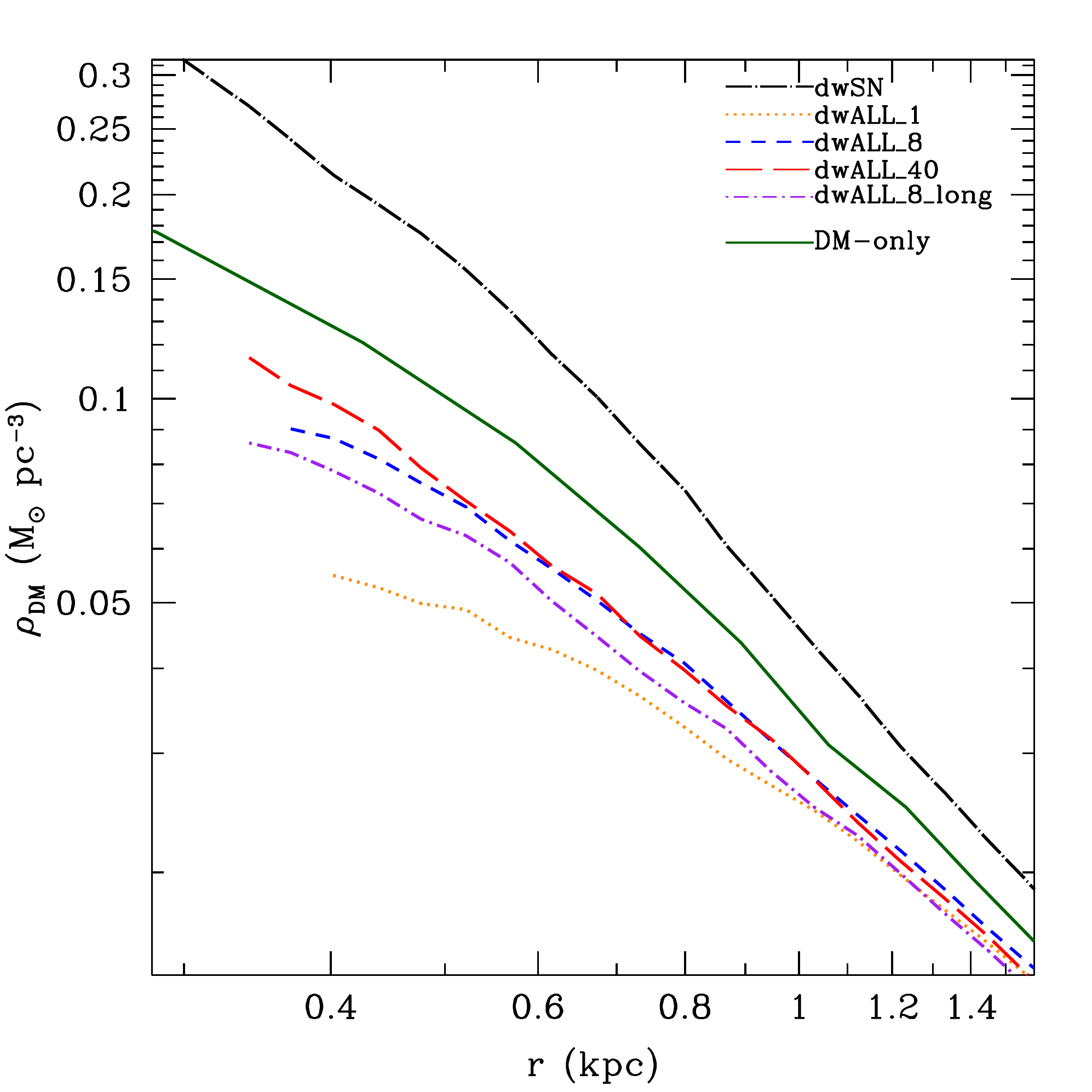} 
 \includegraphics[width=0.49\textwidth]{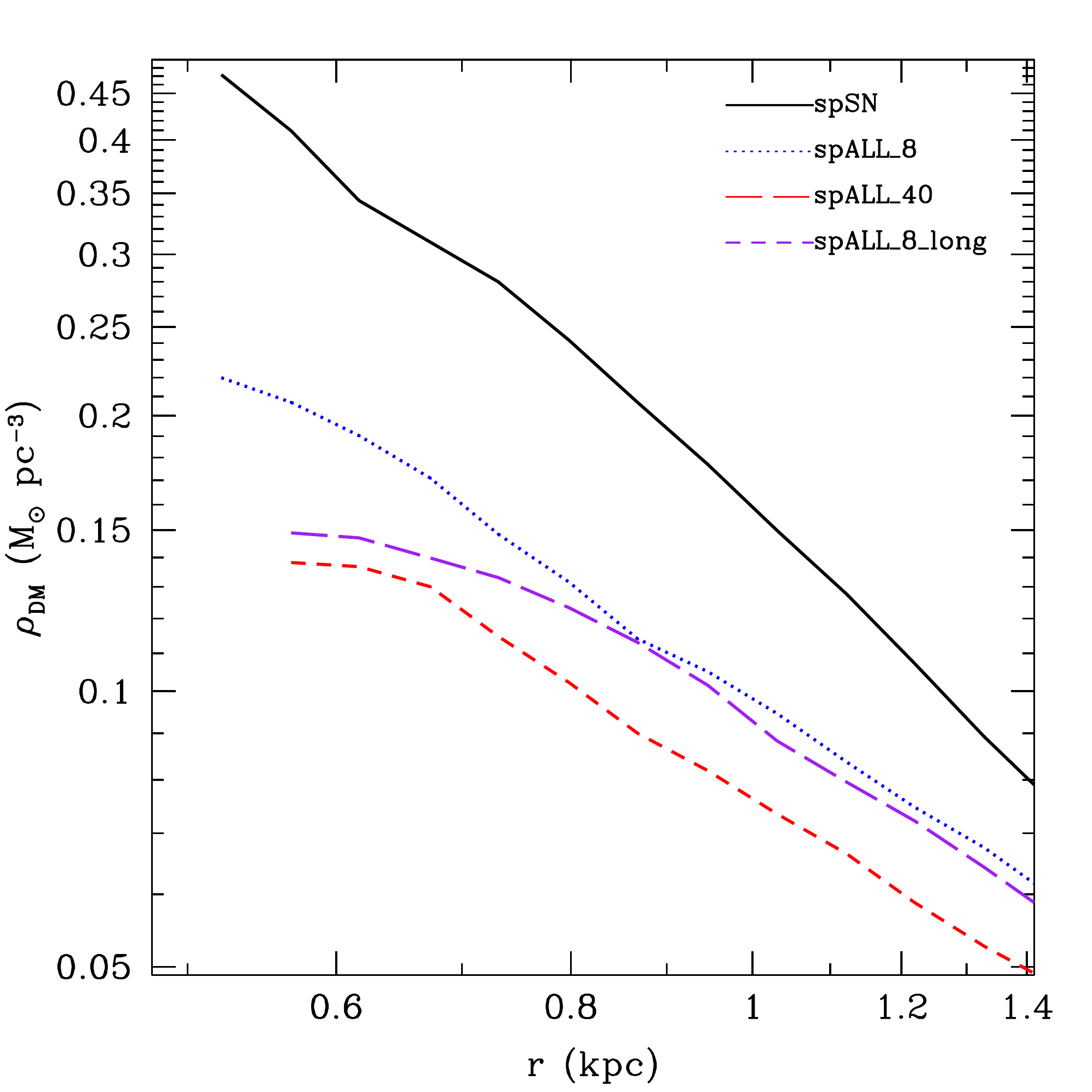} 
 \caption{Inner dark matter density profiles. {\it Left:} \texttt{dwarf} models at $z=0$. The solid line shows the profile of the same simulation run without baryons. {\it Right:} \texttt{spiral} models at $z\approx 1-2$. The central DM mass distribution in models with only supernova explosions is contracted due to the presence of a concentrated massive stellar spheroid, resulting in a steep cusp. Radiation pressure and photoheating from massive stars creates a shallow dark matter core within $\sim 1\kpc$ in the \texttt{dwALL\_1} run as well as in spiral-size haloes with greater photoheating pressure or longer radiative forcing (\texttt{spALL\_40} and \texttt{spALL\_8\_long}).} 
 \label{densitypro}
\end{figure*}

The top panels of Figure~\ref{densitypro} show the inner dark matter density profiles of the simulations.  Several models with full radiative feedback are included to investigate any differences that might arise from the details of the feedback implementation. In addition, for the lower mass halo, we include the profile of the same DM halo simulated without baryons. The profiles are truncated at the radius enclosing $\sim 200$ dark matter particles to avoid numerical artefacts due to resolution in the inner regions \citep{Klypin13}. Clearly, Figure~\ref{densitypro} shows that in our runs with only supernovae feedback, the early runaway star formation in the central kiloparsec increases the central dark matter density compared to the run without baryons. This causes an increase of a factor of $\sim 2$ in the DM density at $700\pc$ and a factor of $\sim 3$ at $400\pc$. Halo contraction has been studied in depth in analytical works as well as simulations \citep{Blumenthal86,Gnedin04,Tissera10,Duffy10,Gnedin11}. It results from baryons in the central regions of the halo dragging the surrounding DM into a more concentrated equilibrium configuration.  In sharp contrast with the SN run, the simulations with full radiative feedback show shallower density profiles. However, measuring the slope of the density profile reveals that only the fiducial radiation pressure and photoheating model, \verb#dwALL_1#, has a slope that is shallower than $-1$ at $r \geq 500\pc$. A different behaviour is observed in the galaxies with $M_{\rm vir} = 2 \times 10^{11}\Msun$, where only the simulations with larger photoheating pressure ($P_{\rm PH}/\rm{k}_B \gtrsim 4\times10^7~\rm{K}~cm^{-3}$) or longer duration of radiative forcing are able to reduce the dark matter density within $700\pc$ considerably by $z\sim 1$. This may be a consequence of the smaller relative amplitude of the star formation bursts or the deeper potential well compared to the dwarf simulations (see Figure~\ref{SFH}). 

Table~\ref{slopestable} shows the logarithmic slope of the dark matter density profile, $\alpha \equiv \frac{ {\rm d} \log \rho }{ {\rm d} \log r  }$. The inner slopes of simulations with radiation feedback are similar to the DM-only run, except in the case of the \verb#dwALL_1# model, where the slope is significantly shallower. We find that as the photoheating pressure is increased, the slopes become progressively steeper and approach or exceed the slope of the run without baryons. \citet{Governato12} suggest that the central dark matter density slope of simulated dwarf galaxies becomes steeper with decreasing stellar mass. 
Table~\ref{slopestable} also shows the slopes of the DM profile at $300< r < 700\pc$ calculated using the \citet{Governato12} relation for the stellar masses of each of our models. In general, stellar radiation feedback yields galaxies with steeper central slopes than the \citet{Governato12} models with effective supernovae blowouts. Moreover, our results are in stark contrast with the effects of the ``early feedback" model used by \citet{Brook12}. Injecting large amounts of SN thermal energy in star-forming regions, they obtain nearly flat DM inner density profiles with more than order of magnitude lower density than the DM-only runs for galaxies with $M_* < 4 \times 10^9 \Msun$. Using a controlled simulation of a $10^{10}\Msun$ dwarf galaxy using thermal supernovae feedback and delayed gas cooling, \citet{Teyssier13} find that star formation bursts with peak-to-trough ratios $\sim 5 - 10$ are necessary to produce the gas blowouts that cause the formation of a shallow DM core. Our dwarf simulations with radiation pressure and photoheating undergo bursts that typically reach amplitudes $\sim 3-5$ relative to the mean SFR (see Figure~\ref{SFH}). Hence, even though our radiative feedback is very effective at regulating the assembly of the stellar mass, our models have steeper dark matter profiles than simulations with extremely efficient SN feedback.

To compare our simulations directly with observations, Table~\ref{slopestable} shows the slope of a power-law fit to the density profile between $300$ and $700\pc$ in our main runs. The table also includes values from mass-modelling of two dwarf irregulars from the THINGS survey, DDO154 and NGC2366, obtained using H\,{\sc i} and $3.6\mu$ observations \citep{Oh11a}. These galaxies were chosen because their $v_{\rm max}$ is close to our dwarf simulations. In the observations the slope was obtained from a fit to the profile within $1\kpc$. Even though the fiducial model with radiative feedback,  \verb|dwALL_1|, has the shallowest profile, it is considerably steeper than observed dwarf irregulars of the same mass. Note, however, that the logarithmic slope contains only partial information about the density profile. A comparison of the actual central density is more meaningful and less sensitive to noise. For instance, the dark matter density of NGC2366 at a distance of 500 pc is $\rho_{\rm DM} \approx 0.04\Msun~\rm{pc}^{-3}$, while for DDO154 the density at the same radius is $\approx 0.05\Msun~\rm{pc}^{-3}$. The only simulated dwarf galaxy that shows a shallow profile, \verb#dwALL_1#, has a central density $\rho_{\rm DM} = 0.052\Msun~\rm{pc}^{-3}$ in excellent agreement with these values.  

\begin{table}
\centering
  \begin{tabular}{@{}lccc@{}}
   \hline \hline
    Galaxy           & $\alpha$ & $\alpha_{\rm G12}$   \\
 \hline \hline
   \verb|dwALL_1|      & $-0.59$ & $-0.46$  \\
   \verb|dwALL_8|      & $-1.10$ & $-0.73$  \\
   \verb|dwALL_40|     & $-1.45$ & $-0.89$  \\
   \verb|dwALL_8_long| & $-0.84$ & $-0.78$  \\
   \verb|dwDM|      & $-1.14$ & -        \\
 \hline
   DDO154            & $-0.29\pm0.15$ & -  \\
   NGC2366           & $-0.32\pm0.10$ & -  \\
 \hline\hline 
\end{tabular}
\caption{Central dark matter profile slope of the simulations compared to the fit by \citet{Governato12} and to the mass models of two observed dwarf irregulars from \citet{Oh11a}. The slope in the simulations was obtained by fitting a power-law to the spherically-averaged density profile at $300<r<700\pc$. In the observations the slope corresponds to a fit for $r<1\kpc$. }
\label{slopestable}
\end{table}

\begin{figure*}
 \includegraphics[width=0.49\textwidth]{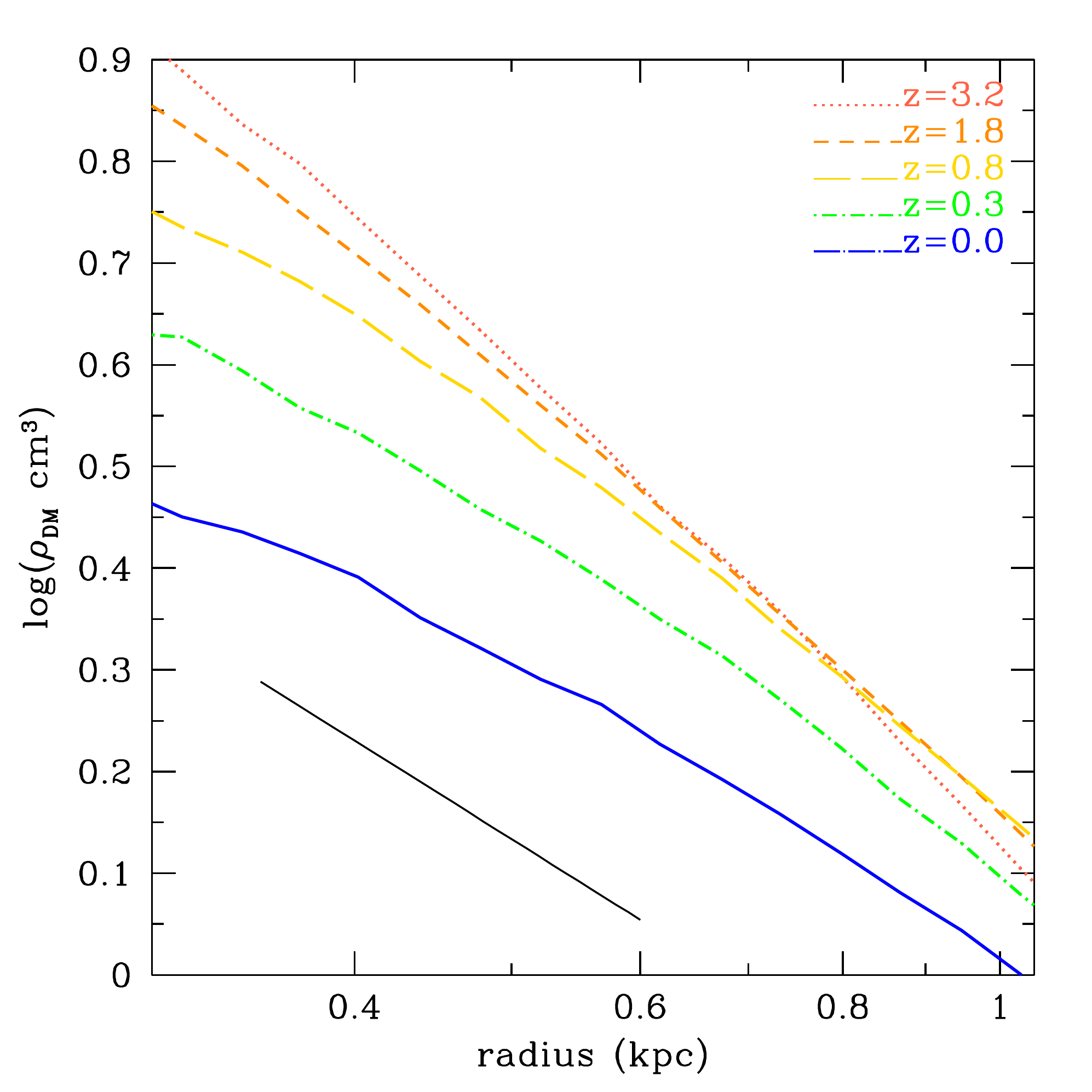} 
 \includegraphics[width=0.49\textwidth]{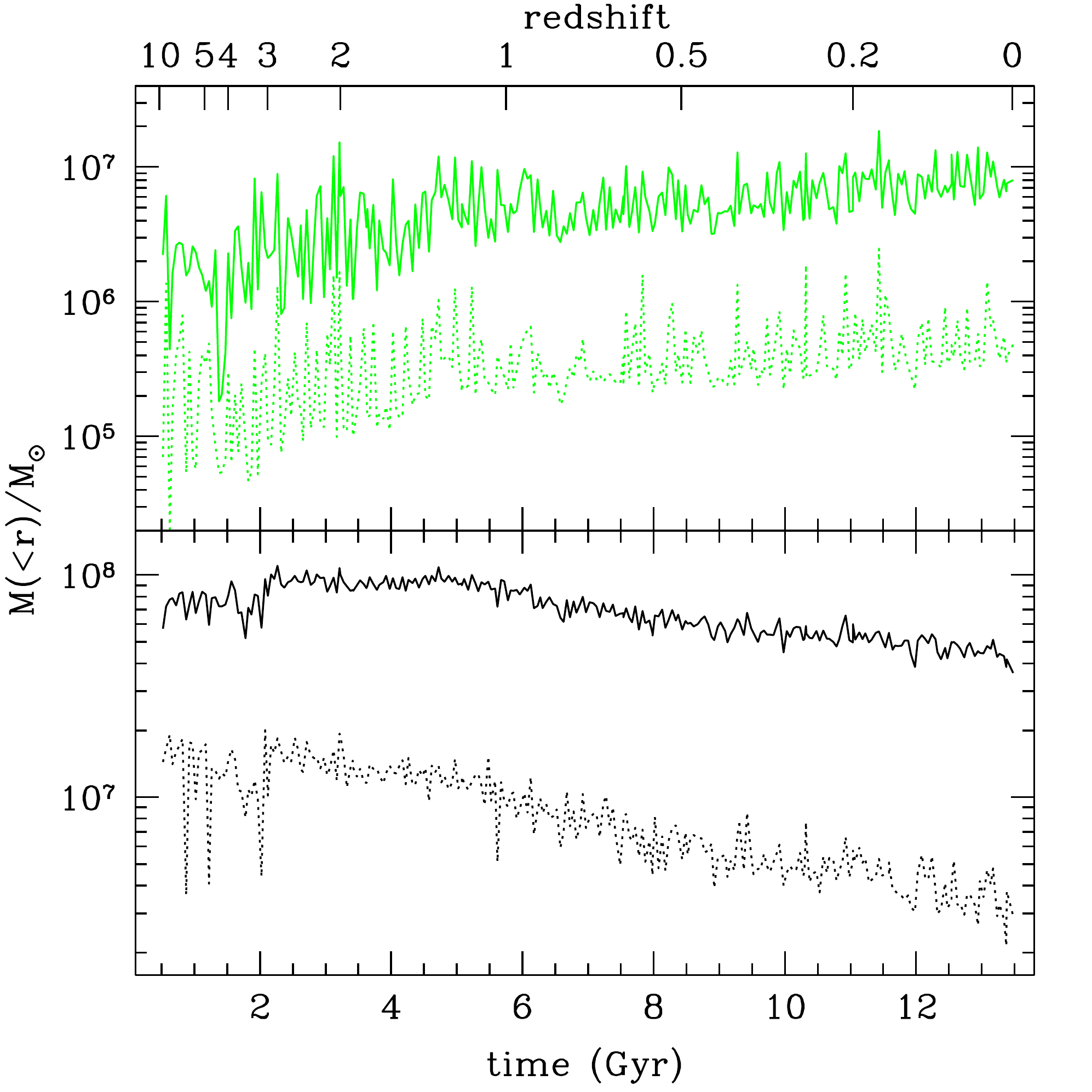} 
 \caption{ Evolution of the inner mass profile of the fiducial dwarf simulation with radiation pressure and photoheating, \texttt{dwALL\_1}. {\it Left:} Evolution of the DM density profile. The straight line corresponds to $\alpha = -1.0$. Note that the slope at high redshift is steeper than $-1$. {\it Top right:} Baryonic mass (gas + stars) contained in the inner $500$ (solid line) and $200\pc$ (dotted line) calculated every $50~\rm{Myr}$. {\it Bottom right:} Same for total mass (baryons + DM). The central DM density decreases significantly after $z = 0.8$. Even though there are order-of-magnitude  variations in the baryonic mass at early times, the dynamical mass only changes by a factor of $\sim 2-3$. } 
 \label{densityproevo}
\end{figure*}

Figure~\ref{densityproevo} shows the evolution of the inner dark matter density profile of \verb|dwALL_1| since $z\sim 3$. There is a fast decrease of the density in the inner kiloparsec after $z\sim 1$, indicating that the DM ``core" begins to form at $z\sim 1$ and slowly grows over $\sim 7~\rm{Gyr}$. By $z=0$ the density $400~\rm{pc}$ from the center has decreased by more than a factor of $2$ campared to the density at $z>2$. Only after $z=0.2$ does the logarithmic slope of the density profile become shallower than $-1$. The right panel of Figure~\ref{densityproevo} also illustrates the evolution of the mass enclosed within the inner regions of the galaxy as a proxy for the time dependence of the gravitational potential. Even though there are large fluctuations in the baryonic mass within the inner kiloparsec, the dynamical mass changes by less than 50 per cent on small timescales. 

\citet{Pontzen12a} discussed the origin of the shallow DM density profile in a simulation of a dwarf with a stellar mass $M_* = 3.7\times 10^{10}\Msun$. At $z=4$ the baryons in the simulation  are dominated by gas, and large supernovae blowouts cause fluctuations in the central gas mass $(r \la 0.5\kpc)$ of nearly an order of magnitude on timescales shorter than the dynamical time of the galaxy. Our fiducial model with radiative feedback shows changes in the baryonic mass of comparable amplitude at high redshift, and decreasing thereafter. However, as shown on the left panel of Figure~\ref{densityproevo}, the simulation does not form a ``core" for another several billion years, at $z\la 0.3$. The right panels of Figure~\ref{densityproevo} show the time evolution of the central mass of the simulation. There are large variations in the central mass at early times. The dynamical mass, which drives the potential, shows fluctuations of only a factor of $\la 2$ occuring preferentially during the second half of cosmic history. This behaviour is a result of the way that star formation proceeds when radiation feedback is included. Since the star formation rate is very low at early times, and the central part of the DM halo is already in place, feedback is insufficient to produce large gas outflows. At lower redshift, the SFR rapidly increases by about an order of magnitude and drives increasingly more feedback energy and momentum into the galactic gas. This results in larger blowouts and stronger fluctuations of the central mass on short timescales which eventually reduce the dark matter density at late times. If, as observations suggest, galaxies of the same mass show a variety of DM central slopes, mergers might play a role in determining which ones evolve the steepest profiles. 

\subsection{The effect of radiation on galaxy morphology}

Another observational constraint that is particularly difficult to reproduce in simulations is the thickness of the stellar distribution. State-of-the-art SPH simulations with ``blastwave" supernovae feedback have recently produced disc-dominated galaxies with small bulge-to-total ratios in MW-mass haloes \citep[e.g.,][]{guedes11,Brook12}. However, about $50$ per cent of observed star-forming galaxies with $M_* = 10^{10}\Msun$ are bulgeless, and this fraction increases at lower luminosities \citep{Dutton09}. No simulation to date has been able to form a bulgeless galaxy of the same mass as the Milky Way. For low-mass galaxies this crisis worsens. Blue galaxies with $r$-band luminosities $M_r > -18$ are essentially all bulgeless discs \citep{blanton03b}. The situation changes for dwarf galaxies ($M_* < 10^9\Msun$) in the field, for which \citet{Geha12} find that essentially all galaxies h  have blue colours and ongoing star formation. These isolated dwarfs have irregular and patchy H\,{\sc i} and UV morphologies with thick stellar distributions and axis ratios $b/a \la 0.5$ \citep{Mateo98,McConnachie12}. In summary, the morphology of field galaxies provides very tight contraints for numerical models. Simulations of isolated dwarfs which include all the relevant physics should \emph{always} produce blue, dispersion-dominated star-forming galaxies. On the other hand, simulated field galaxies with $M_{\rm vir} \sim 10^{11}\Msun$ should be thin, bulgeless discs.  In this section we analyse the stellar distribution of the stars in our models and compare them to observations of nearby galaxies. 

\begin{figure}
 \includegraphics[width=0.49\textwidth]{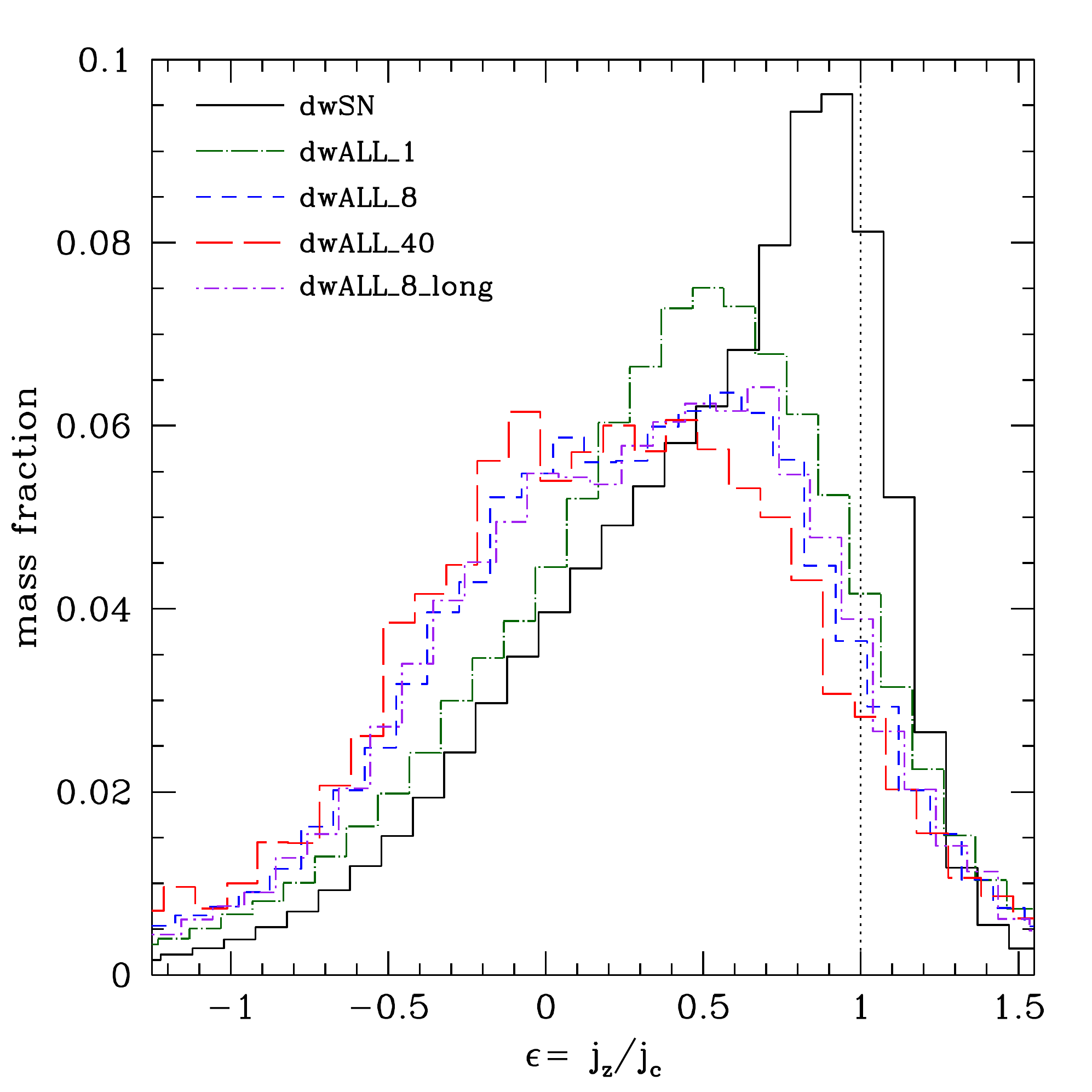} 
 \caption{Mass-weighted distribution of stellar orbital circularity, $\epsilon \equiv j_z/j_c$, where $j_z$ is the specific angular momentum of the stars along the rotation axis of the cold gas disc and $j_c$ is the specific angular momentum of a circular orbit at the same radius. The vertical line marks the location of a circular orbit. Only simulations at $z=0$ are shown. Radiation feedback shifts the peak of the distribution towards $\epsilon \sim 0.5$ and increases the fraction of stars in non-circular and counter-rotating orbits. Runs with larger photoheating values have broader distributions with larger dispersion support.}
 \label{angmom}
\end{figure}

Figure~\ref{angmom} shows the distribution of the angular momentum of the star particles in the \verb|dwarf| simulations. For each particle, the angular momentum is normalized to the angular momentum of a circular orbit at the same radius, $\epsilon \equiv j_z/j_c$. In these distribution, a rotation-dominated galaxy will show a large peak near $\epsilon = 1$, whereas a dispersion supported component would be dominated by stars with $\epsilon < 1$. With only SN feedback, the stars in the dwarf have circularities that peak at $\epsilon \approx 0.9$, with about $18$ per cent of the stellar mass in circular orbits ($0.9 < \epsilon < 1.1$). In the fiducial model with radiative feedback (\verb|dwALL_1|), the peak shifts toward lower values, $\epsilon \approx 0.5$, and the overall distribution of orbital circularity is broader, indicating that the galaxy is less supported by rotation than \verb|dwSN|. The number of stars in nearly circular orbits decreases by $\sim 50$ per cent with radiation pressure and photoheating. In addition, the distribution broadens as the photoheating pressure is increased and the peak shifts to orbits with even lower angular momenta, $j_z/j_c \approx 0.2$, and a larger fraction of counter-rotating orbits as expected in spheroids. 

Heating of stellar orbits is, however, not in conflict with observations of dwarf irregulars since they are characterised by thick stellar discs with small $v/\sigma$ ratios. To quantify the amount of orbital heating due to radiative feedback relative to observed dwarf galaxies, we compare the total angular momentum content of the simulations to robust estimates from observations. \citet{romanowskyfall12} performed a thorough analysis of the total angular momentum content of galaxies as a function of luminosity and morphological type. They find a clear trend of increasing specific angular momentum with increasing stellar mass as well as for later Hubble types at a fixed stellar mass. The data for late type galaxies can be fit using the relation
$\log j_* = \log j_0 + \alpha [\log(M_*/\Msun) - 11.0]$, with $\log j_0 = 3.11$, $\alpha = 0.53$ and $\sigma_{\log j_*} = 0.22$. The total angular momentum within $5\kpc$ in our galaxies simulated with full radiative feedback is $j_* \approx 45 \kpc \kms$ for the fiducial model, $j_* \approx 28 \kpc \kms$ for \verb|dwALL_8|, and $j_* \approx 21 \kpc \kms$ for \verb|dwALL_40|. These are in good agreement with predictions from the fit to observations, $j_* = 38^{+25}_{-15}$, $14.5^{+9.5}_{-5.8} \kpc \kms$ for \verb|dwALL_1| and \verb|dwALL_8| respectively. The simulation with the largest photoheating pressure, \verb|dwALL_40|, contains more angular momentum than the extrapolation from the observations.

\begin{figure}
 \includegraphics[width=0.49\textwidth]{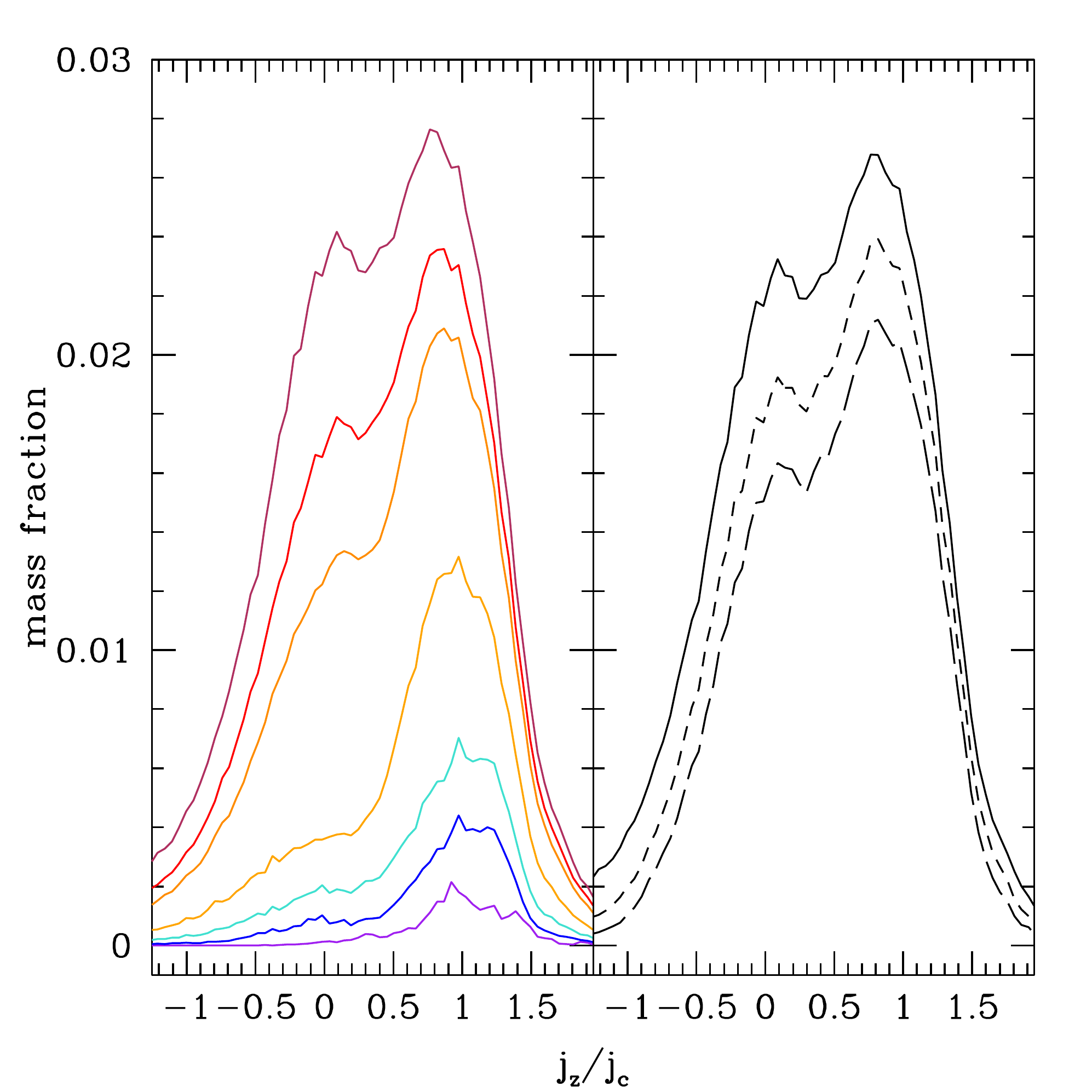}
 \caption{Mass-weighted distribution of stellar orbital circularity for \texttt{spALL\_40} at $z=0.2$. {\it Left:} From top to bottom, the curves show stars with ages $<12, 5, 4, 2, 1, 0.5, 0.05~\rm{Gyr}$. {\it Right:} Effect of removing the innermost stars. All ages are included. Solid, short-dashed and long-dashed curves show only stars with $r > 2, 5, 7~\kpc$ respectively. About $40$ per cent of the stellar mass of the galaxy is in a thickened disc component with non-circular orbits that is composed mostly of stars older than $\sim 4~\rm{Gyr}$. There is no sign of a central dispersion-supported bulge. Only the youngest stars have predominantly circular orbits. }
\label{epsspiral}
\end{figure}

Observations provide much tighter constraints on the stellar distribution of disc galaxies. For the Milky Way, the ratio of disc scaleheight to exponential length is $\sim 0.1$ \citep{Gilmore83}, while nearby late-type galaxies have ratios $\la  0.2$ in the $R$-band \citep{YoachimDalcanton06}.  For the \verb|spALL_40| simulation at $z=0.2$, Figure~\ref{epsspiral} shows the distribution of orbital circularity. In this case, strong radiation feedback causes $60$ per cent of stars to have non-circular orbits. As expected, the left panel of Figure~\ref{epsspiral}, younger stars show a greater degree of rotational support. Typically, in the literature, simulations with a large fraction of stars near $j_{\rm z}/j_{\rm c} \approx 0$ have a prominent bulge which contains essentially all of the low-circularity stars. Interestingly, this is not the case in our simulations with radiative feedback. Instead, in our experiments, the non-rotational component of the galaxy seems to be a part of the stellar disc. This is evident in the right panel of Figure~\ref{epsspiral}, where we compare the circularity distribution of stars located outside a projected radius $2,5,$ and $7\kpc$ to remove any possible contribution from a central bulge. Even when only the disc stars are considered, the pressure supported component remains. 

\begin{figure*}
 \includegraphics[width=0.49\textwidth]{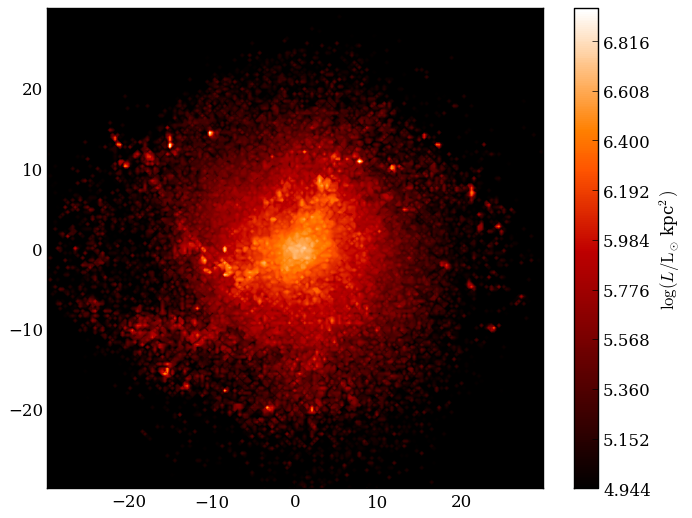}
 \includegraphics[width=0.49\textwidth]{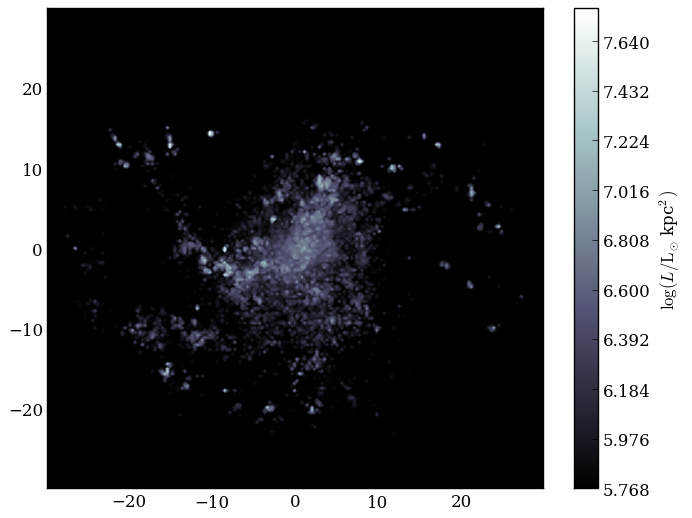}
 \includegraphics[width=0.49\textwidth]{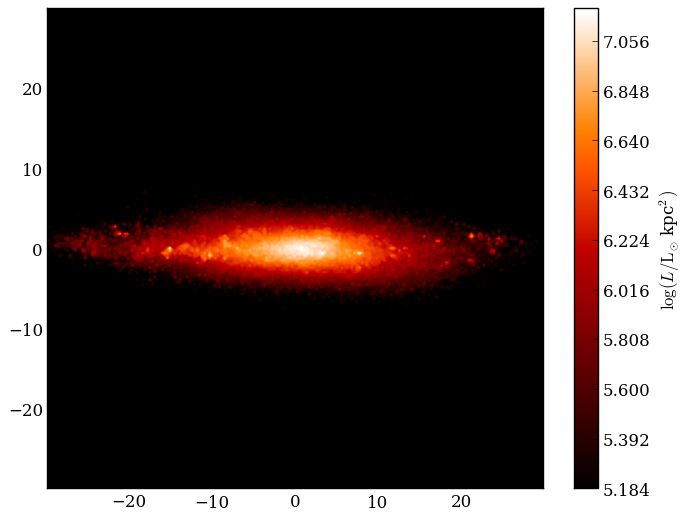}
 \includegraphics[width=0.49\textwidth]{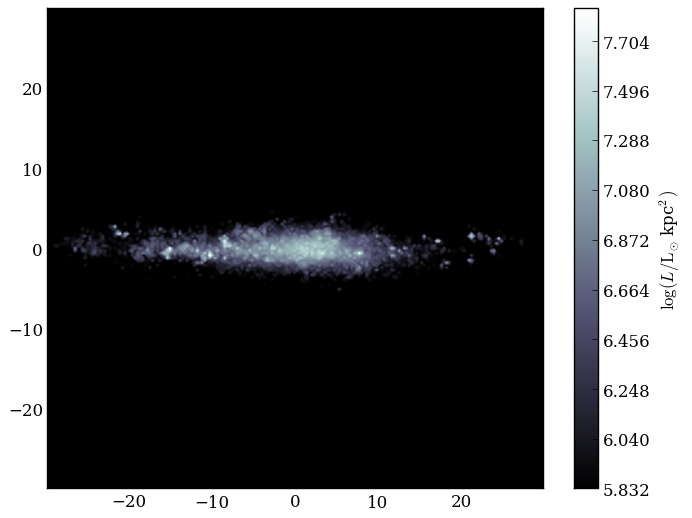}
 \caption{Integrated line-of-sight luminosity for face-on and edge-on projections for \texttt{spALL\_40} at $z=0.2$. {\it Left:} $I$-band. {\it Right:} $U$-band. Scales are in kiloparsecs. Radiation feedback  prevents the formation of a bulge in a low-mass spiral. However, the stellar disc is too large in both radius and height. The thickening is apparent even in the young stars that dominate the UV flux, showing that stars are born far from the plane due to radiation-induced turbulence in the ISM. Luminosities were obtained using the \texttt{CMD2.5} stellar evolution code \citep{Marigo08} and assume solar metallicity.}
 \label{stardensity}
\end{figure*}

Figure~\ref{stardensity} shows the projected luminosity density of the galaxy in both face-on and edge-on projections in the $U$- and $I$-band at $z=0.2$. The mass-to-light ratios were obtained using the \verb|CMD2.5| stellar evolution code \citep{Marigo08}. No central spheroidal component is visible in the infrared edge-on image. 

\begin{figure}
 \includegraphics[width=0.49\textwidth]{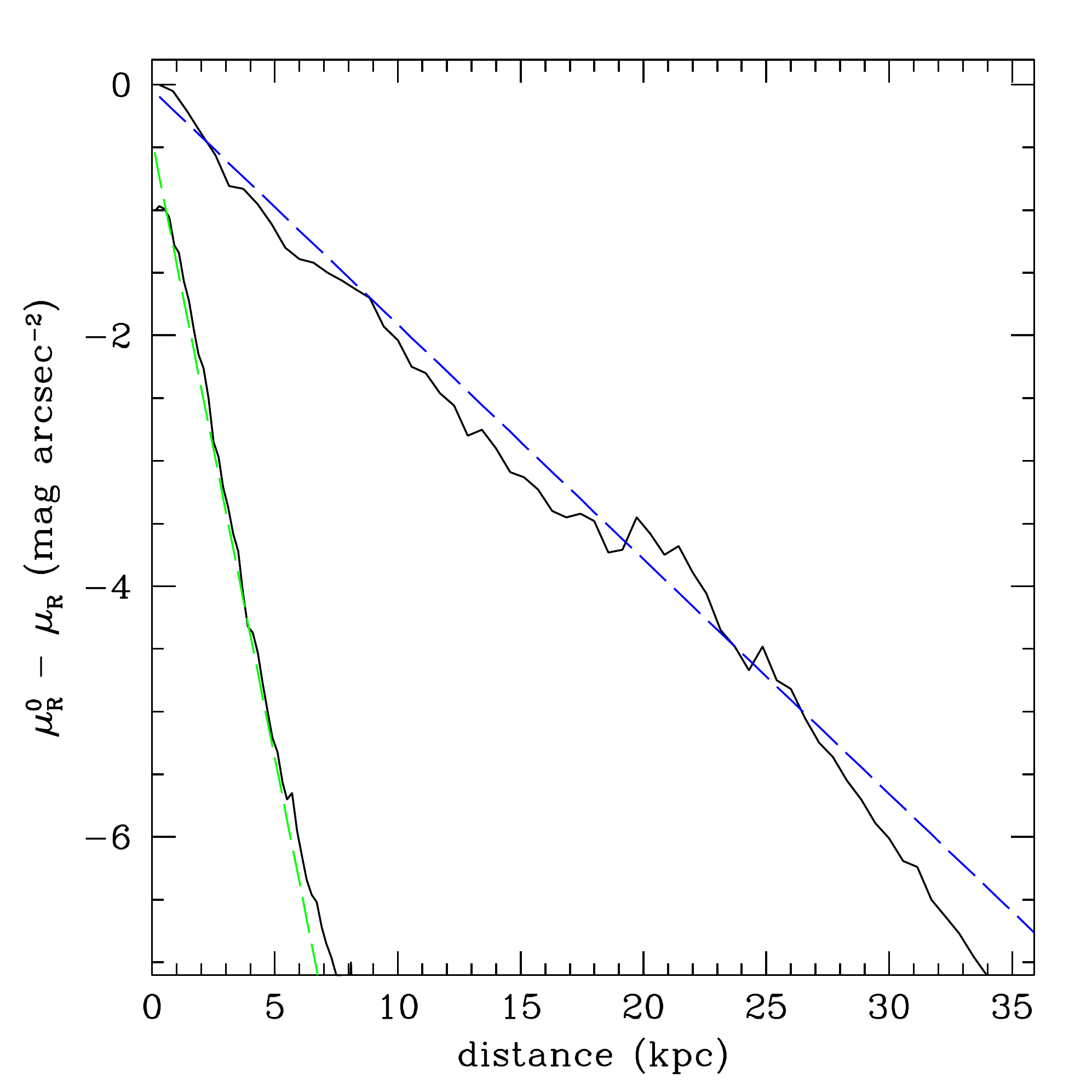}
 \caption{Radial (upper curve) and vertical (lower curve) $R$-band surface brightness profiles for \texttt{spALL\_40}. Dashed lines show exponential fits. The vertical profile is calculated at $6 < r < 9\kpc$. Both the radial and the vertical profiles are well fit by a pure exponential. The disc is truncated at $\approx 25\kpc$.  }
 \label{diskfits}
\end{figure}

To examine the structure of the \verb|spALL_40| model, we fit the radial and vertical surface brightness in the $R$-band with exponential profiles. The results are shown in Figure~\ref{diskfits}. In agreement with the images, the face-on radial profile is fit entirely by a pure exponential, indicating the absence of a bulge. In addition, the disc shows a truncation at $r \approx 25\kpc$. The exponential length in the $R$-band is $r_{\rm d} = 5.8\kpc$, and the scaleheight measured between $6$ and $9\kpc$ is $z_{\rm d} = 1.1\kpc$, giving a ratio $z_{\rm d}/r_{\rm d} = 0.19$. This ratio is consistent with observations and shows that the simulated galaxy is simply more extended than real galaxies in every dimension. We checked the thickness of the star-forming disc in \verb|spALL_40| by fitting the vertical mass profile of the stars with ages $<10~\rm{Myr}$. The star-forming disc has a scaleheight $\approx 400\pc$, or about $5-8$ times thicker than molecular gas in the Milky Way. 

\subsubsection{Discussion}

We have shown that radiation from massive stars significantly affects the distribution of stellar orbits in low-mass galaxies in numerical experiments. However, the effect is opposite to what is observed in other simulations with efficient stellar feedback \citep{governato10,guedes11,Brook12}, where higher resolution and delayed cooling result in more disc-dominated galaxies. When comparing the SN and the fiducial radiation feedback model, we find that the amount of stars in circular orbits with only supernovae feedback is much larger. Radiation pressure increases the amount of random motions in the stellar component and the effect is exacerbated for models with increased photoheating pressure ($P_{\rm PH} > 10^6~\rm{K}~\rm{cm}^{-3}$). In contrast with previous works, our low-mass spiral galaxy with is bulgeless and stars in orbits with low circularity are part of the disc.

In the last decade, many works have focused on the effect of strong supernovae feedback on the morphology of low-mass galaxies. \citet{governato10} show that the distribution of angular momentum in SPH simulations of a dwarf galaxy with $M_* = 4.8\times10^{8}\Msun$ and thermal feedback with delayed cooling agrees well with observational estimates \citep[e.g.,][]{vandenbosch01}. Furthermore, \citet{Christensen12b} show that including a more physical star formation prescription in the same simulations preserves the agreement. Despite the difficulty in directly measuring the distribution of orbital circularity in observations, \citet{Governato07} show that tuning the amount of energy from SN that couples to the gas may dramatically increase the dominance of circular orbits in simulations. 

In a more recent paper, \citet{Roskar13} argue that including radiation feedback in adaptive mesh simulations of the formation of a MW-analog helps to regulate the star formation rate at high redshift. However, this occurs at the expense of creating a kinematically hot disc with a large stellar velocity dispersion. Our results are consistent with \citet{Roskar13} and indicate that there is a limit to the rotational support of the disc when radiation from massive stars is included. For larger forcing due to increased photoheating in H\,{\sc ii} regions, low-mass galaxies show increased support from random motions. This is likely due to the larger velocity dispersion of the star-forming ISM due radiative feedback. This result might indicate the need to include additional physics in the simulations, or it may simply show that our spatial resolution ($50-100\pc$ or about the scale height of the O stars in the Milky Way) is insufficient to produce a thin disc. It may also indicate that large values of radiative forcing do not occur in nature. Our results also indicate that our discs are not only too thick but also over-extended. Star formation at large galactocentric distances could be reduced by, for example, increasing the gas density threshold for star formation. However, a thorough comparison should be carried out before ruling out observational effects. This should include, for instance, performing disc fits of mock observations which include the appropriate observational systematics.


\subsection{How does radiation regulate star formation?}
\label{sec:gasproperties}

\begin{figure*}
 \includegraphics[trim= 1.5cm .6cm 2.0cm 1.5cm,clip,width=0.49\textwidth]{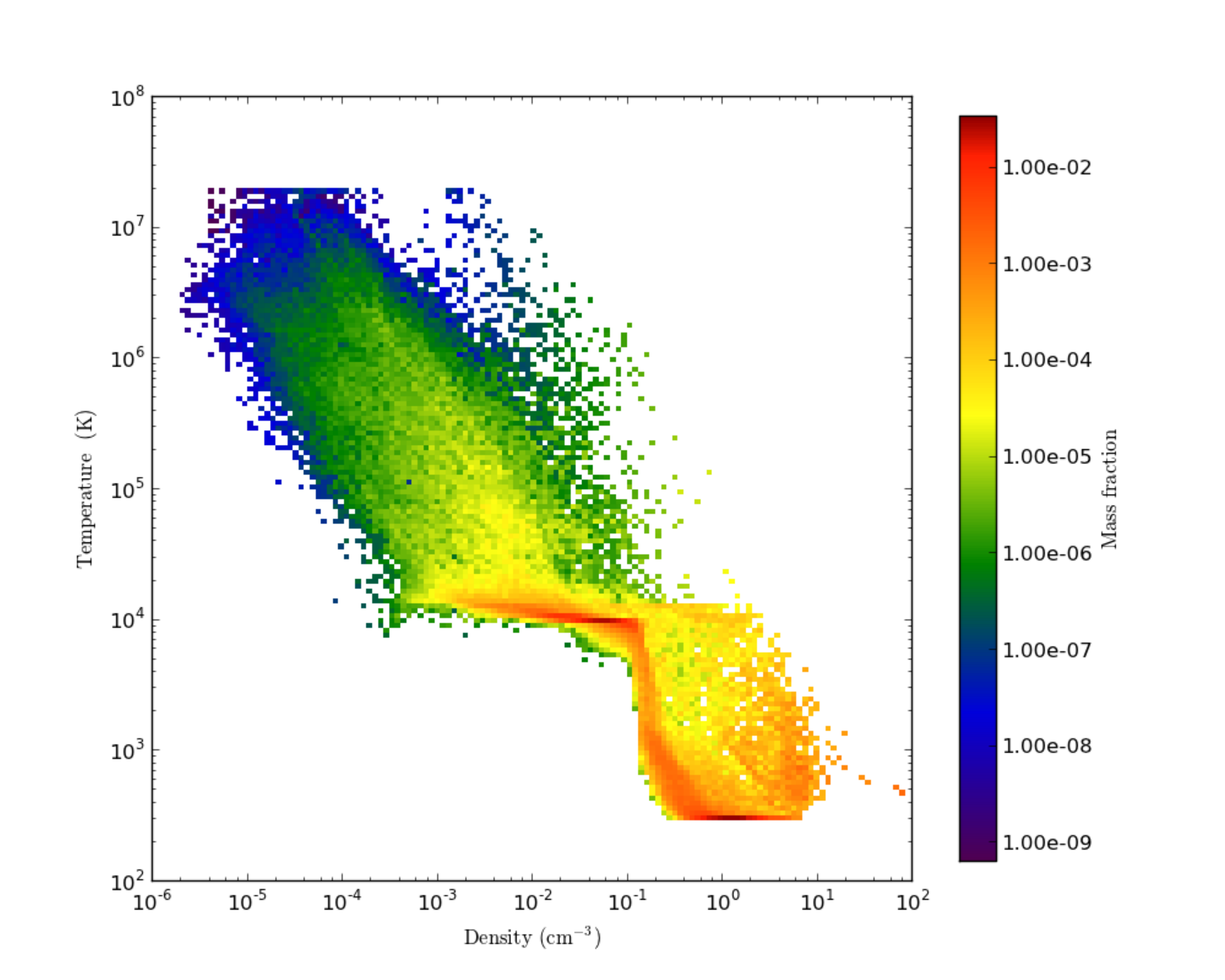}
 \includegraphics[trim= 1.3cm .6cm 2.2cm 1.5cm,clip,width=0.49\textwidth]{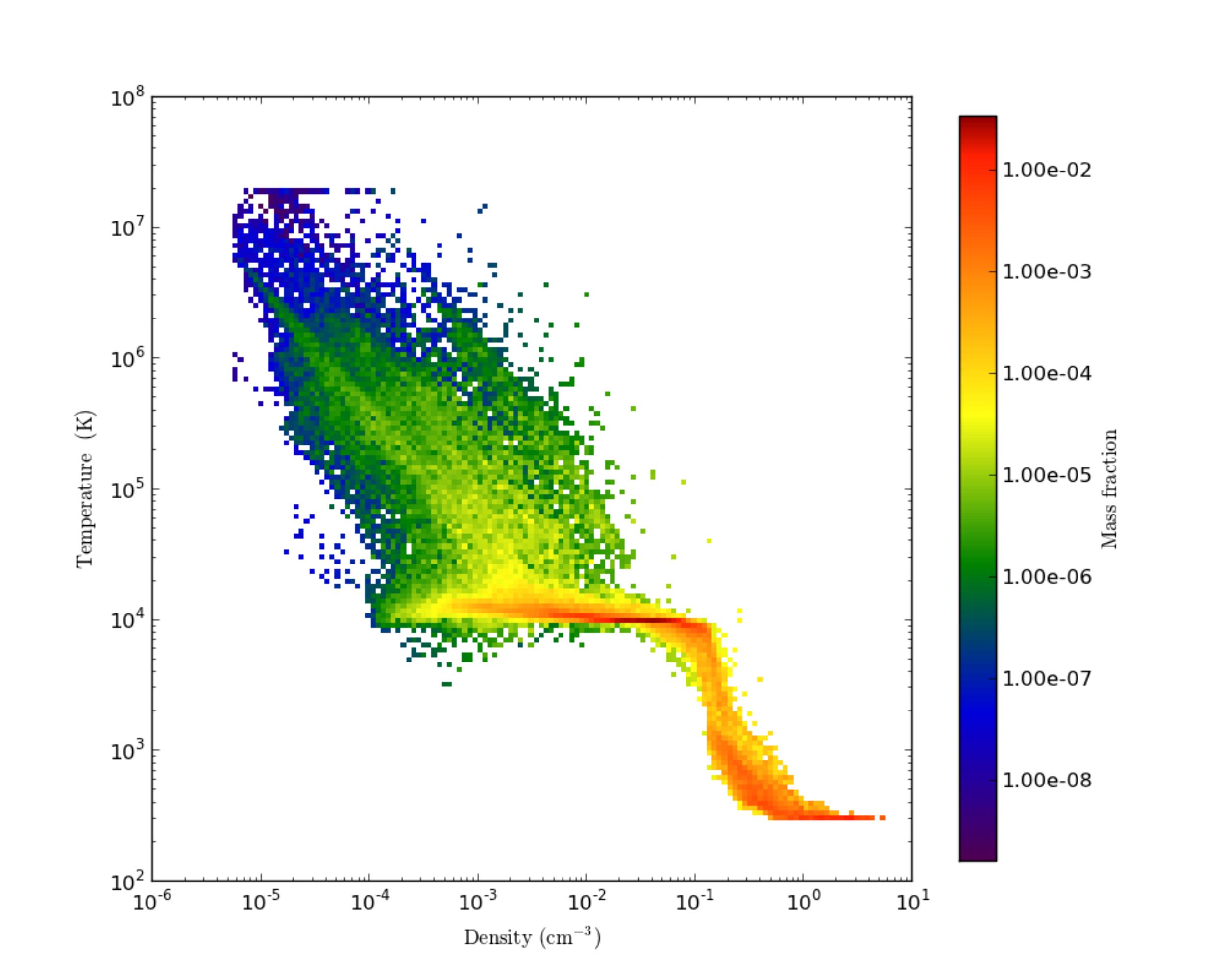} 
 \caption{Density-temperature phase-space distribution of gas mass in dwarf simulations at $z=0$. Only gas contained within a sphere of radius $5\kpc$ around the centre of the galaxy is shown. {\it Left:} \texttt{dwSN}. {\it Right:} \texttt{dwALL\_8}. Radiation feedback from massive stars prevents the accumulation of gas in the cold and high-density tail of the distribution, leading to a reduction of the SFR and ejection of gas from the galaxy. The figure was prepared using the \texttt{yt} analysis and visualization package \citep{yt11}.} 
 \label{phaseplot}
\end{figure*}

In this section, we investigate the properties of the gas in the simulations in order to establish the link between radiative feedback and stellar mass assembly. We begin by calculating the distribution of the gas in the galaxy as a function of density and temperature. Figure~\ref{phaseplot} shows the distribution of gas mass as a function of density and temperature in the ISM of the galaxy $(r<5\kpc)$. The distribution of gas in phase space is similar in all the models, with most of the mass in a narrow locus where feedback heating balances radiative cooling near $10^4~\rm{K}$. Below this temperature, a cold and dense phase with $n>0.1~\rm{cm}^{-3}$ is observed. Above $10^4~\rm{K}$, there is typically a broad tail of warm/hot, very dilute gas. While all the models with radiation pressure and photoheating have similar ISM distributions, simulations with SN feedback contain a large fraction of gas mass in the cold, dense phase where it forms stars actively. This large cold gas reservoir causes the high SFR in Figure~\ref{SFH} as well as the large baryon fraction within the galaxy in Figure~\ref{barfracprofiles}. In sharp contrast, models with full radiation feedback are able to disrupt star-forming gas clouds soon after the first stars are formed, preventing a further increase in the gas density of the ISM of the galaxy. This process greatly reduces the amount of dense and cold gas in the tail of the distribution observed in the fiducial radiation feedback run compared to the \verb|dwSN| model. The phase plot for \verb#dwALL_1# indicates that a large fraction of the cold and dense ISM ($n>0.5~\rm{cm}^{-3}$, $T<10^4~\rm{K}$) gas mass that would otherwise be quickly converted to stars is diluted and expelled from the galaxy.

Radiation from massive stars is capable of dramatically reducing the amount of gas that forms stars by dispersing dense and cold gas clumps that surround star-forming regions. During star formation bursts in dwarf runs the dense ISM is destroyed and star formation is halted in the entire galaxy as observed in Figure~\ref{SFH}. As shown in Figure~\ref{barfracprofiles}, a large fraction of this gas is expelled from the galaxy by means of over-pressured bubbles around young star clusters. In simulations where the photoheating pressure is assumed to be larger, depletion of the ISM is greater across all phases. In a forthcoming paper we discuss the nature of these gas flows and their effect on the circumgalactic medium. 

\subsubsection{Discussion}

In recent works, \citet{Stinson13} and \citet{Brook12} used a combination of delayed cooling and an increase of a factor of $\sim 10$ in the thermal SN energy (compared to the canonical value of $\sim 10^{51}~\rm{erg}$) in cosmological galaxy simulations. This extreme ``early feedback" led to a reduction in the stellar mass growth of galaxies at high redshift. For galaxies with masses $M_{\rm vir} = 3 - 20 \times 10^{10}\Msun$ the simulations presented here have similar star formation histories which are in excellent agreement with semi-empirical models and archaelogical observations of nearby dwarfs. However, as shown in Figure~\ref{phaseplot}, and in Section~\ref{sec:gasproperties}, radiative feedback can effectively disperse and expell cold and dense gas without heating it beyond $\sim 10^5~{\rm K}$. With the delayed cooling used in the ``blastwave" approximation temperatures in star forming gas reach $\sim 2\times10^7 ~\rm{K}$ for several million years. In fact, in controled isolated simulations of a massive spiral galaxy, \citet{Agertz12} demonstrate that thermal energy injection with delayed cooling produces about $100$ times more gas above $10^4~{\rm K}$ in the ISM than momentum from radiation, even if both are able to regulate the star formation rate. 

If hot gas was responsible for the regulation of the SFR it would be ubiquitous around galactic young star clusters. However, observations of nearby evolved H\,{\sc ii} regions detect a negligible volume fraction of hot X-ray emmiting gas \citep{Lopez11,Lopez13}, showing that the effect of shock-heating by SN and stellar winds is short-lived. \citet{Lopez11} and \citet{Lopez13} also estimate the dynamical effect of the observed X-ray gas and find that it is negligible in the context of the dispersal of the molecular cloud. These observations strengthen the argument against the regulation of the star formation in galaxies by large supernovae blastwaves that remain at $T>10^7~{\rm K}$ for several million years. In this paper we have shown that the combination of radiation momentum and photoionization heating is instead a more viable physical mechanism to explain the slow rate of stellar mass growth of low-mass galaxies at $z>1$.

\section{Conclusions}
\label{sec:conclusions}

We have for the first time analyzed the full effects of radiation from massive stars on the formation of low-mass galaxies in cosmological hydrodynamic simulations. In contrast with other works, we explicitly included the important contribution of photoionization heating in H\,{\sc ii} regions in addition to radiation momentum. We find that:

\begin{enumerate}[I.]

\item The local efficiency of star formation observed in nearby star clusters  is not sufficiently high to destroy the parent clouds with energy from supernovae and stellar winds. However, radiation momentum and photoionization heating over-pressure gas and cause it to expand, quickly dispersing the parent gas cloud (Section~\ref{sec:analytics}).



\item The fraction of baryons that condense to form a low-mass galaxy with radiative feedback is $\sim 7$ per cent, about half of the universal fraction, and slightly larger than the baryonic fraction in a galaxy formed with SN feedback alone. This surprising result indicates that stellar radiation may be not be important in ejecting large amounts of gas from the halo. Instead, it acts to regulate the early SFR by dispersing cold and dense gas for several million years (Figure~\ref{phaseplot}).

\item Stellar radiation regulates the star formation in isolated dwarf and low-mass spiral galaxies at all epochs, but especially at $z>1$, where it reduces the SFR by two orders of magnitude compared to SN explosions alone. The star formation histories of simulated dwarfs with radiative feedback are constant or increasing with time, a feature observed in nearby dwarfs that had until now eluded analytical models and simulations. The growth of the stellar mass of a spiral galaxy with $M_{\rm vir} = 2 \times 10^{11}\Msun$ also matches semi-empirical and observational estimates (Figure~\ref{SFH}). 

\item Radiation feedback (mainly due to photoheating) works by delaying the conversion of newly accreted gas into stars, and keeps gas in a reservoir to fuel star formation at $z \la 1$. This mechanism decouples the assembly of the galaxy from the hierarchical growth of the host DM halo (Figure~\ref{SMvstime}). The specific SFR of runs that include radiation momentum and photoheating agree with direct observational estimates at $z=0$ and $z =1$ (Section~\ref{sec:SFRdiscussion}), and with the archaeological data from \citet{Weisz14} (Figure~\ref{weisz14}). Radiation feedback may thus play an important role in producing the phenomenon of galaxy downsizing in sub-$L^*$ galaxies.

\item Low-efficiency star formation reduces the supernovae energy per unit gas mass (Figure~\ref{heating}) and leads to catastrophic overcooling and circular velocity curves that peak at small radii and decline quickly. Pressure from stellar radiation and photoheating overcomes self-gravity and disperses cold gas to prevent overcooling (Figure~\ref{pgradient}). This produces circular velocity curves that rise slowly and are DM-dominated, in excellent agreement with observations of nearby dwarf irregulars (Figure~\ref{vcirc} and Section~\ref{rotcurvetable}). Our simulated spiral galaxy shows no sign of a central bulge in either surface brightness maps (Figures~\ref{stardensity} and \ref{diskfits}), or in the stellar kinematics (Figure~\ref{epsspiral}).

\item Dwarf galaxies simulated with radiative feedback have bursty star formation histories, with larger and more frequent bursts at $z<1$. This is in contrast with models with only SN energy, where the largest bursts occur near the peak of the SFH, at $z \sim 2$ (Figure~\ref{SFH}). Bursty star formation leads to a reduction in the dark matter density in the central $\sim 1 \kpc$ compared to a run without baryons (Figure~\ref{densitypro}). The density and slope agree with estimates from observations of galaxies of the same mass, and confirms the ability of stellar radiation to alter the distribution of DM without the need to artificially increase the SN efficiency. The effect is much smaller in a low-mass spiral galaxy at $z\sim 1$. 

\item Stellar radiation increases the height of the stellar disc in a low-mass spiral. Although the ratio of scaleheight to exponential length fits observations, the scaleheight of the star-forming disc is $5-8$ times greater than in the MW. The thickened disc causes the fraction of stars in circular orbits in the galaxy to decrease by a factor of $\ga 2$ compared to simulations with only supernovae energy. These tensions may be resolved by increasing the SF density threshold in the simulations.

\end{enumerate}

\section{Acknowledgements}

The authors would like to thank the referee for valuable and insightful suggestions which greatly improved the paper. We also thank Mark Krumholz, Oscar Agertz and Andrey Kravtsov for illuminating discussions, as well as Peter Behroozi for providing his data. S.T., A.K. and J.P. were supported by grant STSci/HST-AR-12647.01, NSF grant NSF-AST-1010033, as well as the collaborative grant NSF-AST-1009908. This work was partially supported by MINECO (Spain) - AYA2012-31101 and MICINN (Spain) AYA-2009-13875-C03-02. D.C. is a Juan de la Cierva fellow. K.A. was supported by the National Science Foundation under Grant No. DGE-1144468. All the simulations were performed at the National Energy Research Scientific Computing Center, which is supported by the Office of Science of the U.S. Department of Energy under Contract No. DE-AC02-05CH11231.

\appendix

\section{The Kennicutt-Schmidt relation using local star formation}
\label{sec:appendix}

\begin{figure}
 \includegraphics[width=0.49\textwidth]{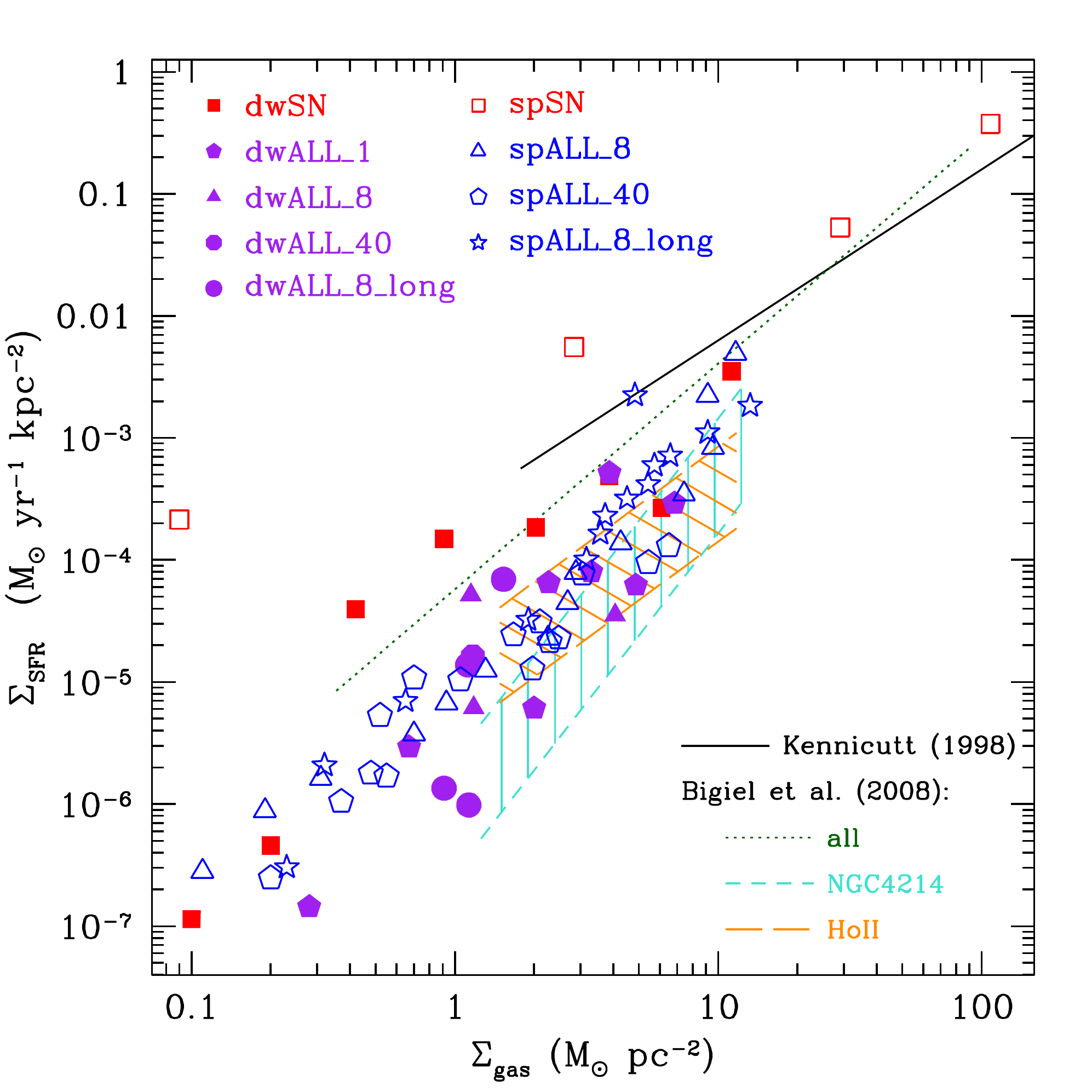} 
 \caption{Surface density of star formation vs. surface density of cold gas with $T<15000~\rm{K}$. Filled symbols represent \texttt{dwarf} models and open symbols represent the \texttt{spiral} runs. The projection is performed onto a plane of thickness $1~\kpc$ defined by the angular momentum vector of the cold gas disc. Different points for each model show the azimuthal averages in $1\kpc$ annuli at various galactocentric distances. The solid line is the fit to local spirals from \citet{Kennicutt98}, while the dotted line show the fit to spirals and irregulars from \citet{bigiel08}, and the hatched areas represent the fits and scatter for two individual dwarf irregular galaxies. Our local model of star formation results in simulated dwarfs that successfully reproduce observations of individual dwarf irregulars. Stellar radiation pressure tends to reduce the SFR surface density compared to supernovae energy alone.} 
 \label{KSrelation}
\end{figure}

The star formation model used here is based on the observed local efficiency of conversion of gas into stars in molecular clouds that are typically less than $100\pc$ in size. This does not guarantee that the global relation between gas density and star formation rate will follow the Kennicutt-Schmidt law \citep{Kennicutt98} on global galactic scales. Figure~\ref{KSrelation} shows the surface density of star formation as a function of cold gas surface density for several runs. Here, gas with temperature $T<15000~\rm{K}$ is considered cold, and density is projected inside a $1\kpc$ tall cylinder oriented along the total angular momentum vector of the young stars in each case. Each symbol in Figure~\ref{KSrelation} represents an azimuthal average in a $1\kpc$ annulus at a different radius. The figure also reproduces the original fit to global spiral and starburst galaxy data by \citet{Kennicutt98} and the recent fit to $750\pc$ patches in spirals and dwarf irregulars by \citet{bigiel08}. In addition, we plot the fit and $1\sigma$ scatter of individual sets of measurements in $750\pc$ apertures throughout two dwarf irregulars, NGC4214 and HoII, from \citet{bigiel08}. 

Qualitatively, all our simulated dwarf galaxies show overall agreement with the normalization and the slope of the Kennicutt-Schmidt relation. Moreover, it is remarkable that using only a local deterministic star formation prescription, we  reproduce the observed Kennicutt-Schmidt relation on galactic scales. For dwarf runs, the effect on the relation of including additional feedback physics is relatively small. However, the relation between SFR and cold gas surface density for the simulated spiral galaxies with only SN energy lies well above the points for models that include radiation feedback from massive stars. The overall slope of the simulations that include radiation is also generally steeper than those with only SN feedback. Star formation in galaxies that incorporate radiation momentum and photoheating pressure seems to proceed at a slower rate for any given value of cold gas column density, and this behaviour is more marked at low gas surface densities near the observational detection thresholds, $\Sigma_{\rm gas} \la 3\Msun~\rm{pc}^{-2}$. In addition to reducing the SFR density, radiation from massive stars also decreases the maximum column density reached by cold gas in the ISM.

\bibliographystyle{mn2e}
\bibliography{merged}

\label{lastpage}

\end{document}